\documentstyle[prb,eqsecnum,aps,epsf]{revtex}
\begin{document}
\draft
\title{Interaction effects and phase relaxation in disordered systems
}
\author{
 I.L. Aleiner$^{1,2}$, B.L. Altshuler$^{3,4}$,
and
M.E. Gershenson$^5$
}
\address{
$^{1}$Ruhr-Universit\"{a}t-Bochum, Bochum, D-44780, Germany\\
$^{2}$Department of Physics and Astronomy,
SUNY at Stony Brook, Stony Brook, NY 11794\\
$^{3}$NEC Research Institute, 4 Independence Way, Princeton, NJ 08540\\
$^{4}$Physics Department, Princeton University, Princeton, NJ 08544\\
$^{5}$Serin Physics
Laboratory, Rutgers University, Piscataway, NJ 08854-8019
\date{\today}
}
\maketitle

\begin{abstract}
This paper is intended to demonstrate that there is no need to revise the
existing theory of the transport properties of disordered conductors in
the so-called weak localization regime. In particular, we demonstrate
explicitly that recent attempts to justify theoretically that the
dephasing rate (extracted from the magnetoresistance) remains finite at
zero temperature are based on the profoundly incorrect calculation.
This demonstration is based on a straightforward evaluation of the effect
of the electron-electron interaction on the weak localization correction
to the conductivity of disordered metals.
Using well-controlled perturbation theory with the inverse conductance $g$
as the small parameter, we show that this effect consists of two
contributions. First contribution comes from the processes with energy
transfer smaller than the temperature. This contribution is responsible
for setting the energy scale for the magnetoresistance. The second
contribution originates from the virtual processes with energy transfer
larger than the temperature. 
It is shown that the latter processes have nothing to do with the
dephasing, but rather manifest the second order (in $1/g$) correction to
the conductance. This correction is calculated for the first time.
The paper also contains a brief review of the existing experiments on the
dephasing of electrons in disordered conductors and an extended
qualitative discussion of the quantum corrections to the conductivity and
to the density of electronic states in the weak localization regime.
\end{abstract}


\section{Introduction}
\label{sec:1}

At low temperatures, the classical conductivity of disordered conductors 
(normal metals and semiconductors) is determined by the scattering of
electrons off a quenched disorder (e.g., impurities and defects). 
This residual conductivity is given by the Drude formula:
\begin{equation}
\sigma = \frac{e^2n\tau}{m}=\frac{e^2nl}{p_F},
\label{eq:1.1}
\end{equation}
where  $v_F$ and $p_F$  are the
Fermi velocity and Fermi momentum of the electrons,  correspondingly,
$m$ and $n$  denote their mass and density,
$\tau$ is the transport elastic mean free time,
while $l=v_F\tau$ is the mean free path of electrons.

This expression can be justified provided the elastic mean free path $l$
is sufficiently large,
since all the corrections to the classical Drude conductivity
(\ref{eq:1.1})
are of the order of $\hbar/(p_Fl)\ll 1$.
However, in low dimensions, these quantum corrections
to the conductivity (QCC)   diverge when temperature $T$ decreases.
Eventually, they drive the system to the insulating regime.

QCC are of a substantial importance even for conductors that are far
from 
the strong localization regime: in a wide range of  parameters QCC,
though being smaller than the classical conductivity (\ref{eq:1.1}),
determine all the temperature and  magnetic field dependences of
the conductivity.
The systematic study of QCC was started almost two decades
ago. The comprehensive review of the status of the problem from both
theoretical and experimental viewpoints can be found in several 
papers\cite{AA,Altshuler87,LeeRama,Bergman84,Altshuler82}.

According to their physical origin QCC, can be divided into two
distinct groups. The correction of the first type, known as the weak
localization (WL) correction, is caused by the quantum interference
effect on the diffusive motion of a single electron. For low
dimensional ($d=1,2$) infinite systems, WL QCC diverges, and this
divergence should be regularized either by magnetic field or by some
other dephasing mechanism.

The second type of QCC, usually referred to as the interaction effects,
are absent in the one-particle approximation; they are 
entirely due to interactions 
between electrons. These corrections can be interpreted as the
scattering of an electron off the inhomogeneous
distribution of the density of the rest of the electrons.
One can attribute this inhomogeneous distribution to Friedel oscillations
produced
by each impurity.
The role of the electron-electron interactions
in this type of QCC
is to produce a static self-consistent
(and temperature-dependent)
potential. Such a potential does not lead to any real transitions
between single-electron quantum states.
Therefore, it does not
break the $T$-invariance of the system and neither affects
nor regularizes the WL corrections.
We will elaborate more on this point in Sec.~\ref{sec:2}.

However, the interactions between electrons are by no means irrelevant
to WL QCC. Indeed, these interactions cause
phase relaxation of the single-electron states,
and thus result in the cut-off the divergences in WL corrections.
This dephasing requires real inelastic collisions between the electrons.
When the temperature $T$ and magnetic field $H$ are low enough, 
this dephasing provides the leading mechanism 
to regularize QCC of the weak localization type.
This means that in this regime interactions between electrons can not be 
taken into account  perturbatively: the higher is the term of the 
perturbation theory expansion the stronger it diverges.
For the reasons, which will become clear below,  in this paper we
restrict ourselves by the lowest order of the perturbation theory in the 
interactions. This is possible, provided the dephasing and, consequently, 
the WL 
effects are mostly determined by the external magnetic field. Under these
conditions interactions between electrons modify the dephasing rate 
slightly,
and thus cause temperature dependence of the magnetoresistance. 
As a result, we end up with two classes of
interaction contributions to the conductivity: corrections to WL QCC due
to the interactions (class I) and the genuine interaction QCC (class II).

Recently, there were several attempts to revise this conventional
picture. Based on their own measurements, 
as well as on experimental results of other groups, 
authors of \cite{Mohanty1,Mohanty2} argued that apparent saturation
of the dephasing rate at $T \rightarrow 0$
has an intrinsic nature -- scattering of the electrons by 
``zero-point fluctuations of electromagnetic
environment''. This suggestion was claimed to be 
put  onto more ``theoretical
grounds'' by Golubev and Zaikin\cite{Zaikin1,Zaikin2}.

Even though these suggestions (as we strongly believe
and aim to explain below) can be rejected using purely physical
arguments, they still generate a sympathy in some part of the physical
community. The situation becomes worrisome, since the paper
\cite{Zaikin2}, at the first glance,
looks like a respectable calculation an experimentalist could
rely on. Since it is much easier to observe the saturation rather than
not to observe it (a possible reason for this saturation is
discussed, e.g.,  in Ref.~\onlinecite{Altshuler98}), one can expect
that in the future, quite a number of  experimental and non-experimental papers
confirming the role of the zero-point motion in the dephasing
mechanism will appear. 
Having all this in mind, we have undertaken in this paper
a straightforward analysis of the effect of 
electron-electron interactions on the weak localization. 
We explicitly demonstrated that the dephasing
(i.e. the characteristic scale of the magnetoresistance) is determined by
the real processes with the energy transfer smaller or of the order of
temperature. The role of the processes with the transmitted energy
larger than temperature (which was referred to as ``zero-point motion'' in
Refs.~\onlinecite{Mohanty1,Mohanty2,Zaikin1,Zaikin2}) is to produce
type two QCC. The explicit formula
for those corrections in the second order 
of the perturbation theory in the inverse dimensionless
conductance has been obtained for the first time.

The paper is organized as follows. Section ~\ref{sec:2} 
reviews qualitatively
physics of the quantum corrections to the conductivity. In the
same section we introduce the notion of the dephasing time $\tau_\varphi$.
 Section ~\ref{sec:3}
briefly outlines the essential content of the ``new
theories''\cite{Zaikin1,Zaikin2} which predict zero-temperature
dephasing. Section \ref{sec:4} contains the main idea of our
calculation and the results, 
which clearly contradict to all the conclusions of
Refs.~\onlinecite{Zaikin1,Zaikin2}.
Technical details of this calculation are described in 
section~\ref{sec:5}.
Section~\ref{sec:6} is devoted to a critical analysis of 
Refs.~\onlinecite{Zaikin1,Zaikin2}. We discuss their results and
possible sources of mistakes  made in these papers.
In the same section we show that the results of these references
disagree with the available experimental data in two- and three-
dimensional systems by several orders of magnitude and the agreement
in one-dimensional case claimed in Ref.~\onlinecite{Zaikin3} is a pure
coincidence in the region of parameters were the weak localization theory
is already not applicable.
Our findings are summarized in Conclusion. 

\section{Quantum corrections to the conductivity: a qualitative picture}
\label{sec:2}

This section contains only rather old results and ideas.  We review
them here in spite of the fact that there is a number of review
articles that discuss in details both qualitative and technical parts
of the theory of disordered conductors in the weak localization
regime\cite{AA,Altshuler87,LeeRama,Bergman84,Altshuler82}.  The main
reason to do this is to make this paper self-contained, so that any
reasonably educated physicist can read it without consulting the
reviews continuously, and to underline the main physical concepts which
are being rejected by
Refs.~\onlinecite{Mohanty1,Mohanty2,Zaikin1,Zaikin2}.  We also propose
somewhat new intuition about the physical origin of the effects of the
electron-electron interaction.  Nevertheless, experts in the field can
skip this section.

\subsection{Weak localization}

\label{sec:2.1}

Anomalous magnetoresistance in disordered conductors (doped
semiconductors and metals) has been recognized for almost 50
years~\cite{MRhistory}.  For a long time this phenomenon has
remained a puzzle.  The theoretical understanding of the
anomalous magnetoresistance emerged as a spinoff of the theory of
Anderson localization. It turned out that the correction to the
conductivity, which is due to quantum interference at large
length scales, is very sensitive to weak magnetic fields. The
quantum correction itself may be much smaller than the classical
conductivity. Nevertheless, the weak field magnetoresistance is
dominated by this correction, and its basic features (its
amplitude, dependence on both magnitude and direction of the
magnetic field, etc.) are very different from that of the
classical magnetoresistance.  Since the quantum correction can
eventually drive the system to the Anderson insulator, the regime
when this correction is small, is called the weak localization
(WL) regime, and the theory of the anomalous magnetoresistance is
now a part of the theory of weak localization.

\begin{figure}
\vspace{0.2 cm}
\epsfxsize=7.7cm
\centerline{\epsfbox{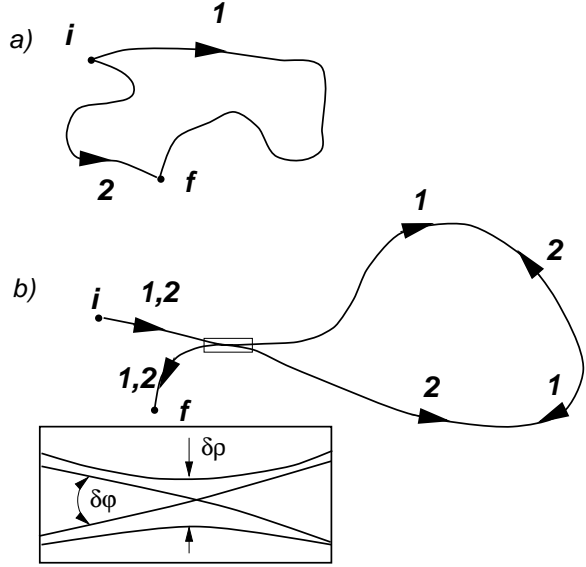}}
\vspace{0.8cm}
\caption{Examples of the classical (a) non-coherent and (b) coherent paths
between points $i$ and $f$. The region of intersection of trajectories
is blown up in the inset.
}
\label{Fig1}
\end{figure}

A qualitative physical interpretation of WL is usually based on
the following arguments, see e.g. Ref.~\onlinecite{ALee}. Consider an
electron diffusing in a good conductor, $p_Fl \gg \hbar$. The 
probability $w$  
for the electron to reach, say, point $i$ starting from point $f$, see
Fig.~\ref{Fig1}, can be obtained by, first, finding the semiclassical
amplitudes $A_\alpha$ for different paths connecting the points, and,
second, by squaring the modulus of their sum:
\begin{equation}
w = \left|\sum_\alpha A_\alpha \right|^2 = \sum_\alpha
\left|A_\alpha\right|^2 + \sum_{\alpha,\beta} A_\alpha A_\beta^\ast.
\label{eq:2.1}
\end{equation}
The first term in Eq.~(\ref{eq:2.1}) is nothing but the sum of the
classical probabilities of the different paths, and the second term is
the quantum mechanical interference of the different amplitudes. For
generic pairs $\alpha, \beta$, the product $A_\alpha A_\beta^\ast$
oscillates  on the scale of the order of $\lambda_F = \hbar/p_F$ as
the function of the position of point $f$. This is because the lengths
of  paths $\alpha$ and $\beta$ are substantially different. Since all
measurable quantities are averaged on the scale much larger than
$\lambda_F$, such oscillating contributions can be neglected. There
are pairs of paths, however, which are coherent. An example of such
paths is shown in Fig.~\ref{Fig1}b. The paths $1$ and $2$ 
coincide almost
everywhere but the loop segment $BEB$ (see Fig.~\ref{Fig1}b) 
is traversed by trajectories $1$ and $2$ in the opposite
directions. In the absence of the magnetic field and spin-orbit
interactions, the phases of the trajectories $1$ and $2$ are equal.
Therefore, the contribution of this paths to the probability $w$
becomes
\begin{equation}
\left|A_1 + A_2\right|^2 = \left|A_1\right|^2 + \left|A_2\right|^2
+2 {\rm Re} A_1A_2^\ast = 4 \left|A_1\right|^2,
\label{eq:2.1a}
\end{equation}
i.e. it is twice as large as the classical probability. Thus, in order
to determine the value of the weak localization correction to the
conductivity $\delta\sigma_{WL}$, one has to determine the classical
probability to find such a self-intersecting trajectory.

In order to find this probability, we label all the
trajectories by the time  $t$ it takes for a classical
particle to go around the loop.  
The classical probability $dP$ that the diffusing 
particle returns into the  phase volume $dV$ at a given time  is
\begin{equation}
dP = \frac{dV}{\left(Dt\right)^{d/2}},
\label{eq:2.2}
\end{equation} 
where $D=v_F^2\tau/d$ is the diffusion constant, and $d$ is the number
of dimensions. The relevant phase volume can be estimated as $v_Fdt
\left(\delta\rho\delta\phi\right)^{d-1}$, where $\delta\rho$ and
$\delta \phi$ characterize the transverse distance between the paths
$1$ and $2$ at the intersection point, see Fig.~\ref{Fig1}b, and
$\delta\phi$ is the intersection angle between them.  For the
interference between paths $1$ and $2$ to be effective, the
uncertainty relation $p_F\delta\phi\delta\rho \simeq \hbar$ should
hold.  We substitute this estimate for $dV$ into Eq.~(\ref{eq:2.2})
and sum up over all the trajectories:
\begin{equation}
\frac{\delta \sigma}{\sigma} \simeq - \int dP \simeq
- v_F \lambda_F^{d-1}\int_\tau^{?} \frac{dt}{\left(Dt\right)^{d/2}},  
\label{eq:2.3}
\end{equation}
where the negative sign is due to the fact that the returning
trajectory should arrive at the intersection point with the momentum
almost opposite to the initial one. At low dimensions $d=1,2$, the
integral in Eq.~(\ref{eq:2.3}) diverges at the upper limit. Below we
will return to
the cut-off of this integral.

The explicit calculation first performed in Ref.~\onlinecite{Gorkov79}
gives the result similar to Eq.~(\ref{eq:2.3}) up to a numerical factor:
\begin{equation}
\delta\sigma_{WL} = - \frac{\sigma}{\pi\nu\hbar}\int_\tau^\infty dt\ {\cal C}
\left(\mbox{\boldmath $r$},\mbox{\boldmath $r$};t\right),
\label{eq:2.4}
\end{equation}
where $\sigma$ is defined in Eq.~(\ref{eq:1.1}) and $\nu$ is the
density of states per one spin.  The retarded classical propagator --
Cooperon is the solution of the equation
\begin{equation}
\left(\frac{\partial}{\partial t} - D\nabla^2_1\right){\cal
C}=\delta\left(\mbox{\boldmath $r$}_1-\mbox{\boldmath
$r$}_2\right)\delta(t).
\label{eq:2.5}
\end{equation}

It follows from the previous discussion, that the weak localization
correction originates from the quantum interference within the pairs
of time reversed paths. 
Therefore, this correction is extremely sensitive to
any violation of the time reversal invariance, {\em e.g.}
magnetic impurities, external magnetic field, external microwave
radiation {\em etc.}  (we will not consider the spin-orbit interaction
in this paper).

Let us discuss first the effect of the magnetic field. The role of the
inelastic processes will be considered in the following subsection.
The effect of the magnetic field on the conductance is a
manifestation of the Aharonov -- Bohm effect~\cite{AB}.  Let us
again consider a pair of the coherent trajectories $1$ and $2$
from Fig.~\ref{Fig1}b. Since those trajectories traverse the loop in
the opposite directions, the Aharonov-Bohm phases acquired by them are
of the same magnitude, but opposite signs:
\begin{equation}
A_1 \to A_1\exp\left(i\frac{e}{c\hbar}\mbox{\boldmath $H S $}\right);
\quad 
A_2 \to A_2\exp\left(-i\frac{e}{c\hbar}\mbox{\boldmath $H S $}\right),
\label{eq:2.6}
\end{equation}
where {\boldmath $S$} is the directed area of the loop and
{\boldmath $H$} is the external magnetic field. As the result,
Eq.~(\ref{eq:2.1a}) should be modified as ${\rm Re} A_1A_2^\ast=
|A_1|^2 \cos\left(2\frac{e}{c\hbar}\mbox{\boldmath $H S $}\right)$
and, accordingly, Eq.~(\ref{eq:2.3}) now reads
\begin{equation}
\frac{\delta \sigma}{\sigma} \simeq 
- v_F \lambda_F^{d-1}\int_\tau^{\infty}
\frac{dt}{\left(Dt\right)^{d/2}}
\langle\cos\left(2\frac{e}{c\hbar}\mbox{\boldmath $H S $}\right)\rangle
\simeq
- v_F \lambda_F^{d-1}\int_\tau^{\infty}
\frac{dt}{\left(Dt\right)^{d/2}}
\exp\left(-2\frac{\langle S^2(t)\rangle}{\lambda_H^4}\right),
\label{eq:2.7}
\end{equation}
where $\lambda_{H} = \sqrt{\hbar c/eH}$ is the magnetic length.  The
last transformation in Eq.~(\ref{eq:2.7}) relies on the fact that
the area $S_\alpha$  swept by the path is the
normally distributed random quantity for
a diffusive system. For the infinite
three-dimensional systems and two-dimensional films in the magnetic
field perpendicular to the film plane, $\langle S^2(t)\rangle \simeq
\left(Dt\right)^2$.  In a 1d case (wire) with the
transverse dimension $a$,\cite{footnote} $\langle S^2(t)\rangle \simeq
\left(Dt\right)a^2$; the same expression holds also for
a two-dimensional film in the magnetic field parallel to the film plane
with $a$ being the film thickness.

As the result, the weak localization correction
is cut-off at the 
large time limit by $\tau_H$ equal to \cite{Altshuler80}
\begin{mathletters}
\begin{eqnarray}
\frac{1}{\tau_H}=\Omega_H=\frac{4DeH}{\hbar c}, &&\quad d=3,\ d=2 \
\left(\mbox{\boldmath $n$} \parallel \mbox{\boldmath $H$}\right);
\label{eq:2.7a}\\
\frac{1}{\tau_H} = \frac{D\left(eHa\right)^2}{3c^2\hbar^2},&&\quad
d=1,
\ d=2 \
\left(\mbox{\boldmath $n$} \perp \mbox{\boldmath $H$}\right),
\label{eq:2.7b}
\end{eqnarray}
where {\boldmath $n$} is the vector normal to the plane of the film.
\end{mathletters} 

More rigorously, the weak localization correction to the
conductivity is still given by Eq.~(\ref{eq:2.4}), however, the
equation for the Cooperon (\ref{eq:2.4})  is modified:
\begin{equation}
\left[\frac{\partial}{\partial t} + 
D\left(-i\mbox{\boldmath $\nabla$}_1 -\frac{2e}{c\hbar}
\mbox{\boldmath $A$}(\mbox{\boldmath $r$}_1)\right)^2\right]{\cal
C}=\delta\left(\mbox{\boldmath $r$}_1-\mbox{\boldmath
$r$}_2\right)\delta(t),
\label{eq:2.8}
\end{equation}
where the vector potential {\boldmath $A$} due to the magnetic field
describes the Aharonov-Bohm phase we have already discussed. For
$d=3$ and $d=2$
$\left(\mbox{\boldmath $n$} \parallel \mbox{\boldmath $H$}\right)$,
this equation is equivalent to the one for 
a charged particle in the magnetic field, 
and $\Omega_H$ from Eq.~(\ref{eq:2.7a}) has a meaning
of the cyclotron frequency. 
For one-dimensional systems and films 
$\left(\mbox{\boldmath $n$} \perp \mbox{\boldmath $H$}\right)$,
it reduces to 
\begin{equation}
\left[\frac{\partial}{\partial t} - 
D\frac{\partial^2}{\partial x^2}+\frac{1}{\tau_H}\right]{\cal
C}=\delta\left(\mbox{\boldmath $r$}_1-\mbox{\boldmath
$r$}_2\right)\delta(t),
\label{eq:2.9}
\end{equation}
with $\tau_H$ being defined by Eq.~(\ref{eq:2.7b}).

Let us now define the dephasing time $\tau_\varphi$ and the corresponding
length scale $L_\varphi=\sqrt{D\tau_\varphi}$ in a
pure phenomenological way. First, we repeat that the phase itself
is not a directly observable quantity, therefore one always has  to specify
the observable which is sensitive to the phase relaxation. As we have
already discussed, the WL correction is cut by the magnetic field. On
the  other hand, it can also be cut by any process that breaks the phase
coherence. We introduce the function
$F_d(x)$ so that
\begin{equation}
\delta\sigma_{WL} = \frac{e^2}{\hbar}\frac{L_\varphi^{2-d}}{d-2} 
F_d\left(\frac{\tau_\varphi}{\tau_H}\right).
\label{eq:2.10}
\end{equation}
Function $F_d(x)$ is different for different dimensionalities $d$, and
it has the asymptotic behavior $F_d(x)= const,\ x \ll 1$ and
$F_d(x)\propto x^{d/2-1},\ x \gg 1$.  Therefore, $\tau_\varphi$ is
defined through the scale of the magnetic fields at which
magnetoresistance occurs.  
Since both the conductivity $\delta\sigma$ and
time $\tau_H$ can be measured with a great
accuracy, one can extract $\tau_\varphi$
by fitting the magnetoresistance.  
In order to define this time with
the numerical coefficient, the explicit form of functions $F_d(x)$ is
needed, which requires some theoretical analysis\cite{AAK}. We
postpone further discussion of this point until the end of the
following subsection.

\subsection{Interaction effects}
\label{sec:2.2}

It was already mentioned in the introduction that effects of the
interaction between the electrons on the conductivity can be
subdivided into two groups. The corrections of the first group are
associated with scattering of a given electron on the static Friedel
-- like oscillations of the charge density caused by disorder.  These
corrections have nothing to do with the energy transfer between the
electrons,
i.e., with inelastic processes. The second group originates entirely
from the inelastic scattering of the electrons. Since the physics
involved is completely different for those two groups, we consider
them separately.

\subsubsection{Correction to the conductivity and tunneling density of
states} 

First, we will discuss the Hartree type corrections to the tunneling
density of states and to the conductivity assuming that the
interaction between electrons is determined
by a short range two-particle potential $V$. 
Later we will show how to generalize obtained
formulas for the Fock correction and for the Coulomb potential.  
The rigorous derivation of the results is given in
Ref.~\onlinecite{AA}. 
Here, our goal is to describe the qualitative physical picture.

Consider the  local density of states of a non-interacting system
\begin{equation}
\nu(\epsilon.\mbox{\boldmath $r$}) \equiv 
\sum_i \left|\psi_i\left(\mbox{\boldmath $r$}\right) \right|^2
\delta\left(\epsilon-\epsilon_i\right),
\label{eq:2.11}
\end{equation}
where $\psi_i$ and $\epsilon_i$ are the eigenfunctions and
eigenenergies of the disordered systems respectively.
Hereafter, all the energies are measured from the Fermi energy.

\begin{figure}
\vspace{0.2 cm}
\epsfxsize=8.5cm
\centerline{\epsfbox{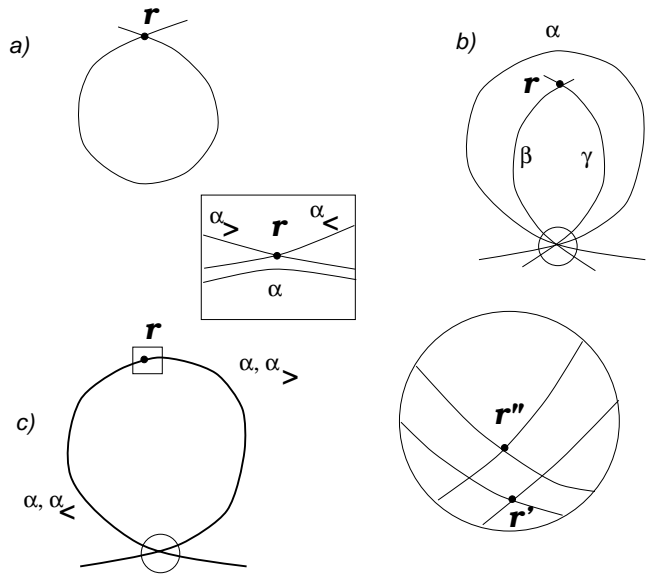}}
\vspace{0.8cm}
\caption{(a) The returning path that contributes to the density of states
$\nu$ and the oscillating part of the electron density $\rho$;
(b) Classical paths responsible for the correction to the amplitude of
the return; (c) Three paths producing non-vanishing correction to the
density of states. The regions of the intersections of the trajectories
are blown up in the insets.  
}
\label{Fig2}
\end{figure}

Semiclassically, the density of states (\ref{eq:2.11}) can be presented in
the form\cite{Gutzwiller}
\begin{equation}
\nu(\epsilon,\mbox{\boldmath $r$}) \approx const + {\rm Im}
\sum_\alpha A_\alpha(\epsilon,\mbox{\boldmath $r$}), 
\label{eq:2.12}
\end{equation}
where $A_\alpha(\epsilon,\mbox{\boldmath $r$})$ is the quantum
mechanical amplitude corresponding to the path $\alpha$ starting at
and returning to the point {\boldmath $r$}, see Fig.~\ref{Fig2}a, and
``$const$'' stands instead of a function of energy that is smooth on
the scale of the Fermi energy.

The second term in Eq.~(\ref{eq:2.12}) oscillates on the scale of the
order of $\lambda_F = \hbar/p_F$ as the function of the position of the
point {\boldmath $r$}, because the lengths of the paths
$\alpha,\beta,\gamma$ are 
much larger than $\lambda_F$. As the result, this contribution
vanishes upon disorder averaging and there is no interference
contribution to the density of states of non-interacting electrons in
the disordered metals. 

Let us demonstrate now that the interactions between electrons change
the situation drastically. In the Hartree approximation, the only
effect of the interactions is reduced to the effective one-electron
potential $V_H$, which should be considered in addition to the random
potential of impurities
\begin{equation}
V_H\left(\mbox{\boldmath $r$}\right) =
\int d\mbox{\boldmath $r$}^\prime 
\rho (\mbox{\boldmath $r$}^\prime)
V\left(\mbox{\boldmath $r$}-
\mbox{\boldmath $r$}^\prime\right).
\label{eq:2.13}
\end{equation}
Here $\rho(\mbox{\boldmath $r$})$ is the electron density and
$V\left(\mbox{\boldmath $r$}-\mbox{\boldmath $r$}^\prime\right)$ is
the potential of the interaction between two electrons located at the
points $\mbox{\boldmath $r$}$ and $\mbox{\boldmath $r$}^\prime$.  
We assume that this potential is weak and short-range on the scale
of the order of the elastic mean free-path $l$. In a clean infinite
system, the electron density is homogeneous, and the Hartree
correction (\ref{eq:2.13}) leads only to a uniform (and thus not
observable) shift of the chemical potential.  In disordered or finite
systems, however, the electron density has an inhomogeneous term
\begin{equation}
\rho (\mbox{\boldmath $r$}) = \int d\epsilon\ n_F(\epsilon)
\ {\rm Im}
\sum_\alpha A_\alpha(\epsilon;\mbox{\boldmath $r$})
\label{eq:2.14} 
\end{equation}
which oscillates on the scale of the order of $\lambda_F$; for a single
impurity these oscillations correspond to the well-known Friedel
oscillations. Here, $n_F(\epsilon) = 1/(1+e^{\epsilon/T})$ is the
Fermi-Dirac distribution function.

Scattering of an electron by these density oscillations results in
the singularity in the average tunneling density of states.  To
demonstrate this, consider the scattering off the Hartree potential in
the first-order  perturbation theory.  According to the usual rules
of quantum mechanics, the correction to the return amplitude $\delta
A$ is given by
\begin{equation}
\delta A(\epsilon; \mbox{\boldmath $r$}) \approx
\int d\mbox{\boldmath $r$}^\prime\ V_H
\left(\mbox{\boldmath $r$}^\prime\right)
\sum_{\beta,\gamma} 
A_\beta\left(\mbox{\boldmath $r$},
\mbox{\boldmath $r$}^\prime;\epsilon\right)
A_\gamma
\left(\mbox{\boldmath $r$}^\prime,
\mbox{\boldmath $r$};\epsilon\right), 
\label{eq:2.15}
\end{equation}
where $A_\alpha\left(\mbox{\boldmath $r$},
\mbox{\boldmath $r$}^\prime;\epsilon\right)$ is the quantum mechanical
amplitude for the electron to get from point $\mbox{\boldmath $r$}$ to
point  $\mbox{\boldmath $r$}^\prime$ using the classical  path $\alpha$, 
see Fig.~\ref{Fig2}b; $ A_\alpha\left(\epsilon; \mbox{\boldmath $r$}\right) 
\equiv
A_\alpha\left(\mbox{\boldmath $r$},
\mbox{\boldmath $r$};\epsilon\right)$. 

Substituting Eqs.~(\ref{eq:2.13}) and (\ref{eq:2.14}) into
Eq.~(\ref{eq:2.15}), we obtain with the help of Eq.~(\ref{eq:2.12})
\begin{equation}
\delta\nu(\epsilon.\mbox{\boldmath $r$}) \approx 
 \int d\mbox{\boldmath $r$}^\prime d\mbox{\boldmath
 $r$}^{\prime\prime}
V\left(\mbox{\boldmath $r$}^\prime -\mbox{\boldmath
 $r$}^{\prime\prime} \right) \int d\epsilon^\prime 
n_F\left(\epsilon^\prime\right)  
\sum_{\alpha, \beta, \gamma}
\left[{\rm Im} A_\alpha(\epsilon^\prime,\mbox{\boldmath
 $r$}^\prime)\right]
\left[{\rm Im}
A_\beta\left(\mbox{\boldmath $r$},
\mbox{\boldmath $r$}^{\prime\prime};\epsilon\right)
A_\gamma
\left(\mbox{\boldmath $r$}^{\prime\prime},
\mbox{\boldmath $r$};\epsilon\right)
\right].
 \label{eq:2.16}
\end{equation}
Contribution of a typical set of paths 
$(\alpha,\beta,\gamma)$ to the sum 
(\ref{eq:2.16}) oscillates as the function of {\boldmath $r$}.
However, similarly to the case of weak localization, 
there exist special sets
of coherent paths, with non-vanishing contributions to the sum. 
An example of such sets is shown in
Fig.~\ref{Fig2}c. In this case, 
paths $\gamma$ and $\beta$ are chosen to be
parts of the trajectory 
$\alpha$. We will denote them $\alpha_<$ and $\alpha_>$. 
Let us neglect in the sum in Eq.~(\ref{eq:2.16}) 
all the terms except the contributions of the sets 
 $(\alpha,\beta,\gamma)$ of this type:
$(\beta = \alpha_<$, $\gamma = \alpha_>$  
The phase of the 
quantum amplitude $A_\alpha$ of a path $\alpha$ is
connected with 
the classical action ${\cal S}_\alpha$  which corresponds to this path: 
$A_\alpha \propto e^{i{\cal S}_\alpha}/\hbar$.
Note that difference between the energies $\epsilon$ and 
$\epsilon^\prime$ is small, and that points
$\mbox{\boldmath $r$}^\prime$  and
$\mbox{\boldmath $r$}^{\prime\prime} $ are close due to
the short range nature of the potential $V$.
Expansion of ${\cal S}_\alpha$ in coordinate 
and energy with the use of identities
\[
-\frac{\partial
{\cal S}_\alpha\left(\mbox{\boldmath $r$}_1,
\mbox{\boldmath $r$}_2;\epsilon\right)}{\partial \epsilon}
= t_\alpha; \quad 
\frac{\partial
{\cal S}_\alpha\left(\mbox{\boldmath $r$}_1,
\mbox{\boldmath $r$}_2;\epsilon\right)}{\partial \mbox{\boldmath
$r$}_1}
= \mbox{\boldmath $p$}_\alpha^f; \quad
\frac{\partial
{\cal S}_\alpha\left(\mbox{\boldmath $r$}_1,
\mbox{\boldmath $r$}_2;\epsilon\right)}{\partial \mbox{\boldmath
$r$}_2}
= \mbox{\boldmath $p$}_\alpha^i,
\]
($t_\alpha$ is the time it takes for an electron with energy
$\epsilon$ to go along the path $\alpha$ 
from the initial point $\mbox{\boldmath $r$}_2$ 
to the final point $\mbox{\boldmath $r$}_1$, while
$\mbox{\boldmath $p$}_{(1,2)}^{(\alpha)}$ are respectively 
its final and initial momenta) allows to
rewrite $A_\alpha(\epsilon^\prime,\mbox{\boldmath $r$}^\prime)$ as
\[
A_\alpha(\epsilon^\prime,\mbox{\boldmath $r$}^\prime) =
A_\alpha(\epsilon, \mbox{\boldmath $r$}^{\prime\prime})
e^{it_\alpha (\epsilon-\epsilon^\prime)} 
\exp\left[i\left( \mbox{\boldmath $r$}^\prime
- \mbox{\boldmath $r$}^{\prime\prime}\right) \cdot
\left(\mbox{\boldmath $p$}_2^{(\alpha)}
\mbox{\boldmath $p$}_1^{(\alpha)}\right)\right].
\]
As a result, Eq.~(\ref{eq:2.16}) acquires the form
\begin{equation}
\delta\nu(\epsilon.\mbox{\boldmath $r$}) \approx 
\sum_{\alpha}
 \int d\mbox{\boldmath $r$}^\prime 
V\left(\mbox{\boldmath $r$}^\prime \right) 
\exp\left[i\mbox{\boldmath $r$}^\prime 
\left( \mbox{\boldmath $p$}_\alpha^f -\mbox{\boldmath $p$}_\alpha^i\right)
\right]
{\rm Re}
\int d\mbox{\boldmath
 $r$}^{\prime\prime}
\int d\epsilon^\prime 
n_F\left(\epsilon^\prime\right)  e^{it_\alpha (\epsilon-\epsilon^\prime)}
 A_\alpha(\epsilon,\mbox{\boldmath
 $r$}^{\prime\prime})
A_{\alpha_<}^\ast\left(\mbox{\boldmath $r$},
\mbox{\boldmath $r$}^{\prime\prime};\epsilon\right)
A_{\alpha_>}^\ast
\left(\mbox{\boldmath $r$}^{\prime\prime},
\mbox{\boldmath $r$};\epsilon\right).
 \label{eq:2.17}
\end{equation}
To further simplify equation
(\ref{eq:2.17}), we notice that for
the diffusive system, directions of the final and initial momenta are
not correlated. Therefore, the first factor in
Eq.~(\ref{eq:2.17}) equals to
\begin{equation}
 \int d\mbox{\boldmath $r$}
V\left(\mbox{\boldmath $r$} \right) 
\exp\left[i\mbox{\boldmath $r$}
\left( \mbox{\boldmath $p$}_\alpha^f -\mbox{\boldmath
$p$}_\alpha^i\right)
\right] \to \overline{ \tilde{V}
\left( \mbox{\boldmath $p$}_1 -\mbox{\boldmath
$p$}_2\right)},
\label{eq:2.18} 
\end{equation}
where $\tilde{V}$ is the Fourier transform of the interaction
potential and the bar denotes the averaging over the Fermi surface.
Also, since the averaged density of states is translationally
invariant, 
one can replace the integration over the
interaction point $\mbox{\boldmath $r$}^{\prime\prime}$ in 
Eq.~(\ref{eq:2.17}) by the integration over the observation point 
$\mbox{\boldmath $r$}$.
We use the identity
\[
\int d \mbox{\boldmath $r$}A_{\alpha_<}^\ast\left(\mbox{\boldmath $r$},
\mbox{\boldmath $r$}^{\prime\prime};\epsilon\right)
A_{\alpha_>}^\ast
\left(\mbox{\boldmath $r$}^{\prime\prime},
\mbox{\boldmath $r$};\epsilon\right)
= it_\alpha A_{\alpha}^\ast
\left(\mbox{\boldmath $r$}^{\prime\prime};\epsilon\right),
\]
which can be derived using the exact relation for the Green function.
The fact that the result is proportional to the length of the
trajectory $t_\alpha$ can be easily understood since the observation
point $\mbox{\boldmath $r$}$ can be inserted anywhere along the
path $\alpha$.
 
After the integration over $\epsilon^\prime$, 
we obtain from Eq.~(\ref{eq:2.17})
\begin{equation}
\delta\nu(\epsilon) \approx 
\overline{ \tilde{V}
\left( \mbox{\boldmath $p$}_1 -\mbox{\boldmath
$p$}_2\right)}\
{\rm Re}
\sum_{\alpha}\frac{\pi Tt_\alpha \ e^{i\epsilon t_\alpha}}
{\sinh\pi Tt_\alpha} 
\left|A_\alpha(\epsilon,\mbox{\boldmath
 $r$}^{\prime\prime})\right|^2.
 \label{eq:2.19}
\end{equation}
As a result, the correction to the density of states is expressed in
familiar terms of the return probability, which we have already dealt
with in a previous subsection!  Substituting $P(t) \simeq
1/(Dt)^{d/2}$ for the return probability, we estimate the correction
to the tunneling density of states as
\begin{equation}
\frac{\delta\nu(\epsilon)}{\nu} \approx 
\overline{ \tilde{V}
\left( \mbox{\boldmath $p$}_1 -\mbox{\boldmath
$p$}_2\right)}\
\int_\tau^\infty dt \cos \left(\epsilon t\right)
\frac{\pi Tt }{\sinh\pi Tt} 
\frac{1}{\left(Dt\right)^{d/2}}.
 \label{eq:2.20}
\end{equation}
Therefore, the one-particle density of states $\nu(\epsilon)$ gets
significantly modified due to the scattering of a given electron off
the charge density oscillations of the rest of the electron gas.  At
low dimensions, the correction $\delta\nu(\epsilon)$ (\ref{eq:2.20})
diverges at large times when $T = 0$ and $\epsilon = 0$.  One can
interpret the divergence as the Bragg scattering of the incident electron
off Friedel oscillations.  Finite electron energy $\epsilon$ sets the
upper limit for the integration because the scattering remains
resonant only as long as the length of trajectory $v_Ft$ times the
difference $\hbar v_F/\epsilon$ between the momentum becomes of the
order of unity, which corresponds to
the condition $\epsilon t \lesssim\hbar$. 
A finite temperature $T$ also cuts the correction: it smears
out the oscillating contributions related to the trajectories longer
than $\hbar v_F/T$.

One can obtain a more familiar and convenient form \cite{AA} of
(\ref{eq:2.20}) by making Fourier transform in time
\begin{equation}
\frac{\delta \nu^H\left(\epsilon\right)}{\nu} = {\rm Im}
\int_0^\infty \frac{d\omega}{2\pi} 
\left[\tanh\left(\frac{\omega +\epsilon}{2T}\right)
+\tanh\left(\frac{\omega -\epsilon}{2T}\right)\right]  
\int\frac{d^dQ}{\left(2\pi\right)^d}
\frac{2\bar{V}}{\left(-i\omega +DQ^2\right)^2}.
\label{eq:2.21}
\end{equation}
Here we introduced the short-hand notation $\bar{V}$ for the constant
defined in Eq.~(\ref{eq:2.18}).

As far as the correction to the conductivity for a short-range
interaction is concerned, it is directly related to the
correction to the density of states.  For instance, one can determine
Hartree QCC directly from Eq.~(\ref{eq:2.21})
\begin{equation}
\delta\sigma^H = \sigma \int d\epsilon
\frac{dn_F(\epsilon)}{d\epsilon}
\frac{\delta \nu^H\left(\epsilon\right)}{\nu} = 
\frac{\sigma}{2\pi} {\rm Im}
\int {d\omega} 
 \frac{\partial }{\partial \omega}
\left(\omega \coth\frac{\omega}{2T}\right)
\int\frac{d^dQ}{\left(2\pi\right)^d}
\frac{2\bar{V}}{\left(-i\omega +DQ^2\right)^2}.
\label{eq:2.22}
\end{equation}
Inclusion of the Fock part of the self-consistent potential is also
straightforward. It amounts to the addition of the Fourier transform of
the interaction potential $- \tilde{V}(q=0)$ to the potential
$2\bar{V}$ in Eqs.~(\ref{eq:2.21}) and (\ref{eq:2.22}).  The negative
sign reflects the fermionic statistics of electrons.

For the Coulomb interaction between electrons $\tilde{V}(q=0)$  diverges. 
This indicates that the screening should be taken into account. 
As a result, the
correction to the tunneling density of states acquires the form
\begin{equation}
\frac{\delta \nu^C\left(\epsilon\right)}{\nu} =- {\rm Im}
\int_0^\infty \frac{d\omega}{2\pi} 
\left[\tanh\left(\frac{\omega +\epsilon}{2T}\right)
+\tanh\left(\frac{\omega -\epsilon}{2T}\right)\right]  
\int\frac{d^dQ}{\left(2\pi\right)^d}
\frac{U(Q,\omega)}{\left(-i\omega +DQ^2\right)^2},
\label{eq:2.23}
\end{equation}
where the dynamically screened interaction potential is given by
\begin{equation}
U(Q,\omega) = \left\{\frac{1}{V(Q)}+ 2\nu 
\frac{DQ^2}{-i\omega+DQ^2}\right\}^{-1}.
\label{eq:2.24}
\end{equation}

For a long range interaction, there is no straightforward relation 
between the density of states and the conductivity as Eq.~(\ref{eq:2.22}),
since the large part of such interaction should be gauged out. One finds
(see Ref.~\onlinecite{AA} and Sec.~\ref{sec:5} of the present paper
for the further technical details)
\begin{equation}
\delta \sigma_{C} = -\frac{2\sigma}{\pi\hbar d}{\rm Im}\left[
\int d\omega \frac{\partial }{\partial \omega}
\left(\omega \coth\frac{\omega}{2T}\right)
\int\frac{d^dQ}{\left(2\pi\right)^d}
U\left(Q,\omega\right)\frac{DQ^2}{\left(-i\omega +DQ^2\right)^3}
\right].
\label{eq:2.25}
\end{equation}
For the further references, we show here the results of integration in
Eq.~(\ref{eq:2.25}):
\begin{mathletters}
\label{eq:2.250}
\begin{eqnarray}
\delta\sigma_C(T) = - \frac{e^2}{2\pi^2\hbar}
\ln \left(\frac{\hbar}{T\tau}\right)
;\quad d=2;
\label{eq:2.250a}\\
\delta\sigma_C(T) = - \frac{e^2}{\pi\hbar}
\sqrt{\frac{\hbar D}{2\pi T}}
\left(\frac{3 \zeta\left(3/2\right)}{2}\right)
;\quad d=1,
\label{eq:2.250b} 
\end{eqnarray}
where $\zeta(x)$ is the Riemann zeta-function, $\zeta (3/2) = 2.612\dots$.
\end{mathletters}

It is crucial (and we will keep emphasizing it) that the
frequency $\omega$ in Eqs.~(\ref{eq:2.23}), transfered by the Coulomb
interaction and (\ref{eq:2.25}) is {\em larger} than the temperature
$T$. We saw that physical meaning of this correction is the {\em
elastic} scattering of the electron from the potential (Hartree or
Fock) created by the rest of electrons. These processes do not cause
any phase breaking since they do not violate the time reversal symmetry.

\subsubsection{Energy and phase relaxation}

We turn  now to discussion of the second type of effects due to the
electron-electron interaction -- the inelastic scattering.

In early papers (see {\em e.g.} Ref.~\onlinecite{Thouless}) on the
theory of localization, the dephasing rate $1/\tau_{\varphi}$ was
considered to be of the same order as the inelastic collision
rate in clean conductors.  The latter can be expressed
as the sum of the electron - phonon ($1/\tau_{e-ph}\simeq
T^{3}/\Theta_{D}^{2}$) and electron - electron ($1/\tau_{e-e}\simeq
T^{2}/E_{F} $) contributions, where $E_{F}$ and $\Theta_{D}$ are
the Fermi and Debye energies correspondingly (here and almost
everywhere below we put $ \hbar =1$ and $k_{B} = 1$).  Even under
this assumption, the $e-e$ contribution dominates at low enough
temperatures.  It became clear later that the static disorder
enhances strongly  the $e-e$ contribution to the inelastic scattering
rate \cite{Schmid,AA}, while $1/\tau_{e-ph}$ is less affected
\cite{Reyzer}.  As the result, both dephasing and energy
relaxation rates at low temperatures are governed by collisions
between electrons.

To recall the main results on the $e-e$ dephasing rate, let us
start with a single electron excitation, assuming that $T = 0$,
and the rest of the electron gas occupies states below the Fermi
level.  The dependence of $1/\tau_{e-e}$ on the excitation energy
$\epsilon$ 
(the energy of an electron counted from the Fermi level)
can be determined in a perturbative calculation~\cite
{Schmid,AA,inelastic}.  The result [Eq.(4.4) of
Ref.~\onlinecite{AA}] can be rewritten through the dimensionless
conductance $g(L)$ (the conductance measured in units of $e^{2}/h
\simeq 1/25.8k\Omega $) of a d - dimensional cube of the size $L$
\begin{equation}
\frac{1}{\tau_{e-e}(\epsilon)} = C_{d} \frac{\epsilon}{g(L_{\epsilon})},
\quad L_{\epsilon} \equiv \sqrt{D/\epsilon},  \label{eq:2.26}
\end{equation}
where $C_{d}$ is the dimension-and-coupling-constant-dependent
coefficient.  For a weakly interacting $1d$ electron gas $C_{d} =
\sqrt{2}$. Equation~(\ref{eq:2.26}) can be also rewritten as
\begin{equation}
\frac{1}{\tau_{e-e}(\epsilon)} = C_{d} \delta_{1}(L_{\epsilon}),
\label{eq:2.27}
\end{equation}
where $\delta_{1} (L) = (L^{d} \nu)^{-1}$ is the one-particle mean
level spacing in a $d$ - dimensional cube of the size $L$ 
and $\nu$  is the one-particle density of states.

There are several interpretations of this result.  One of them
\cite{AA} is based on the concept of the interaction time which
becomes much longer in the disordered case due to diffusive
rather than ballistic motion of electrons. It is also possible to
appeal to statistical properties of exact one-electron wave
functions~\cite{AGKL,AP}, and we outline this
interpretation below.

The inelastic rate $1/\tau_{e-e} (\epsilon)$ is determined by a pair of
collisions between electrons with all four energies - two initial
($\epsilon_\alpha $ and $\epsilon_\gamma$) and two final
($\epsilon_\beta$ and $ \epsilon_\delta$), all the energies here are
counted from the Fermi level.

Given the typical absolute value $M_{\alpha\beta\gamma\delta}$ of the
matrix element for such a collision in a sample of the size $L$, the
inelastic rate can be estimated~\cite{Aronov,AGKL} with the help of
the Fermi Golden Rule
\begin{equation}
\frac{\hbar}{\tau_{e-e}} \simeq \int\frac{d\epsilon_\beta \
d\epsilon_\gamma \ d\epsilon_\delta }{\left[\delta_{1} (L)\right]^3}
\left| M_{\alpha\beta\gamma\delta}\right|^2
\delta\left(\epsilon_\alpha - \epsilon_\beta + \epsilon_\gamma
-\epsilon_\delta\right) \left[1-n_F(\epsilon_\beta)\right] 
n_F(\epsilon_\gamma)\left[1-n_F(\epsilon_\delta)\right] 
.
\label{eq:2.28}
\end{equation}

The matrix elements can be represented as integrals of products of
four exact one-particle wave functions. In a disordered system, these
wave functions oscillate randomly in space, and are only weakly
correlated with each other.  As a result, the matrix elements are
random and for $L$ smaller than $ L_{\epsilon}$ ($0d$ case), their
typical absolute value $M(L)$ turns out to be of the order of
$\delta_{1}(L)/g(L)$, where the small factor $g^{-1}$ reflects the
weakness of the correlation between the wave functions. (In the limit
$g \rightarrow \infty$ Random Matrix Theory is valid; according to
this theory, there is no correlation at all between different
eigenvectors and the non-diagonal matrix elements vanish.) As a
result, integration in Eq.~(\ref{eq:2.28}) for $0d$ case can be easily
performed\cite{AP} and for $T=0$
\begin{equation}
\frac{\hbar}{\tau_{e-e}} \simeq \frac{\epsilon^{2}}{g^{2} \delta_{1}(L)}.
\label{eq:2.29}
\end{equation}

This rate increases with $\epsilon$, and at $L = L_{\epsilon}$, 
it becomes of the order of
$\delta_1(L_{\epsilon})$.  This estimate corresponds exactly to
Eqs.~(\ref{eq:2.26}) and (\ref{eq:2.27}), and it remains valid even
for large samples, $L > L_{\epsilon}$, since $ 1/\tau_{e-e} $ cannot
depend on $L$ in this limit.

This observation enables us to estimate the dependence of the matrix
elements in a $d$ -- dimensional sample on the energy transfer
$\omega=\epsilon_\alpha-\epsilon_\beta$, when $\omega$ is larger than
the Thouless energy.
Comparison of Eqs.~(\ref{eq:2.27}) and (\ref{eq:2.29}) 
yields for this dependence 
\begin{equation}
  |M_{\alpha\beta\gamma\delta}|^2 \sim \frac{\delta_{1} (L)^{3}\delta_{1}
    (L_\omega)}{\omega^2} = \frac{
    \delta_{1}(L)^{4}L^{d}}{\omega^2 L_{\omega }^{d}} \propto
  \omega^{-2+d/2}, \quad \omega = |\epsilon_\alpha - \epsilon_\beta|.
\label{eq:2.30}
\end{equation}
This energy dependence of the matrix elements reflects the
properties of noninteracting disordered system and is not
sensitive to the energy distribution of the electrons.

It follows from Eq.~(\ref{eq:2.30}) that the matrix elements diverge when
$d < 4$ and $\omega \rightarrow 0$. At $T=0$ this divergence is not
dangerous because of the two integrations in Eq.~(\ref{eq:2.28}).
However, the situation changes when the temperature is finite.
Assuming that the energy transfer $\omega$ is much smaller than the
temperature $T$, we perform the integration over $\epsilon_\gamma$ and
$\epsilon_\delta$ in Eq.~(\ref{eq:2.28}) and obtain with the help of
Eq.~(\ref{eq:2.30})
\begin{equation}
\frac{1}{\tau_{e-e}} = \int d\omega {\cal R}(\omega);\quad
{\cal R}(\omega)=\frac{T}{\omega g(L_\omega)}.
\label{eq:2.31}
\end{equation}
The integral over $\omega$ diverges in the infrared limit
for $d=1,2$.  Therefore, $1/\tau_{e-e}$ is ill-defined at finite
temperatures and in low dimensions\cite{AndersonLee}.

This is not a catastrophe, though: $1/\tau_{e-e}$ itself has no
physical meaning.  When the energy relaxation rate
$1/\tau_{\epsilon}$, (i.e. the inverse time of
thermalization of an excitation with energy $ \epsilon$ much
larger than temperature $T$) is considered, the quasielastic
processes are not important. To be more precise, for the energy
diffusion the relevant quantity is
\[
\langle\frac{\omega^2}{\tau_{e-e}}\rangle = 
\int_0^T d\omega {\omega^2}{\cal R}(\omega),
\]
which is perfectly convergent at the lower limit.
Therefore Eqs.~(\ref{eq:2.26}) and
(\ref{eq:2.27}) give a good estimate of $1/\tau_{\epsilon}$.

\begin{figure}
\vspace{0.2 cm}
\epsfxsize=8.7cm
\centerline{\epsfbox{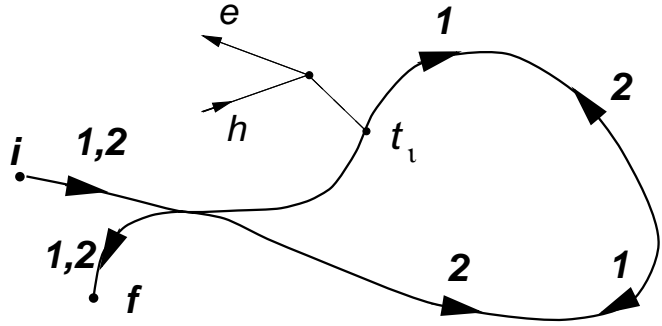}}
\vspace{0.5cm}
\caption{(a) Example of the coherent paths
involving excitation of the electron hole pair at point $t_i$.
The phase difference between paths $1$ and $2$ is 
$\delta\phi_{12} =\omega(t-2t_i)$,
where $\omega$ is the energy of the excited pair.
}
\label{Fig3}
\end{figure}

As it was mentioned in the end of  Subsection~\ref{sec:2.1}, the
characteristic scale of the magnetic fields in the weak localization
is determined by the dephasing rate $1/\tau_\phi$. It turns out that
this {\em dephasing rate at low dimensions is governed by the
quasielastic processes}.

Let us illustrate the microscopic mechanism of the dephasing. Consider
a pair of coherent paths $1$ and $2$ from Fig.~\ref{Fig1}b. Suppose
now that an electron during its motion along  path $1$ emits (or
absorbs) an electron-hole pair with the energy $\omega$, see
Fig.~\ref{Fig3}. Clearly, the interference term will be lost, unless
the electron going through path $2$  emits (absorbs)
the same electron-hole pair. In order to evaluate, e.g., the
conductivity, one has to sum up over all the final states. The
contribution of a final state with the extra electron-hole pair
contains the interference part from  trajectories $2$ and $1$ with
the emission of the very same pair.  The only difference between paths $1$
and $2$ is the following: for path $1$ the electron has momentum $k_i=k_F
+ \epsilon/v_F$ during the time $t_i$, see Fig.~\ref{Fig3}, and
then it has momentum $k_i - \omega/v_F$ during the remaining time
$t-t_i$; the electron going along path $2$ has momentum $k_i$
during the time $t-t_i$ and momentum $k_i - \omega/v_F$ during time
$\tau_i$.  As a result, the phases acquired by the trajectories are no
longer the same: 
\begin{equation}
\delta\varphi_{12}=\int{\bf k}_1d{\bf l}-\int{\bf k}_2d{\bf l}=
\omega (t-2t_i).
\label{eq:2.32}
\end{equation}
Since the effect of the dephasing is significant only for
$\delta\varphi_{12} \simeq 1$, the processes with the frequency, smaller
than the inverse length of the path do not lead to the dephasing, 
and thus the relevant quantity for the path with length $t$ is not the total
electron-electron relaxation rate (\ref{eq:2.31}) but rather
\begin{equation}
\frac{\hbar}{\tau_{\varphi}(t)} = 
\int_{1/t}^\infty d\omega {\cal R}(\omega)=
\frac{T}{g\left(\sqrt{Dt}\right)}.
\label{eq:2.33}
\end{equation}
It is this quantity that determines the dependence of the weak
localization correction on  temperature. If the magnetic field is
strong enough, $\tau_H \lesssim \tau_\varphi(\tau_H)$, 
see Eqs.~(\ref{eq:2.7}), the dephasing
introduces a small correction of the order of
$\tau_H/\tau_\varphi(\tau_H)$. On the other hand, if the magnetic field
is weak, the length of the relevant trajectories should be determined 
self-consistently by putting $t=\tau_\varphi$ in the right-hand side of 
Eq.~(\ref{eq:2.33})
\begin{equation}
\frac{\hbar}{\tau_{\varphi}} = 
\frac{T}{g\left(\sqrt{D\tau_\varphi}\right)}.
\label{eq:2.34}
\end{equation}
Solution of Eq.~(\ref{eq:2.34}) yields\cite{AA} 
\begin{mathletters}
\label{eq:2.35}
\begin{eqnarray}
\frac{1}{\tau_\varphi}=
\left(\frac{e^2T\sqrt{\hbar D}}
{\hbar\sigma_1}\right)^{2/3}; 
\quad  d=1,
\label{eq:2.35a}\\
\frac{1}{\tau_\varphi} = \frac{T}{\hbar} \frac{R_\Box e^2}{2\pi\hbar}
\ln \left(\frac{\pi\hbar}{e^2R_\Box}\right)
; \quad  d=2,
\label{eq:2.35b}
\end{eqnarray}
where $\sigma_1$ is the conductivity per unit length of a
one-dimensional system (wire) and $R_\Box$ is the sheet resistance of the
two-dimensional film. 
\end{mathletters}

An expression for $1/\tau_\varphi$ with the
exact numerical coefficient (as
Eqs.~(\ref{eq:2.35})) rather than an order of magnitude estimate
is meaningful only if it is supplied with the explicit dependencies of
the conductivity on $1/\tau_\varphi$ and $H$.  In order to evaluate these
dependencies, one notices that the dephasing is dominated by the
processes with the energy transfers much smaller than temperature. In
this case, instead of considering the interaction between electrons,
one can think about an electron moving in a {\em classical} electric
field, created by the rest of the electron gas. Since this field
fluctuates in time and space, the calculation of observable quantities
involves averaging over these fluctuations.  The program was realized
in Ref.~\onlinecite{AAK}, where it was shown that the weak
localization correction is still given by expression
similar to Eq.~(\ref{eq:2.4})
\begin{equation}
\delta\sigma_{WL} = - 
\frac{2\sigma}{\pi\nu\hbar}\int_0^\infty dt\ \langle {\cal C}
\left(\mbox{\boldmath $r$},\mbox{\boldmath
$r$};t.-t\right)\rangle_A,
\label{eq:2.36}
\end{equation} 
where the Cooperon in the external vector potential {\boldmath $A$}
is given by expression similar to Eqs.~(\ref{eq:2.5}) and (\ref{eq:2.8})
\begin{equation}
\left[\frac{\partial}{\partial t} + 
D\left(-i\mbox{\boldmath $\nabla$}_1 -\frac{e}{c\hbar}
\mbox{\boldmath $A$}(\mbox{\boldmath $r$}_1,t/2)
- \frac{e}{c\hbar}
\mbox{\boldmath $A$}(\mbox{\boldmath $r$}_1,-t/2)
\right)^2\right]{\cal
C}\left(\mbox{\boldmath $r$}_1,\mbox{\boldmath
$r$}_2;t,t^\prime\right)
=\delta\left(\mbox{\boldmath $r$}_1-\mbox{\boldmath
$r$}_2\right)\delta(t-t^\prime),
\label{eq:2.37}
\end{equation}
where $\mbox{\boldmath $A$}(\mbox{\boldmath $r$},t)$ is the
fluctuating vector potential created due to the emission and
adsorption of the real electron-hole pairs. Those fields are Gaussian
ones with the propagator defined as
\begin{equation}
\langle A^iA^j\rangle_{Q,\omega} = -\frac{2T}{\omega}\
\frac{c^2Q^iQ^j}{e^2\omega^2}\ {\rm Im}\ U(Q,\omega) =
\frac{2T}{\sigma_d}\frac{c^2}{\omega^2}\frac{Q^iQ^j}{Q^2},
\label{eq:2.38}
\end{equation}
where $U(Q,\omega)$ is the screened interaction potential defined in
Eq.~(\ref{eq:2.23}), and $\sigma_d$ is the conductivity in $d$
dimensions. The last expansion in this formula is possible since the
relevant frequencies and wavevectors are such that $Q \simeq
\sqrt{\omega/D} \simeq 1/L_\varphi \ll 1/a_s$, where $a_s$ is the screening
radius. Formula (\ref{eq:2.38}) is consistent with Eq.~(4.21) of
Ref.~\onlinecite{AA}. The first term in the latter formula describes
the transverse fluctuation of the electromagnetic field which is
usually small and we will not take them into account. Calculation
described in great details in Sec.~4.2 of Ref.~\onlinecite{AA} yields
\begin{mathletters}
\label{eq:2.39}
\begin{eqnarray}
\delta\sigma_{WL}= -\frac{e^2}{2\pi^2\hbar} \left[\ln
\frac{1}{\tau\Omega_H}- \Psi\left(\frac{1}{2}
+\frac{1}{\tau_\varphi\Omega_H}\right)\right],\quad d=2,\
\left(\mbox{\boldmath $n$} \parallel \mbox{\boldmath $H$}\right)
;
\label{eq:2.39a}
\\
\delta\sigma_{WL}= -\frac{e^2}{2\pi^2\hbar} \ln
\frac{\tau_H\tau_\varphi}{\tau \left(\tau_H +\tau_\phi\right)}
,\quad d=2,\
\left(\mbox{\boldmath $n$} \perp \mbox{\boldmath $H$}\right);
\label{eq:2.39a1}\\
\delta\sigma_{WL}= -\frac{e^2\sqrt{ D\tau_\varphi}}{\pi\hbar}
\frac{1}{\left[- \ln{\rm Ai}\left(\tau_\varphi/\tau_H\right)
\right]^{\prime}} ,\quad d=1,
\label{eq:2.39b}
\end{eqnarray}
where the dephasing time $\tau_\varphi$ and the time  $\tau_H$ 
associated 
with magnetic field  are
given by Eqs.~(\ref{eq:2.35}) and (\ref{eq:2.7}) respectively 
for different dimensions $d$,
${\rm Ai}(x)$ is the Airy function, and $\Psi(x)$ is the digamma-function.
Forms of Eqs.~(\ref{eq:2.39}) are consistent with the general relation
(\ref{eq:2.10})\cite{footnote2}.
\end{mathletters}

It is important to emphasize that derivation of Eqs.~(\ref{eq:2.36})
- (\ref{eq:2.38}) is neither an assumption nor a phenomenological recipe
-- it is the result of the {\em parametrically justified procedure}. All the
corrections to these formulas are small as $1/(\tau_\varphi T)$. As such,
this calculation is already self-contained, however
some people find it useful to interpret the results
appealing to  Gedanken
experiments similar to that of Ref.~\onlinecite{Stern}.

 For further considerations  it is useful to rewrite Eq.~(\ref{eq:2.39})
 in a slightly different form.
First, we perform a gauge transformation of Eq.~(\ref{eq:2.37}):
\begin{equation}
\left[\frac{\partial}{\partial t} + 
D\left(-i\mbox{\boldmath $\nabla$}_1 -\frac{2e}{c\hbar}
\mbox{\boldmath $A$}(\mbox{\boldmath $r$}_1)\right)^2
+ i\left(\chi(\mbox{\boldmath $r$}_1,t/2)-
\chi(\mbox{\boldmath $r$}_1,-t/2)\right)
\right]{\cal
C}\left(\mbox{\boldmath $r$}_1,\mbox{\boldmath
$r$}_2;t,t^\prime\right)
=\delta\left(\mbox{\boldmath $r$}_1-\mbox{\boldmath
$r$}_2\right)\delta(t-t^\prime),
\label{eq:2.40}
\end{equation}
where $\mbox{\boldmath $A$}(\mbox{\boldmath $r$})$ represents external
magnetic field, while the fluctuating electric field caused by the
interaction between electrons enters through a fluctuating {\em
scalar} potential $\chi(\mbox{\boldmath $r$},t)$. According to
Eq.~(\ref{eq:2.38}), correlation function of this scalar potential in
$(\omega,Q)$ representation equals to
\begin{equation}
\langle\chi^2\rangle_{Q,\omega} = -\frac{2T}{\omega}\
 {\rm Im}\ U(Q,\omega) =
\frac{2Te^2}{\sigma_dQ^2}.
\label{eq:2.41}
\end{equation}
Secondly, for the regime $\tau_H \ll \tau_\varphi$, the fluctuating
potential $\chi$ can be taken into account by perturbation theory.
This yields the correction to the conductivity of the form
\begin{eqnarray}
\delta\sigma_{WL} &=& - \frac{\sigma}{\pi\nu\hbar}
\left[
\int \frac{d^d Q}{\left(2\pi\right)^d}
{\cal C}(Q,0)
- \int \frac{d^d Qd^d
q}{\left(2\pi\right)^{2d}} \int_{-T}^{T}\frac{d\omega}{2\pi}
\left(\frac{4Te^2}{\sigma_dq^2}\right)
\left(
\left[{\cal C}(Q,0)\right]^2
{\cal C}(\mbox{\boldmath $Q$}+\mbox{\boldmath $q$},\omega)
-\left|{\cal C}(Q,\omega)\right|^2
{\cal C}(\mbox{\boldmath $Q$}+\mbox{\boldmath $q$},0)
\right)
\right],
\nonumber\\
{\cal C}(Q,\omega)&=&\frac{1}{-i\omega+DQ^2+\frac{1}{\tau_H}},
\label{eq:2.42}
\end{eqnarray}
where the first term in brackets is just a weak-localization
correction for non-interacting electrons and the second term is the
perturbative effect of dephasing on the weak localization.  Since in
$d=1$ the integral over $\omega$ is determined by its lower limit, the
upper limit $|\omega|=T$ is not important. In $d=2$ the integral is
logarithmical and the exact definition of the cut-off is also not
necessary.  Integration in Eq.~(\ref{eq:2.42}) gives exactly the same
result as the direct expansion of Eqs.~(\ref{eq:2.39}) in the small
parameter $\tau_H/\tau_\varphi$. Here, and through the rest of this
paper we will consider only one-dimensional wires and the films in the
parallel magnetic field to avoid unnecessary technical complications.

\subsection{Summary of this section.}
This section is a review of the results on the effects of
electron-electron interaction on the conductivity of disordered
metals. We tried to give the physical interpretation to all of the
rigorous formulas. The main message is that the microscopic mechanism of
dephasing is the excitation or absorption of {\em real} electron-hole
pairs.  That is why the frequency transferred through the interaction
propagator in the formulas for dephasing can not be much larger than
temperature, since excitations with higher energy are forbidden by
Pauli principle.  As a result the dephasing rate $1/\tau_\phi$ {\em
vanishes} when the temperature tends to zero (see Eqs.~(\ref{eq:2.35})).
The interaction correction associated with higher frequency {\em are
not due to the real processes} but rather due to the scattering of
the electron off the {\em static} self-consistent potential created by
all the other electrons. Therefore, this scattering has nothing to do
with the dephasing. In the next section, we outline ``new theories'' of
dephasing which attempt to revise this physical picture.

\section{Recent ``theories'' of zero-temperature dephasing}
\label{sec:3}

Recently, Mohanty and
Webb\cite{Mohanty2} suggested the following expression [formula (4) of
Ref.~\onlinecite{Mohanty2}] for the
dephasing time
\begin{equation}
\frac{1}{\tau_p} = \frac{e^2}{\sigma \hbar^2}\int du
\int \frac{d^d k}{\left(2\pi\right)^d}\int \frac{d\omega}{2\pi}
\hbar\omega \coth\left(\frac{\hbar \omega}{2T}\right) k^{-2}
\exp\left(-Dk^2|u|-i\omega u\right).
\label{eq:3.1}
\end{equation}
We will denote this time $\tau_p$ to distinguish it from the results
of previous section. We also use integration $\frac{d^d
k}{\left(2\pi\right)^d}$ instead of their $\frac{d^2
k}{\left(2\pi\right)^2}$.  ( We do not quite understand why the
authors of Ref.~\onlinecite{Mohanty2} conclude that ``In a quasi-1D
system, for any two interfering paths the wave vector {\bf k} must be
a two-dimensional vector'').  Formula (\ref{eq:3.1}) was obtained as
an extension of the derivation of Refs.~\onlinecite{AAK,AA} in the
form popularized in Ref.~\onlinecite{Chakravarty}. This extension can
be hardly called a derivation.

Golubev and Zaikin\cite{Zaikin1,Zaikin2} claimed that they derived
Eq.~(\ref{eq:3.1}). To obtain Eq.~(\ref{eq:3.1}), one needs just to
substitute Eq.(74) of Ref.~\onlinecite{Zaikin2} into Eq.~(71) of the
same paper, see also Eqs.~(7) and (8) of Ref.~\onlinecite{Zaikin1}.
 We will discuss derivation of Ref.~\onlinecite{Zaikin2}
in more details later.

Beforehand we would like to to emphasize the differences between 
Eq.~(\ref{eq:3.1}) and the results of previous section. 
Though authors do not write it explicitly, the magnetic field dependence of 
the conductivity in Refs.~\onlinecite{Mohanty1,Mohanty2,Zaikin1,Zaikin2} is 
presumably given by
\begin{equation}
\delta\sigma_{WL} = - \frac{\sigma}{\pi\nu\hbar}
\int \frac{d^d Q}{\left(2\pi\right)^d}
\frac{1}{DQ^2+\frac{1}{\tau_H} +\frac{1}{\tau_p}}
\label{eq:3.2}
\end{equation}
Consider the strong magnetic field $\tau_H \ll\tau_p$
limit.  As it was already mentioned, in this limit one can expand
Eq.~(\ref{eq:3.2}) in small parameter $\tau_H/\tau_p$.  Let us
restrict ourselves to the first order of this expansion, substitute
$\tau_p$ from Eq.~(\ref{eq:3.1}) and integrate over $u$ in
Eq.~(\ref{eq:3.1}). The result is
\begin{equation}
\delta\sigma_{WL} = - \frac{\sigma}{\pi\nu\hbar}
\left[
\int \frac{d^d Q}{\left(2\pi\right)^d}
{\cal C}(Q,0)
- \int \frac{d^d Qd^d
q}{\left(2\pi\right)^{2d}} \int_{-1/\tau}^{1/\tau}\frac{d\omega}{2\pi}
\left(\frac{2Te^2}{\sigma_dq^2}\right)
\left(\hbar\omega\coth\frac{\hbar\omega}{2T}\right)
\left[{\cal C}(Q,0)\right]^2
\frac{1}{-i\omega + Dq^2}
\right]
\label{eq:3.3}
\end{equation}   
where ${\cal C}(Q,\omega)$ is defined in Eq.~(\ref{eq:2.42}).

Let us now compare Eq.~(\ref{eq:3.3}) with the conventional result
(\ref{eq:2.42}).  The first terms of both equations describe weak
localization of non-interacting correction, and they simply coincide.
There is no dramatic difference in the contributions to the second
terms from small energy transfers. Indeed, the integrand in the second
term of Eq.~(\ref{eq:3.3}) at $\hbar\omega \ll T$ is of the same order
of magnitude as the one in the second term of Eq.~(\ref{eq:2.42}).  In
order to avoid an unphysical conclusion that some dephasing can be
caused by the fluctuations with $\omega=0$ or $q=0$, i.e., by a static
or uniform potential, the authors of
Refs.~\onlinecite{Mohanty1,Mohanty2,Zaikin1,Zaikin2}
introduced an {\em ad hoc} infrared cut-off at $\omega \simeq 1/\tau_p$.

The difference at high frequencies is much more serious. While
Eq.~(\ref{eq:2.42}) limits the frequency integration over the energies
smaller than temperature, Eq.~(\ref{eq:3.3}) extends the integration
over the whole range of frequencies. As a result of this extension,
the dephasing rate becomes finite at $T\to 0$.  Thus, according to
Eq.~(\ref{eq:3.3}) {\em the processes with the energy transfer much
larger than temperature do break the time reversal symmetry and do cause
dephasing}. On the other hand, all the physical arguments of
Sec.~\ref{sec:2} apparently contradict to this statement.  

We emphasize that it is extremely difficult to accept the idea of the
zero-point motion dephasing from the qualitative point of
view. Consider a quantum particle which moves in the environment of
the harmonic oscillators: each oscillator is characterized by its
frequency $\omega$. The result of the collision of the particle with
an oscillator depends on the relation between $\hbar\omega$ and the
temperature.  If $\hbar\omega \ll T$, the oscillator  is in an 
excited state with a high
probability. The particle energy, counted from
the Fermi level, is also of the order of $T$, i.e. much larger than
$\hbar\omega$. As the result, the energy conservation does not prevent
the energy transfer between the particle and the oscillator, i.e. the
probability for the inelastic collision is substantial. 

The situation for $\hbar\omega \gg T$ is quite different. Indeed, up
to the exponentially small terms, the oscillator is in the ground
state, and the particle has the energy smaller than
$\hbar\omega$. Therefore, the energy transfer is forbidden by the
energy conservation and the collision is {\em elastic}. Therefore,
there is no difference whatsoever between the collision with such an
oscillator and with the quenched disorder, which definitely does not
dephase. As to the zero-point motion, their energy is simply the
difference between the ground state energy of the oscillator and the
bottom of the harmonic potential. Since, this is the energy of the
ground state, it can not be absorbed, and the smearing of the
wave-function of the oscillator is just a renormalization of the
cross-section for the {\em elastic} processes.

However,
maybe our intuition is wrong? We did not find any convincing physical
argument which would require to reconsider the common understanding of
dephasing in disordered systems in
Refs.~\onlinecite{Mohanty1,Mohanty2,Zaikin1,Zaikin2}.

The only theoretical argument in favor of the zero-temperature dephasing
presented up to now is the calculation of Ref.~\onlinecite{Zaikin2}.
We do not want to reject these calculations entirely on the basis of
the qualitative picture. The history of Physics knows examples when a
calculation preceded an understanding. We are going to demonstrate
that it is not the case in the present situation. As we explain in the
next section, the results of Golubev and Zaikin, being purely
perturbative ($1/\tau_p$ is proportional to the first power of the
fluctuation propagator), explicitly contradict to the straightforward
calculation based on the perturbation theory.

\section{Main idea of the calculation and results.}
\label{sec:4}

\begin{figure}
\vspace{0.2 cm}
\epsfxsize=10.7cm
\centerline{\epsfbox{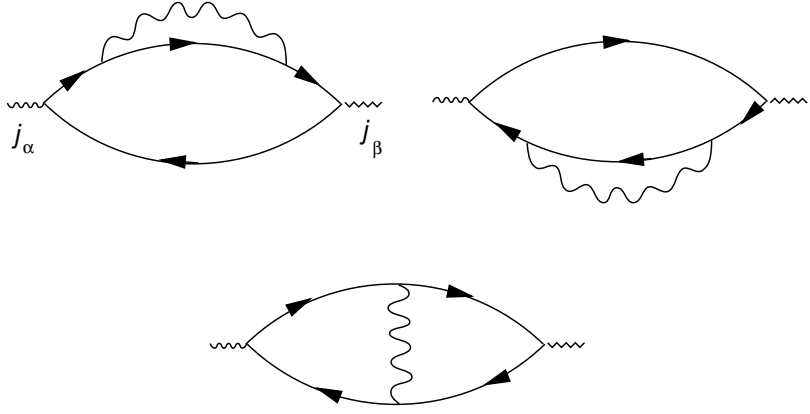}}
\vspace{0.8cm}
\caption{Diagrams for the first order corrections to the
conductivity. Electron Green functions (solid lines) are exact for the given
disordered potential. Wavy lines stand for the interaction propagator
with Keldysh components ${\cal L}^R$, ${\cal L}^A$, ${\cal L}^K$,
see Sec.~\protect\ref{sec:5}.
}
\label{Fig4}
\end{figure}

The main idea of the calculation we are about to describe is the
observation that in the strong magnetic field both results of the
conventional theory (\ref{eq:2.42}) and of the ``new theory''
(\ref{eq:3.3}) are purely perturbative in the interaction propagator
$U(Q,\omega)$, see Eq.~(\ref{eq:2.23}). Therefore, the truth may be
revealed just by straightforward calculation of the interaction
correction to the conductivity, see Fig.~\ref{Fig4}.  We believe that
we understand the qualitative physical picture.  Nevertheless, we will
not use this understanding in the calculation, trying, e.g., in the
very beginning to identify ``diagrams relevant for inverse inelastic
time $1/\tau_i$ \dots'' or to introduce cut-offs in the integrations
using some ideas based on the qualitative understanding.  Instead, we
will take into account {\em all} the diagrams of the first order in
the interaction and integrate over the {\em whole} interval of the
energy transfer. The only small parameter we will afford to use is the
inverse dimensionless conductance of the system.

As usual, we start with expressing the correction to the conductivity
through electronic and interaction propagators in a given disorder
realization.  This expression should be then averaged over disorder.
For our purposes this averaging should be performed with taking the
weak localization correction into account. It means that {\em both
zero and first order} in the inverse dimensionless conductance $1/g$
should be kept while we average. Therefore, after the averaging and
integration over the energies and momenta transfers, we will obtain
corrections to the conductivity which are of the {\em first and second
} order in $1/g$ .

It is important to emphasize that we  evaluate the interaction
correction to the observable (and thus gauge invariant) quantity --
conductivity, rather than to a gauge noninvariant object like
Cooperon.

Before we dig into details of the calculation, it is useful to realize
what kind of correction we are interested in. 
In Sec.~\ref{sec:2.2} the two kind of interaction effects -- dephasing and
interaction correction -- were mentioned. 
Is it possible to separate  the two corrections, when they are of
the same order in $1/g$? 
Fortunately, one can propose a procedure, 
which allows to perform this  separation unambiguously.

Indeed, let us write down the total correction to the
conductivity in the form:
\begin{equation}
\delta\sigma = \delta\sigma_{WL}+ \delta\sigma_{C}
+ \delta\sigma_{\rm deph} + \delta\sigma_{\rm CWL} + \dots
\label{eq:4.1}
\end{equation}
where $\dots$ stand for the corrections of the order higher than the
second order in $1/g$.  Here, leading weak localization and interaction
corrections to the conductivity are given by, see Eqs.~(\ref{eq:2.39})
and Eqs.~(\ref{eq:2.250})
\begin{mathletters}
\label{eq:4.2}
\begin{eqnarray}
&\displaystyle{\delta\sigma_{WL}= -\frac{e^2}{2\pi^2\hbar}\ln
\frac{1}{\tau\Omega_H};
\quad
\delta\sigma_C(T) = - \frac{e^2}{2\pi^2\hbar}
\ln \left(\frac{\hbar}{T\tau}\right);\quad d=2;}&\\
&\displaystyle{\delta\sigma_{WL}=-\frac{e^2\sqrt{ D\tau_H}}{\pi\hbar}
;
\quad
\delta\sigma_C(T) = - \frac{e^2}{\pi\hbar}
\sqrt{\frac{\hbar D}{2\pi T}}
\left(\frac{3 \zeta\left(3/2\right)}{2}\right)
;\quad d=1,}&
\end{eqnarray}
and time $\tau_H \gg \hbar/T$ is given by Eqs.~(\ref{eq:2.7}). 
\end{mathletters}
The term $\sigma_{\rm deph}$ 
appears as the first order correction to the time
scale cutting off the weak localization correction. In accordance with
the previous discussion, it should have the form 
\begin{mathletters}
\label{eq:4.3}
\begin{eqnarray}
&\displaystyle{\delta\sigma_{\rm deph}= \frac{e^2}{2\pi^2\hbar}
\frac{\tau_H}{\tau_\varphi};
\quad d=2;}&\\
&\displaystyle{\delta\sigma_{\rm deph}=\frac{e^2\sqrt{
D\tau_H}}{\pi\hbar}
\frac{\tau_H}{\tau_\varphi}
;\quad d=1,}&
\end{eqnarray}
where the scale $1/\tau_\varphi$ is to be determined. If the calculations
of Golubev and Zaikin were correct, this scale  would coincide with
that given by Eqs.~(77) and (81) of Ref.~\onlinecite{Zaikin2}.
\end{mathletters}

The last term in Eq.~(\ref{eq:4.1}) characterizes the effect of weak
localization on the interaction correction and vice versa, which do not
break the time reversal invariance. On purely dimensional grounds they
can be written as
\begin{mathletters}
\label{eq:4.4}
\begin{eqnarray}
&\displaystyle{\delta\sigma_{\rm CWL}= \frac{e^2}{2\pi^2\hbar}
 \frac{R_\Box e^2}{2\pi^2\hbar}\left(
\alpha_1\ln^2 T\tau + \alpha_2\ln^2 \Omega_H\tau +
\alpha_3 \ln T\tau\ln \Omega_H\tau
\right);
\quad d=2;}&\\
&\displaystyle{\delta\sigma_{\rm CWL}=\alpha
\frac{e^2\sqrt{
D\tau_H}}{\pi\hbar} \left(\frac{e^2 }{\pi \hbar \sigma}\sqrt{
\frac{\hbar D}{T}}\right)
;\quad d=1,}&
\end{eqnarray}
where $\sigma_1$ is the conductivity per unit length of a
one-dimensional system (wire), $R_\Box$ is the sheet resistance of the
two-dimensional film, and $\alpha_i$ are numerical coefficients to be
found.  One can separate contributions (\ref{eq:4.3}) and
(\ref{eq:4.4}) using the apparent difference in the magnetic field
dependence of these two terms.
\end{mathletters}

Now it is time to outline our strategy. We calculate the correction to
the conductivity in the first order in fluctuation propagator. Next
step is to average the result over disorder, taking weak localization
into account in each term. Finally, we separate the result into terms
of the form (\ref{eq:4.3}) and (\ref{eq:4.4}), thus evaluating the
dephasing rate $1/\tau_\varphi$.

This program is described in details in Sec.~\ref{sec:5}, but we would
like to present and discuss the results right now:
\begin{enumerate}
\item All the terms which scale according to Eq.~(\ref{eq:4.3}) are
produced by the expression of the form similar to Eq.~(\ref{eq:2.42})
\begin{equation}
\delta \sigma_{\rm deph} =  \frac{\sigma}{\pi\nu\hbar}
\int \frac{d^d Qd^d
q}{\left(2\pi\right)^{2d}} 
\int\frac{d\omega}{2\pi}f\left(\frac{\omega}{T}\right)
\left(\frac{4Te^2}{\sigma_dq^2}\right)
\left(
\left[{\cal C}(Q,0)\right]^2
{\cal C}(\mbox{\boldmath $Q$}+\mbox{\boldmath $q$},\omega)
-\left|{\cal C}(Q,\omega)\right|^2
{\cal C}(\mbox{\boldmath $Q$}+\mbox{\boldmath $q$},0)
\right),
\label{eq:4.5}
\end{equation}
where Cooperon ${\cal C}$ is defined by Eq.~(\ref{eq:2.42}) and
dimensionless function $f$ is given by 
\begin{equation}
f\left(\frac{\omega}{T}\right)= \left(\frac{\omega}{4T}\right)
\int d\epsilon \left(\frac{d}{d\epsilon} \tanh
\frac{\epsilon}{2T}\right)
\left(\coth \frac{\omega}{2T} + \tanh \frac{\epsilon-\omega}{2T}\right)=
\left(\frac{\omega}{2T}\right)^2 \frac{1}{\sinh^2\frac{\omega}{2T}}.
\label{eq:4.6}
\end{equation}
\item
For the cross term of interaction and weak localization of the type
(\ref{eq:4.4}) we found
\begin{equation}
\delta\sigma_{CWL}=\frac{e^2}{\pi^2\hbar} 
\frac{e^2}{\pi^2\hbar\sigma_d} 
\left(\frac{T}{\hbar D}\right)^{d-2}
\frac{1}{2^d}
\left[ \left(2\pi T\tau_H\right)^{1-d/2}
\frac{\ (4-d)}{d}\Gamma^2(2-d/2)
\zeta\left(2-d/2\right) - 
\frac{2}{2-d}
\zeta\left(3-d\right) \right],
\label{eq:4.7}
\end{equation}
where $\Gamma(x)$ is the Gamma function and $\zeta (x)$ is the Riemann
zeta-function. Here, we retained only most singular at $d \to 2$
contribution into the second not dependent on magnetic field term.
\end{enumerate}

Note, that in Eq.~(\ref{eq:4.7}) the ultraviolet convergence is
provided by the function $f\left(\omega/T\right)$ from
Eq.~(\ref{eq:4.6}), which falls down exponentially at $\omega \gg T$
(rather than by the limit $\simeq T$ as in Eq.~(\ref{eq:2.42})).  The
structure of $f\left(\omega/T\right)$ deserves some
discussion. The $coth$ term is the only term which was taken into
account in Ref.~\onlinecite{Mohanty2,Zaikin1,Zaikin2}. This term
indeed dominates for $\omega \ll T$.  At high frequencies, however,
this term is canceled out almost exactly by the second $tanh$
term\cite{Reyzerphi}.  It is this term, neglected in
Ref.~\onlinecite{Mohanty2,Zaikin1,Zaikin2}, that is responsible for
the Pauli principle. The clearly indicates the inelastic origin
$\tau_\phi$ in accord with the arguments of Sec.~\ref{sec:2}. The
reason why the authors of Ref.~\onlinecite{Zaikin1,Zaikin2} missed
this term is, as it is explained in Sec.~\ref{sec:6}, in the incorrect
procedure of the disorder average.

Since Eq.~(\ref{eq:4.5}) is essentially the same as
Eq.~(\ref{eq:2.42}) in the frequency domain $\omega \ll T$, which gives
the main contribution in low dimensional systems, this is not
surprising that the results of the integration are perfectly
consistent with Refs.~\onlinecite{AA,AAK}. In the leading order in
$T\tau_H$, we found
\begin{mathletters}
\begin{eqnarray}
&\displaystyle{\delta \sigma_{\rm deph}= \frac{e^2}{2\pi^2\hbar}
 \frac{T\tau_H}{\hbar} \frac{R_\Box e^2}{2\pi\hbar}
\ln \left(\frac{T\tau_H}{\hbar}\right)
; \quad d=2;}&
\\
&\displaystyle{\delta\sigma_{\rm deph}=\frac{e^2\sqrt{
D\tau_H}}{\pi\hbar}
{\tau_H}^{3/2}D^{1/2}\frac{e^2T}{4\hbar^2\sigma_1}
;\quad d=1,}&
\end{eqnarray}
\label{eq:4.8}
which according to Eq.~(\ref{eq:4.3}) means that
\end{mathletters}
\begin{mathletters}
\begin{eqnarray}
&\displaystyle{\frac{1}{\tau_\varphi}= 
 \frac{T}{\hbar} \frac{R_\Box e^2}{2\pi\hbar}
\ln \left(\frac{T\tau_H}{\hbar}\right)
; \quad d=2;}&
\\
&\displaystyle{\frac{1}{\tau_\varphi}=
\frac{T}{4\hbar}
\frac{e^2\sqrt{D\tau_H}}{\hbar\sigma_1}
;\quad d=1.}&
\end{eqnarray}
\label{eq:4.9}
\end{mathletters}
Equations (\ref{eq:4.9}) are applicable provided that $\tau_H \ll
\tau_\varphi$; in the opposite case $1/\tau_H$ in the right-hand side of
these equations, should be substituted by $1/\tau_\varphi$ and the
procedure of Ref.~\onlinecite{AAK} is required for finding the
temperature dependence for the magnetoresistance.

The cross term of weak localization and interaction (\ref{eq:4.7}) is
smaller than the dephasing term. We obtain
for a one dimensional case from Eq.~(\ref{eq:4.7})  
\begin{eqnarray}
\label{eq:4.10}
&\displaystyle{\delta\sigma_{\rm CWL}=
\frac{e^2\sqrt{
D\tau_H}}{2\pi\hbar} \left(\frac{e^2 }{2\pi \hbar \sigma}\sqrt{
\frac{\hbar D}{2\pi T}}\right) {3\zeta(3/2)}
;\quad d=1,}&
\end{eqnarray}
where $\sigma_1$ is the conductivity per unit length of
one-dimensional system (wire), and $\zeta(x)$ is the Riemann
zeta-function, $\zeta (3/2) = 2.612\dots$.

Smallness of the term (\ref{eq:4.10}) in
comparison with the dephasing term (\ref{eq:4.9}) one more time illustrates
the simple fact that the procedure of Ref.~\onlinecite{AAK} is the
parametrically justified program of summation of most divergent term
in each order of perturbation theory in the interaction.

The resulting conductivity is given by the sum of Eq.~(\ref{eq:4.5})
and Eq.~(\ref{eq:4.7}). However, keeping the dephasing term in the
leading approximation (\ref{eq:4.8}) and taking into account the
correction (\ref{eq:4.7}) would be overstepping the accuracy of the
approximation, since the contribution from $\omega \simeq T$ in
Eq.~(\ref{eq:4.5}) is of the same order as
Eq.~(\ref{eq:4.7}). Therefore, the calculation of the integral in
Eq.~(\ref{eq:4.5}) should be performed with the same accuracy as
Eq.~(\ref{eq:4.7}). It gives
\begin{eqnarray}
\delta\sigma_{deph}&=&\frac{e^2}{\pi^2\hbar} 
\frac{e^2}{\pi\hbar\sigma_d}
\frac{4\Gamma^2\left(2-d/2\right)}{2^d\left(2-d\right)} 
\nonumber\\
&\times&
\left\{
\left[\frac{T\tau_H}{\hbar}\right]
\left[
\frac{1}{2(3-d)}
\left(\frac{1}{2\pi D\tau_H}\right)^{d-2}
+
\left(\frac{d \cos \frac{\pi d}{4}}{2\pi}\right)
\zeta\left(\frac{d}{2}\right)
\Gamma\left(\frac{d}{2}\right)
\left(\frac{1}{2\pi D\tau_H}\right)^{d/2-1}
\left(\frac{T}{2\pi\hbar D}\right)^{d/2-1}
\right]
\right. 
\nonumber\\
 &+&\left.
\frac{1}{2\pi}
\left(\frac{T}{\hbar D}\right)^{d-2}
\left[
\zeta\left(2-\frac{d}{2}\right)
\frac{(2-d)(3-d)}{4}
 \left(2\pi T\tau_H\right)^{1-d/2}
 +
\frac{2d}{6-d}
\frac{\Gamma (3-d) \Gamma (d/2)}{\Gamma (2-d/2)}
\zeta\left(3-d\right) \right]
\right\}.
\label{eq:4.11}
\end{eqnarray}

Summing Eq.~(\ref{eq:4.7}) and Eq.~(\ref{eq:4.11}), we obtain the final
result for the overall effect of the electron-electron interaction on
the weak localization correction
\begin{eqnarray}
\delta\sigma_{I\times WL}&=&\frac{e^2}{\pi^2\hbar} 
\frac{e^2}{\pi\hbar\sigma_d}
\frac{4\Gamma^2\left(2-d/2\right)}{2^d} 
\nonumber\\
&\times&
\left\{
\frac{1}{2-d}
\left[\frac{T\tau_H}{\hbar}\right]
\left[
\frac{1}{2(3-d)}
\left(\frac{1}{2\pi D\tau_H}\right)^{d-2}
+
\left(\frac{d \cos \frac{\pi d}{4}}{2\pi}\right)
\zeta\left(\frac{d}{2}\right)
\Gamma\left(\frac{d}{2}\right)
\left(\frac{1}{2\pi D\tau_H}\right)^{d/2-1}
\left(\frac{T}{2\pi\hbar D}\right)^{d/2-1}
\right]
\right. 
\nonumber\\
 &+&\left.
\frac{1}{2\pi}
\left(\frac{T}{\hbar D}\right)^{d-2}
\left[
\zeta\left(2-\frac{d}{2}\right)
\frac{8+d-d^2}{4d}
 \left(2\pi T\tau_H\right)^{1-d/2}
 +
\frac{1}{2-d}
{\cal O}(1)
 \right]
\right\}.
\label{eq:4.12}
\end{eqnarray}
Taking the last term in brackets into account would be beyond the
accuracy of the theory since the second order interaction correction
gives the similar contribution.

\begin{mathletters}
Let us write down the final results for the correction (\ref{eq:4.12})
in different dimensionalities
\begin{eqnarray}
\delta\sigma_{I\times WL}&=&\frac{e^2}{\pi\hbar} 
\frac{e^2}{\hbar\sigma_d}
\left\{
D\tau_H
\left(\frac{T\tau_H}{4\hbar}\right)
\left[1+\zeta\left(\frac{1}{2}\right)
\sqrt{\frac{2\hbar}{\pi T\tau_H}}
\right]
+
\frac{\zeta\left(\frac{3}{2}\right)}{\pi}
\sqrt{\frac{ \hbar D^2\tau_H}{2\pi T}}
\right\}, \quad d = 1,\\
\delta\sigma_{I\times WL}&=&
\frac{e^2}{2\pi^2\hbar} \frac{R_\Box e^2}{2\pi^2\hbar}
\left\{
 \frac{\pi T\tau_H}{\hbar}
\left[
\ln \left(\frac{T\tau_H}{\hbar}\right)
+1
\right]
+\frac{3}{2}\ln \left(\frac{\tau_H}{\tau}\right)
+{\cal O}\left[\ln \left(T\tau/\hbar\right)\right]
\right\}
,
\quad d=2
\end{eqnarray}
\label{eq:4.13}
\end{mathletters}
where $\zeta (1/2) = -1.461\dots$, $\zeta (3/2) = 2.612\dots$.

To conclude, the results of the well-controlled perturbation theory
contradict to those of Ref.~\onlinecite{Zaikin2}. Thus, the
statement made in this paper that the ``analysis should in principle
include {\em all} diagrams'' is false.  The theoretical conclusion
about saturation of the dephasing rate at $T \to 0$ contradicts not
only the physical intuition but also the straightforward calculation.

\section{Calculation of the interaction correction to the
conductivity}
\label{sec:5}
\subsection{Conductivity for a given disorder realization}
In this section we will use units $\hbar =c =1$.  We start with the
conventional expression for the current density $\mbox{\boldmath
$j$}(\mbox{\boldmath $r$},t)$ in terms of the electron Green functions
in Keldysh technique\cite{Keldysh}
\begin{equation}
\mbox{\boldmath $j$}(\mbox{\boldmath $r$},t) = \left.-\frac{1}{4m}{\rm Tr}^K
\left(\hat{\tau}_2 -2\hat{\tau}_1\right)
\left[
\mbox{\boldmath $\nabla$}_{\mbox{\boldmath $r$}_1}
- \mbox{\boldmath $\nabla$}_{\mbox{\boldmath $r$}_2}
 - 2ie
\mbox{\boldmath $A$}(\mbox{\boldmath $r$})
\right]
\hat{G}\left(\mbox{\boldmath $r$}_1, t_1;\ 
\mbox{\boldmath $r$}_2,t_2\right)\right|_{\matrix{\mbox{\boldmath
$r$}_1=
\mbox{\boldmath $r$}_2 \cr
t_1 = t_2 + 0 =t
}}
\label{eq:5.1}
\end{equation}
where the trace is performed in the Keldysh space, 
the matrix Green function is defined as
\begin{equation}
\hat{G}(1,2) = 
\left(
\matrix{G^R(1,2) & G^K(1,2) \cr 0 & G^A(1,2)}
\right),
\label{eq:5.2}
\end{equation}
and the Pauli matrices in the Keldysh space are
\begin{equation}
\hat{\tau}_1 =
\left(
\matrix{1 & 0 \cr 0 & 1}
\right),
\quad
\hat{\tau}_2 =
\left(
\matrix{0 & 1 \cr 1 & 0}
\right).
\label{eq:5.3}
\end{equation}
Entrees in Eq.~(\ref{eq:5.2}) are given by
\begin{mathletters}
\begin{eqnarray}
iG^R(1,2) = \theta(t_1 - t_2)\langle \hat{\psi}(1)
\hat{\psi}^\dagger(2) 
+
\hat{\psi}^\dagger(2)\hat{\psi}(1) 
\rangle
\label{eq:5.4a}
\\
iG^A(1,2) = - \theta(t_2 - t_1)\langle \hat{\psi}(1)
\hat{\psi}^\dagger(2) 
+
\hat{\psi}^\dagger(2)\hat{\psi}(1) 
\rangle\label{eq:5.4b}\\
iG^K(1,2) = \langle \hat{\psi}(1)
\hat{\psi}^\dagger(2) 
-
\hat{\psi}^\dagger(2)\hat{\psi}(1) 
\rangle
\label{eq:5.4c}
\end{eqnarray}
where $\langle \dots\rangle$ means averaging over quantum mechanical
states (do not confuse with the ensemble averaging !), $1$ is the short hand
notation for $(\mbox{\boldmath $r$}_1, t_1)$, $\hat{\psi}(\mbox{\boldmath
$r$}, t)$ is the fermionic operator in the Heisenberg representation,
and $\theta (x)$ is the step function. Spin indices are omitted and
summation over them is implied whenever it is necessary.
\label{eq:5.4}
\end{mathletters}

Let us now expand the Green function up to the 
first order in the interaction propagator
\begin{equation}
\hat{{\cal L}}(1,2) =
\left(
\matrix{{\cal L}^K(1,2) & {\cal L}^R(1,2) \cr  {\cal L}^A(1,2) & 0}
\right).
\label{eq:5.6}
\end{equation} 
In the short-hand notation $\int d3\equiv \int dt_3
d\mbox{\boldmath $r$}_3$ the connection between Green functions in the presence
($\hat{G}$) and in the absence ($\hat{G}^{(n)}$) of the interaction 
 can be written as  
\begin{equation}
i\hat{G}(1,2)=
i\hat{G}^{(n)}(1,2) +\frac{i^3}{2}\int d3\int d4\sum_{i,j=1,2}
\hat{G}^{(n)}(1,3)\hat{\tau}_i\hat{G}^{(n)}(3,4)\hat{\tau}_j\hat{G}^{(n)}(4,2)
{\cal L}_{ij}(3,4),
\label{eq:5.5}
\end{equation}

For our purposes it will be sufficient to use the equilibrium values of
the matrix elements of the interaction propagator $\hat{{\cal
L}}(1,2)$.  In this case these matrix elements do not depend on $t_1 +
t_2$.  Performing Fourier transform in the difference of the times
$t_- = t_1 - t_2$
\[
\hat{{\cal L}}(\omega) = 
\int d t_- \hat{{\cal L}}( t_-) \exp\left(i\omega t_-\right),
\]
one obtains the following connection between the matrix elements\cite{Keldysh}
\begin{equation}
 {\cal L}^K\left(\omega\right) = \coth \frac{\omega}{2T} 
\left[
{\cal L}^R\left(\omega\right) -  {\cal L}^A\left(\omega\right)
\right].
\label{eq:5.7}
\end{equation}

In the leading order, the retarded and advanced parts of the 
interaction propagator depend on only on the 
difference of its coordinates. For
the Coulomb interaction they are related to the screened interaction
potential by
\begin{equation}
{\cal L}^R\left(\omega,Q\right)=  {\cal L}^A\left(-\omega,Q\right)
= - U(\omega,Q) \approx - \frac{1}{2\nu}\left[\frac{-i\omega
+D(\omega)Q^2}{D(\omega) Q^2}\right].
\label{eq:5.8}
\end{equation}
Here $\nu$ is the density of states per one spin and $U(\omega,Q)$ is
the dynamically screened interaction potential (\ref{eq:2.24}). 
To calculate  the cross-term $\delta\sigma_{CWL}$ correctly, 
one has to include the weak localization correction to the 
diffusion coefficient $D$ (see below). This is the reason why the diffusion
coefficient $D(\omega)$ acquired the frequency dependence. 
The mesoscopic fluctuations of the propagator $\hat{{\cal L}}(\omega)$
are disregarded in the present calculation, since they
lead to the higher order in $1/g$  corrections to the conductivity.

Now, we substitute Eq.~(\ref{eq:5.5}) into Eq.~(\ref{eq:5.1}).  In the
equilibrium, the longitudinal current vanishes, and therefore all the
Green functions should be expanded up to the first order in the
external vector potential $\mbox{\boldmath $A$}^{\rm ext}$ which
represents the applied electric field $\mbox{\boldmath
$E$}=\partial_t\mbox{\boldmath $A$}^{\rm ext}$:
\begin{equation}
\hat{G}^{(n)}(1,2)=\hat{G}^{(0)}(1,2) + \int d3\
\hat{G}^{(0)}(1,3)\
\hat{\mbox{\boldmath $j$}}
\mbox{\boldmath $A$}^{\rm ext}(t_3)
\ \hat{G}^{(0)}(3,2).
\label{eq:5.9}
\end{equation}
The current operator in the coordinate representation is defined as
\begin{equation}
f_1(\mbox{\boldmath $r$})\hat{\mbox{\boldmath $j$}}f_2(\mbox{\boldmath
$r$})= \frac{ie}{2m}
\left[\left({\mbox{\boldmath $\nabla$}} f_1\right)f_2
- f_1{\mbox{\boldmath $\nabla$}} f_2
\right]
 - \frac{e \mbox{\boldmath $A$}(\mbox{\boldmath $r$})}{m}
f_1(\mbox{\boldmath $r$})f_2(\mbox{\boldmath
$r$})
\label{eq:5.10}
\end{equation}
for any two functions $f_{1,2}$. 

The time Fourier transform of the matrix
elements of equilibrium Green functions $\hat{G}^{(0)}$ in
Eq.~(\ref{eq:5.9}) are given by
\begin{mathletters}
\label{eq:5.11}
\begin{eqnarray}
G^K\left(\epsilon; \mbox{\boldmath $r$}_1, \mbox{\boldmath $r$}_2
\right) = \tanh \frac{\epsilon}{2T} \left[
G^R\left(\epsilon; \mbox{\boldmath $r$}_1, \mbox{\boldmath $r$}_2
\right) -
G^A\left(\epsilon; \mbox{\boldmath $r$}_1, \mbox{\boldmath $r$}_2
\right)
\right]
\label{eq:5.11a}\\
G^R\left(\epsilon; \mbox{\boldmath $r$}_1, \mbox{\boldmath $r$}_2
\right) = \frac{1}{\epsilon +\mu - \hat{H} +i0 }, \quad
G^A\left(\epsilon; \mbox{\boldmath $r$}_1, \mbox{\boldmath $r$}_2
\right) = \frac{1}{\epsilon +\mu - \hat{H} -i0 }, 
\label{eq:5.11b}
\end{eqnarray}
where $\hat{H}$ is the exact one-electron Hamiltonian for a given
system, and $\mu$ is the chemical potential. In Eq.~(\ref{eq:5.11})
and in all  subsequent formulas we will omit superscript $(0)$
since only such equilibrium Green functions appear in all further
calculations.
\end{mathletters}

Substituting Eqs.~(\ref{eq:5.9}) into Eq.~(\ref{eq:5.5}) and the
result into Eq.~(\ref{eq:5.1}) we obtain after simple algebra the Ohm
law $ \mbox{\boldmath $j$} = \hat{\sigma}\mbox{\boldmath $E$} $ with
the conductivity tensor
\begin{mathletters}
\label{eq:5.12}
\begin{eqnarray}
&&\displaystyle{\sigma_{\alpha\beta} = \sigma^{\rm ni}_{\alpha\beta} + 
\delta\sigma^{\rm deph}_{\alpha\beta} + 
\delta\sigma^{\rm int}_{\alpha\beta}}
\label{eq:5.12a}\\
&&\displaystyle{\sigma^{\rm ni}_{\alpha\beta}
=\int \frac{d\mbox{\boldmath $r$}_1d \mbox{\boldmath
$r$}_2}{\cal V}
\int\frac{d\epsilon}{8\pi} \left(\frac{d}{d\epsilon}
\tanh \frac{\epsilon}{2T}\right)
\left[
G^R_{12}\left(\epsilon\right)-
G^A_{12}\left(\epsilon\right)
\right]\hat{j}_\alpha
\left[
G^A_{21}\left(\epsilon\right)-
G^R_{21}\left(\epsilon\right)
\right]\hat{j}_\beta}
\label{eq:5.12b} \\
&&\displaystyle{\sigma^{\rm deph}_{\alpha\beta}
=-\frac{i}{16}\int 
\frac{
d\mbox{\boldmath $r$}_1
d\mbox{\boldmath $r$}_2
d\mbox{\boldmath $r$}_3
d\mbox{\boldmath $r$}_4
}
{\cal V}
\int\frac{d\epsilon}{2\pi}\frac{d\omega }{2\pi} 
\left(\frac{d}{d\epsilon}
\tanh \frac{\epsilon}{2T}\right) 
\left(\coth \frac{\omega }{2T} +
\tanh \frac{\epsilon-\omega }{2T}\right)
\left[{\cal L}^R_{34}\left(\omega\right)-
{\cal L}^A_{34}\left(\omega\right)
\right] \times}
\nonumber\\
&&\displaystyle{\quad\left\{2\hat{j}_\alpha\left[
G^R_{12}\left(\epsilon\right)-
G^A_{12}\left(\epsilon\right)
\right]\hat{j}_\beta
\left[
G^A_{23}\left(\epsilon\right)
G^A_{34}\left(\epsilon - \omega\right)
G^A_{41}\left(\epsilon\right)
-
G^R_{23}\left(\epsilon\right)
G^R_{34}\left(\epsilon - \omega\right)
G^R_{41}\left(\epsilon\right)\right]\right.}
\nonumber\\
&&\displaystyle{\quad\left.
+
\hat{j}_\alpha\left[
G^R_{13}\left(\epsilon\right)G^R_{32}\left(\epsilon - \omega\right)-
G^A_{13}\left(\epsilon\right)G^A_{32}\left(\epsilon - \omega\right)
\right]\hat{j}_\beta
\left[
G^A_{24}\left(\epsilon - \omega\right)
G^A_{41}\left(\epsilon\right)
-
G^R_{24}\left(\epsilon - \omega\right)
G^R_{41}\left(\epsilon\right)
\right] 
+ \alpha \leftrightarrow \beta
\right\}}
\label{eq:5.12c} \\
&&\displaystyle{\sigma^{\rm int}_{\alpha\beta}
=-\frac{i}{8}\int 
\frac{
d\mbox{\boldmath $r$}_1
d\mbox{\boldmath $r$}_2
d\mbox{\boldmath $r$}_3
d\mbox{\boldmath $r$}_4
}
{\cal V}
\int\frac{d\epsilon}{2\pi}\frac{d\omega }{2\pi} 
\left(\frac{d}{d\epsilon}
\tanh \frac{\epsilon}{2T}\right) 
\tanh \frac{\epsilon-\omega }{2T}
\times}
\nonumber\\
&&\displaystyle{\quad\left\{\hat{j}_\alpha\left[
G^R_{12}\left(\epsilon\right)-
G^A_{12}\left(\epsilon\right)
\right]\hat{j}_\beta
\left[
G^A_{23}\left(\epsilon\right)G^A_{41}\left(\epsilon\right)
-G^R_{23}\left(\epsilon\right)G^R_{41}\left(\epsilon\right)
\right]
\left[
G^R_{34}\left(\epsilon - \omega\right){\cal L}^A_{34}\left(\omega\right)
-G^A_{34}\left(\epsilon - \omega\right){\cal L}^R_{34}\left(\omega\right)
\right]
\right.}
\nonumber\\
&&\displaystyle{\quad\left.
+
\left[
G^R_{13}\left(\epsilon\right)-
G^A_{13}\left(\epsilon\right)
\right]\hat{j}_\alpha
\left[
G^A_{14}\left(\epsilon\right)-
G^R_{14}\left(\epsilon\right)
\right]
\left[
G^R_{42}\left(\epsilon - \omega\right)\hat{j}_\beta
G^R_{23}\left(\epsilon - \omega\right)
{\cal L}^A_{34}\left(\omega\right)
-G^A_{42}\left(\epsilon - \omega\right)\hat{j}_\beta
G^A_{23}\left(\epsilon - \omega\right)
{\cal L}^R_{34}\left(\omega\right)
\right]
\right.}\nonumber\\
&&\displaystyle{\left.\quad\quad\quad
+ \alpha \leftrightarrow \beta
\right\}}
\label{eq:5.12d}
\end{eqnarray}
where ${\cal V}$ is the volume of the system,  $\hat{j}_\alpha$ is
the component of current operator defined in Eq.~(\ref{eq:5.10}), and
the short-hand notation $G_{ij}\left(\epsilon\right) \equiv
G\left(\epsilon; \mbox{\boldmath $r$}_i, \mbox{\boldmath
$r$}_j\right)$ is introduced.  In Eqs.~(\ref{eq:5.12b}) -
(\ref{eq:5.12c}) only the symmetric part of the conductivity tensor is
retained.  We emphasize that all these formulas are exact in the
first order of perturbation theory, i.e. we made no assumptions about
the behavior of the Green functions or the relevant frequencies in the
interaction propagator. At this stage separation of the interaction
correction into two pieces (\ref{eq:5.12c}) and (\ref{eq:5.12d}) is
just a matter of convenience physical meaning of which will become
clear shortly.
\end{mathletters}

Equations (\ref{eq:5.12}), which express the conductivity in terms of
the exact retarded and advance Green functions for a given realization
of potential, are the main results of this subsection.  In order to
find the physical conductivity of a large sample, one has to perform
the disorder averaging. This is the subject of the following
subsection.

\subsection{Averaging over disorder}

Let us start the disorder averaging of Eqs.~(\ref{eq:5.12b}) -
(\ref{eq:5.12d}) with the observation that any mean product of the
electron Green functions does not depend on their common energy
$\epsilon$. This fact allows to perform the energy integration in
Eqs.~(\ref{eq:5.12b}) - (\ref{eq:5.12d}).  It is also important and
useful to notice that after the averaging, the products containing only
retarded or only advanced Green functions vanish. Using this feature,
we simplify Eqs.~(\ref{eq:5.12}) to the form
\begin{mathletters}
\label{eq:5.13}
\begin{eqnarray}
&&\displaystyle{\sigma^{\rm ni}_{\alpha\beta}
=\frac{1}{4\pi}\int \frac{d\mbox{\boldmath $r$}_1d \mbox{\boldmath
$r$}_2}{\cal V}
\langle\langle
G^R_{12}\left(\epsilon\right)\hat{j}_\alpha
G^A_{21}\left(\epsilon\right)\hat{j}_\beta
\rangle\rangle + \alpha \leftrightarrow \beta;  
}
\label{eq:5.13a} \\
&&\displaystyle{\sigma^{\rm deph}_{\alpha\beta}
=-\frac{i}{8\pi}\int 
\frac{
d\mbox{\boldmath $r$}_1
d\mbox{\boldmath $r$}_2
d\mbox{\boldmath $r$}_3
d\mbox{\boldmath $r$}_4
}
{\cal V}
\int\frac{d\omega }{2\pi}
\left(\frac{\omega}{2T\sinh^2\frac{\omega}{2T}}\right) 
\left[{\cal L}^R_{34}\left(\omega\right)-
{\cal L}^A_{34}\left(\omega\right)
\right] \times}
\nonumber\\
&&\displaystyle{\quad\left\{
\langle\langle
\hat{j}_\alpha
G^R_{12}\left(\epsilon\right)
\hat{j}_\beta
G^A_{23}\left(\epsilon\right)
G^A_{34}\left(\epsilon - \omega\right)
G^A_{41}\left(\epsilon\right)
\rangle\rangle
+
\langle\langle
\hat{j}_\alpha
G^A_{12}\left(\epsilon\right)
\hat{j}_\beta
G^R_{23}\left(\epsilon\right)
G^R_{34}\left(\epsilon - \omega\right)
G^R_{41}\left(\epsilon\right)
\rangle\rangle
\right.}
\nonumber\\
&&\displaystyle{\quad\quad\quad\quad\left.
+
\langle\langle
\hat{j}_\alpha
G^R_{13}\left(\epsilon\right)G^R_{32}\left(\epsilon - \omega\right)
\hat{j}_\beta
G^A_{24}\left(\epsilon - \omega\right)
G^A_{41}\left(\epsilon\right)
\rangle\rangle
+ \alpha \leftrightarrow \beta
\right\};}
\label{eq:5.13b} \\
&&\displaystyle{\sigma^{\rm int}_{\alpha\beta}
=-\frac{1}{4\pi}{\rm Im}\int 
\frac{
d\mbox{\boldmath $r$}_1
d\mbox{\boldmath $r$}_2
d\mbox{\boldmath $r$}_3
d\mbox{\boldmath $r$}_4
}
{\cal V}
\int
\frac{d\omega }{2\pi} 
\left[\frac{d}{d\omega}\left(\omega
\coth \frac{\omega}{2T}\right)\right] {\cal L}^A_{34}\left(\omega\right)
\times}
\nonumber\\
&&\displaystyle{\quad\left\{
\langle\langle
\hat{j}_\alpha
G^A_{12}\left(\epsilon\right)
\hat{j}_\beta
G^R_{23}\left(\epsilon\right)
G^R_{34}\left(\epsilon - \omega\right)
G^R_{41}\left(\epsilon\right)
\rangle\rangle +
\langle\langle
\hat{j}_\alpha
G^R_{12}\left(\epsilon\right)
\hat{j}_\beta
G^A_{23}\left(\epsilon\right)
G^R_{34}\left(\epsilon - \omega\right)
G^A_{41}\left(\epsilon\right)
\rangle\rangle
\right.}
\nonumber\\
&&\displaystyle{\quad\left.
-\langle\langle G^A_{31}\left(\epsilon\right)\hat{j}_\alpha
G^A_{12}\left(\epsilon\right)\hat{j}_\beta
G^A_{24}\left(\epsilon \right)
G^R_{43}\left(\epsilon - \omega\right)\rangle\rangle
-\langle\langle G^A_{31}\left(\epsilon\right)\hat{j}_\alpha
G^A_{14}\left(\epsilon\right)
G^R_{42}\left(\epsilon - \omega\right)\hat{j}_\beta
G^R_{23}\left(\epsilon - \omega\right)\rangle\rangle
\right.}\nonumber\\
&&\displaystyle{\left.\quad
+ 2\langle\langle G^R_{31}\left(\epsilon\right)\hat{j}_\alpha
G^A_{14}\left(\epsilon\right)
G^R_{42}\left(\epsilon - \omega\right)\hat{j}_\beta
G^R_{23}\left(\epsilon - \omega\right)\rangle\rangle
+ \alpha \leftrightarrow \beta
\right\}}
\label{eq:5.13c}
\end{eqnarray}
where $\langle\langle \dots \rangle\rangle$ stand for the disorder average.
\end{mathletters}

Equations (\ref{eq:5.13}) express the average conductance in terms of
the average of products of Green functions of a non-interacting system.
Those averages should be performed with taking the weak localization
correction into account. We will perform the average using the
diagrammatic technique\cite{AA,AGD}; the same results can be obtained
with the help of supersymmetric non-linear $\sigma$-model\cite{Efetov}.

\begin{figure}
\vspace{0.2 cm}
\epsfxsize=13.7cm
\centerline{\epsfbox{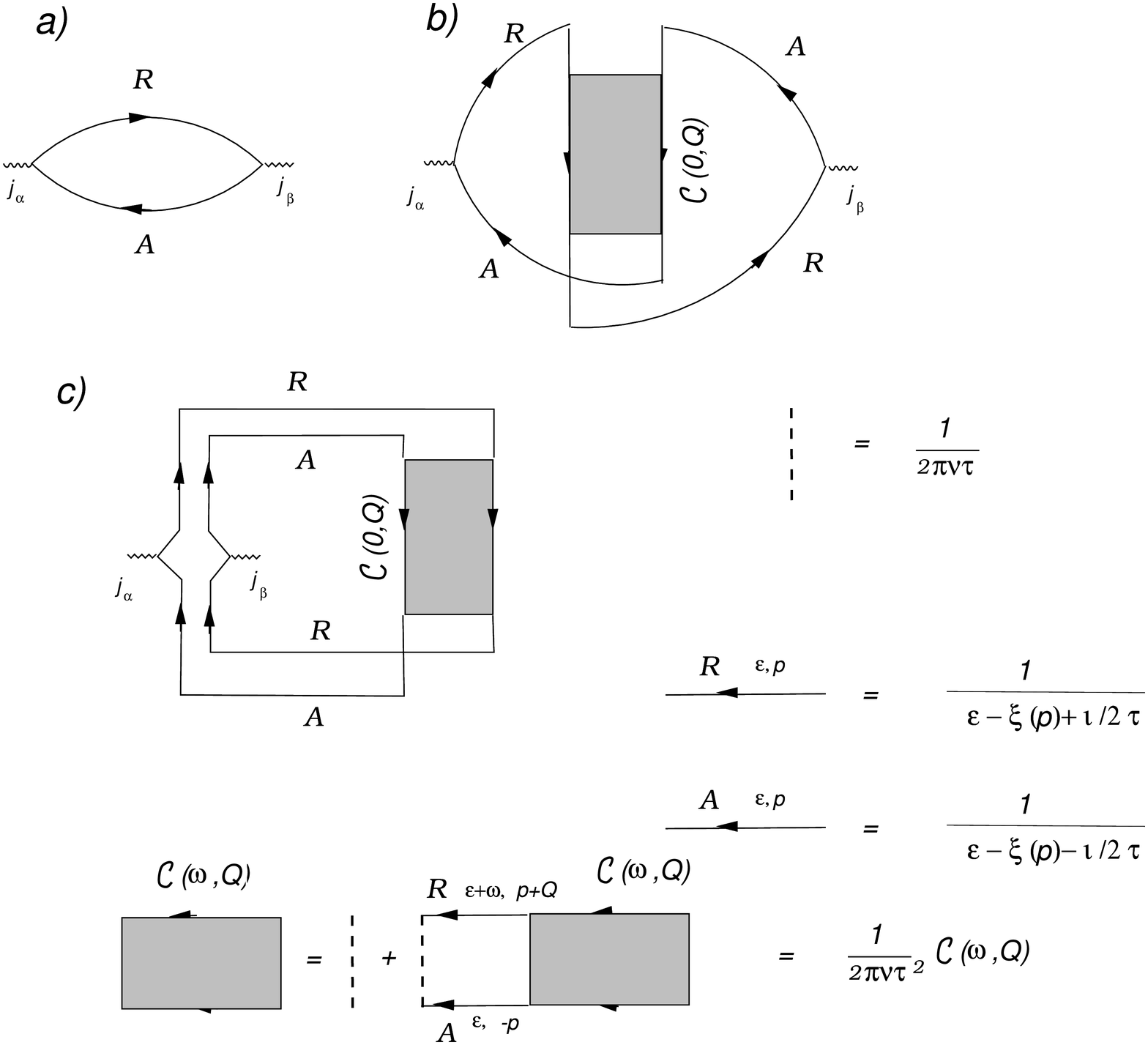}}
\vspace{0.8cm}
\caption{Diagrams for the averaging of $\sigma^{ni}$. (a) Drude
conductivity. (b) Weak localization correction\protect\cite{Gorkov79};
(c) More convenient for the further usage representation of diagram (b).
Hereafter, solid lines denotes disorder average Green functions and
dashed lines denote disorder.
}
\label{Fig5}
\end{figure}

Evaluation of the mean conductivity of {\em non-interacting} electrons
 $\sigma^{ni}$, i.e., averaging of the equation (\ref{eq:5.13a}), 
is now a textbook problem.
Summation of the diagrams in Fig.~\ref{Fig5} gives $\sigma_{\alpha\beta}
=\delta_{\alpha\beta}
\left(
\sigma + \delta\sigma_{WL}\right)$
where Drude conductivity $\sigma$ and the weak localization
correction given by (\ref{eq:1.1}) and the weak localization correction
 $\delta\sigma_{WL}$ are given by (\ref{eq:1.1}) and (\ref{eq:2.4})
respectively. 

Before averaging the interaction correction,
 it is convenient to Fourier transform Eqs.~(\ref{eq:5.13b})
and (\ref{eq:5.13c})
 over the difference $\mbox{\boldmath $r$}_3 - 
\mbox{\boldmath $r$}_4$. 
We obtain from Eq.~(\ref{eq:5.13b}) 
\begin{eqnarray}
&&\displaystyle{\sigma^{\rm deph}_{\alpha\beta}
=-\frac{1}{4\pi} 
\int\frac{d\omega }{2\pi}\int\frac{d^dq}{\left(2\pi\right)^d}
\left(\frac{\omega}{2T\sinh^2\frac{\omega}{2T}}\right) 
\left[{\rm Im} {\cal L}^A\left(\omega,q\right)
\right] \left\{
\langle\langle
\hat{j}_\alpha
G^R_{12}\left(\epsilon\right)
\hat{j}_\beta
G^A_{23}\left(\epsilon\right)
G^A_{34}\left(\epsilon - \omega\right)
G^A_{41}\left(\epsilon\right)
\rangle\rangle_{34}
\right.}
\label{eq:5.14a}\\
&&\displaystyle{\quad\left.
+
\langle\langle
\hat{j}_\alpha
G^A_{12}\left(\epsilon\right)
\hat{j}_\beta
G^R_{23}\left(\epsilon\right)
G^R_{34}\left(\epsilon - \omega\right)
G^R_{41}\left(\epsilon\right)
\rangle\rangle_{34}
+
\langle\langle
\hat{j}_\alpha
G^R_{13}\left(\epsilon\right)G^R_{32}\left(\epsilon - \omega\right)
\hat{j}_\beta
G^A_{24}\left(\epsilon - \omega\right)
G^A_{41}\left(\epsilon\right)
\rangle\rangle_{34}
+ \alpha \leftrightarrow \beta
\right\},}
\nonumber
\end{eqnarray}
where hereafter we use the short hand notation
\[
\langle\langle F^{(1)}_{12}
F^{(2)}_{23}\dots F^{(n)}_{n1}\rangle\rangle_{ij} \equiv
\int\frac{
d\mbox{\boldmath $r$}_1
\dots 
d\mbox{\boldmath $r$}_n
}
{\cal V} \langle\langle F^{(1)}\left(\mbox{\boldmath $r$}_1,
\mbox{\boldmath $r$}_2 \right)
F^{(2)}\left(\mbox{\boldmath $r$}_2,
\mbox{\boldmath $r$}_3 \right) \dots
 F^{(n)}\left(\mbox{\boldmath $r$}_n,
\mbox{\boldmath $r$}_1 \right)\rangle\rangle
\exp\left[i\mbox{\boldmath $q$}\left(\mbox{\boldmath $r$}_i-
\mbox{\boldmath $r$}_j \right)\right]
\]
for any functions $F^{(i)}$.

Three last terms in brackets of Eq.~(\ref{eq:5.13c}) can be simplified
with the help of the identities
\begin{eqnarray}
\label{identities}
\label{ID}
\int
d\mbox{\boldmath $r$}_3 
G^{R}\left(\epsilon, \mbox{\boldmath $r$}_1,
\mbox{\boldmath $r$}_3 \right)\hat{j}_\alpha
G^{R}\left(\epsilon, \mbox{\boldmath $r$}_3,
\mbox{\boldmath $r$}_2 \right)&=& - i e 
\left(\mbox{\boldmath $r$}_1 -\mbox{\boldmath $r$}_2\right)_\alpha
G^{R}\left(\epsilon, \mbox{\boldmath $r$}_1,
\mbox{\boldmath $r$}_2 \right);\\
\int
d\mbox{\boldmath $r$}_3 d\mbox{\boldmath $r$}_4
G^{R}\left(\epsilon, \mbox{\boldmath $r$}_1,
\mbox{\boldmath $r$}_3 \right)\hat{j}_\alpha
G^{R}\left(\epsilon, \mbox{\boldmath $r$}_3,
\mbox{\boldmath $r$}_4 \right)
\hat{j}_\beta 
G^{R}\left(\epsilon, \mbox{\boldmath $r$}_4,
\mbox{\boldmath $r$}_2 \right)
\hat{j}_\beta 
&=&
\nonumber\\
&&\hspace*{-4cm} - \frac{e^2}{2} 
\left(\mbox{\boldmath $r$}_1 -\mbox{\boldmath $r$}_2\right)_\alpha
\left(\mbox{\boldmath $r$}_1 -\mbox{\boldmath $r$}_2\right)_\beta
G^{R}\left(\epsilon, \mbox{\boldmath $r$}_1,
\mbox{\boldmath $r$}_2 \right)
+\frac{e^2\delta_{\alpha\beta}}{2m}
\frac{\partial}{\partial \epsilon}
G^{R}\left(\epsilon, \mbox{\boldmath $r$}_1
\mbox{\boldmath $r$}_2 \right),
\nonumber
\end{eqnarray}
which follow from the gauge invariance of the theory and do not
require any translational invariance, see Appendix. We find
\begin{eqnarray}
&&\displaystyle{\sigma^{\rm int}_{\alpha\beta}
=-\frac{1}{4\pi}{\rm Im}
\int
\frac{d\omega }{2\pi} \int\frac{d^dq}{\left(2\pi\right)^d}
\left[\frac{d}{d\omega}\left(\omega
\coth \frac{\omega}{2T}\right)\right] {\cal L}^A\left(\omega,q\right)
\left\{
\langle\langle
\hat{j}_\alpha
G^A_{12}\left(\epsilon\right)
\hat{j}_\beta
G^R_{23}\left(\epsilon\right)
G^R_{34}\left(\epsilon - \omega\right)
G^R_{41}\left(\epsilon\right)
\rangle\rangle_{34}\right.} 
\nonumber\\
&&\displaystyle{\left.
+
\langle\langle
\hat{j}_\alpha
G^R_{12}\left(\epsilon\right)
\hat{j}_\beta
G^A_{23}\left(\epsilon\right)
G^R_{34}\left(\epsilon - \omega\right)
G^A_{41}\left(\epsilon\right)\rangle\rangle_{34}+
\frac{e^2}{2}\frac{\partial^2}{\partial q_\alpha\partial q_\beta}
\langle\langle G^A_{12}\left(\epsilon\right)
G^R_{21}\left(\epsilon - \omega\right)\rangle\rangle_{12}
\right.}\nonumber\\
&&\displaystyle{\left.\quad
+ 2e\frac{\partial}{\partial q_\beta}
\langle\langle G^R_{31}\left(\epsilon\right)\hat{j}_\alpha
G^A_{12}\left(\epsilon\right)
G^R_{23}\left(\epsilon - \omega\right)\rangle\rangle_{23}
+ \alpha \leftrightarrow \beta
\right\}}.
\label{eq:5.14b}
\end{eqnarray}
Deriving Eq.~(\ref{eq:5.14b}) we used the fact that
\begin{eqnarray*}
&&{\rm Im}\int
{d\omega } 
\left[\frac{d}{d\omega}\left(\omega
\coth \frac{\omega}{2T}\right)\right] {\cal L}^A\left(\omega,q\right)
\langle\langle
G^R_{21}\left(\epsilon - \omega\right)
\partial_\epsilon G^A_{12}\left(\epsilon\right)
\rangle\rangle_{12}
\nonumber\\
&=&
\int
{d\omega } 
\left[\frac{d}{d\omega}\left(\omega
\coth \frac{\omega}{2T}\right)\right] 
\left[
{\cal L}^A\left(\omega,q\right)
\langle\langle
G^R_{21}\left(\epsilon - \omega\right)
\partial_\epsilon G^A_{12}\left(\epsilon\right)
\rangle\rangle_{12}
-
{\cal L}^R\left(\omega,q\right)
\langle\langle
G^A_{12}\left(\epsilon - \omega\right)
\partial_\epsilon G^R_{21}\left(\epsilon\right)
\rangle\rangle_{12}
\right]\\
&=&
\int
{d\omega } 
\left[\frac{d}{d\omega}\left(\omega
\coth \frac{\omega}{2T}\right)\right] 
{\cal L}^A\left(\omega,q\right)
\left[
\langle\langle
G^R_{21}\left(\epsilon - \omega\right)
\partial_\epsilon G^A_{12}\left(\epsilon\right)
\rangle\rangle_{12}
+
\langle\langle
G^A_{12}\left(\epsilon + \omega\right)
\partial_\epsilon G^R_{21}\left(\epsilon\right)
\rangle\rangle_{12}
\right]\\
&&=\int
{d\omega } 
\left[\frac{d}{d\omega}\left(\omega
\coth \frac{\omega}{2T}\right)\right] 
{\cal L}^A\left(\omega,q\right)
\left[
\partial_\epsilon
\langle\langle
G^R_{21}\left(\epsilon - \omega\right)
 G^A_{12}\left(\epsilon\right)
\rangle\rangle_{12}
+
\langle\langle
G^A_{12}\left(\epsilon + \omega\right)
\partial_\epsilon G^R_{21}\left(\epsilon\right)
\rangle\rangle_{12} -
\langle\langle
G^A_{12}\left(\epsilon\right)
\partial_\epsilon G^R_{21}\left(\epsilon - \omega\right)
\rangle\rangle_{12}
\right]
\\&=& 0.
\end{eqnarray*}

\begin{figure}
\epsfxsize=10.7cm
\centerline{\epsfbox{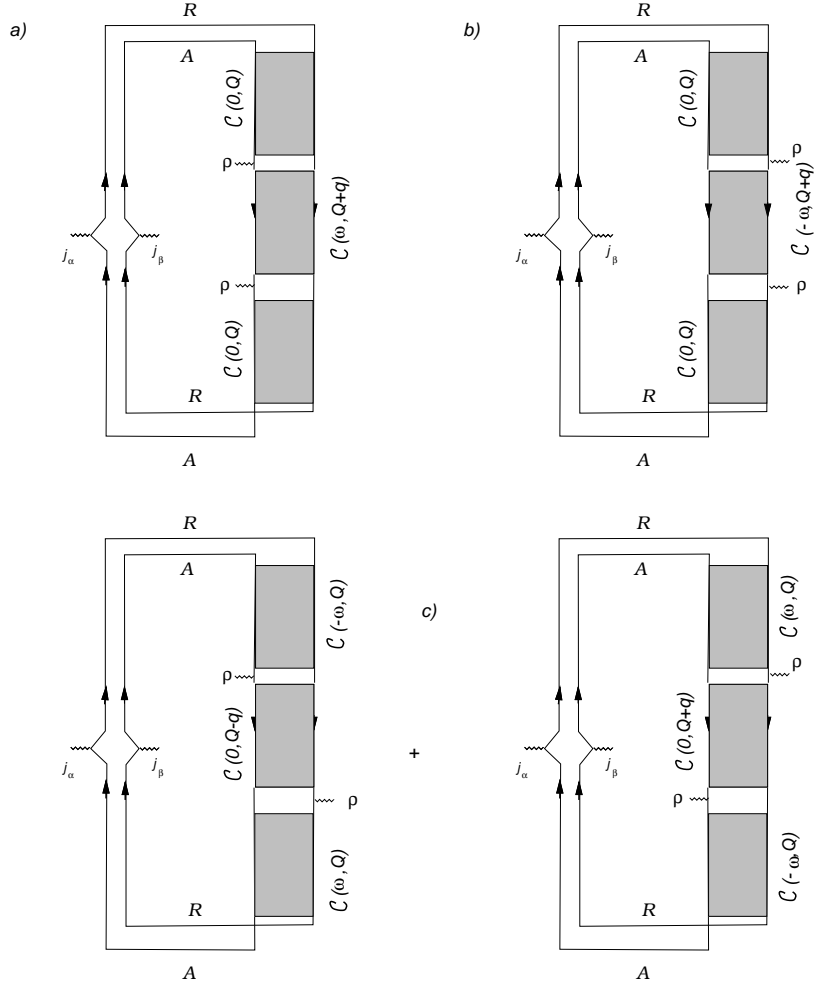}}
\vspace{0.8cm}
\caption{Diagrams for the averaging of products (a)
$\langle\langle
\hat{j}_\alpha
G^R_{12}
\hat{j}_\beta
G^A_{23}
G^A_{34}
G^A_{41}
\rangle\rangle_{34}$;
 (b) $\langle\langle
\hat{j}_\alpha
G^A_{12}
\hat{j}_\beta
G^R_{23}
G^R_{34}
G^R_{41}
\rangle\rangle_{34}$; and (c)
$\langle\langle
\hat{j}_\alpha
G^R_{13}G^R_{32}
\hat{j}_\beta
G^A_{24}
G^A_{41}
\rangle\rangle_{34}$
}
\label{Fig6}
\end{figure}

Now we are in the position to perform actual disorder average of the
products entering into Eqs.~(\ref{eq:5.14a}) and
(\ref{eq:5.14b}). Diagrams for all the terms from Eq.~(\ref{eq:5.14a})
are shown in Fig.~\ref{Fig6}.
They can be obtained by cutting the corresponding Green functions on the
diagrams of Fig.~\ref{Fig5}b. The analytic expression for these diagrams is
\begin{mathletters}
\label{eq:5.15}
\begin{eqnarray}
\langle\langle
\hat{j}_\alpha
G^R_{12}\left(\epsilon\right)
\hat{j}_\beta
G^A_{23}\left(\epsilon\right)
G^A_{34}\left(\epsilon - \omega\right)
G^A_{41}\left(\epsilon\right)
\rangle\rangle_{34}&=&
\langle\langle
\hat{j}_\alpha
G^A_{12}\left(\epsilon\right)
\hat{j}_\beta
G^R_{23}\left(\epsilon\right)
G^R_{34}\left(\epsilon - \omega\right)
G^R_{41}\left(\epsilon\right)
\rangle\rangle_{34}\label{eq:5.15a}\\
&=&\delta_{\alpha\beta}
\frac{2\sigma}{\nu}
\int\frac{d^dQ}{\left(2\pi\right)^d}
{\cal C}^2(0, \mbox{\boldmath $Q$})
{\cal C}(-\omega, \mbox{\boldmath $Q$}+ \mbox{\boldmath $q$})
\nonumber\\
\langle\langle
\hat{j}_\alpha
G^R_{13}\left(\epsilon\right)G^R_{32}\left(\epsilon - \omega\right)
\hat{j}_\beta
G^A_{24}\left(\epsilon - \omega\right)
G^A_{41}\left(\epsilon\right)
\rangle\rangle_{34}&=&-
\delta_{\alpha\beta}
\frac{4\sigma}{\nu}
\int\frac{d^dQ}{\left(2\pi\right)^d}
\left|{\cal C}(\omega, \mbox{\boldmath $Q$})\right|^2
{\cal C}(0, \mbox{\boldmath $Q$}+ \mbox{\boldmath $q$})
\label{eq:5.15b}
\end{eqnarray}
where the Cooperon propagator is defined by Eq.~(\ref{eq:2.42}).
Substitution of Eqs.~(\ref{eq:5.15}) and (\ref{eq:5.8}) 
into Eq.~(\ref{eq:5.14a}) immediately yields Eq.~(\ref{eq:4.5}).
\end{mathletters}

\begin{figure}[t]
\vspace{3.3cm}
\epsfxsize=14.0cm
\centerline{\epsfbox{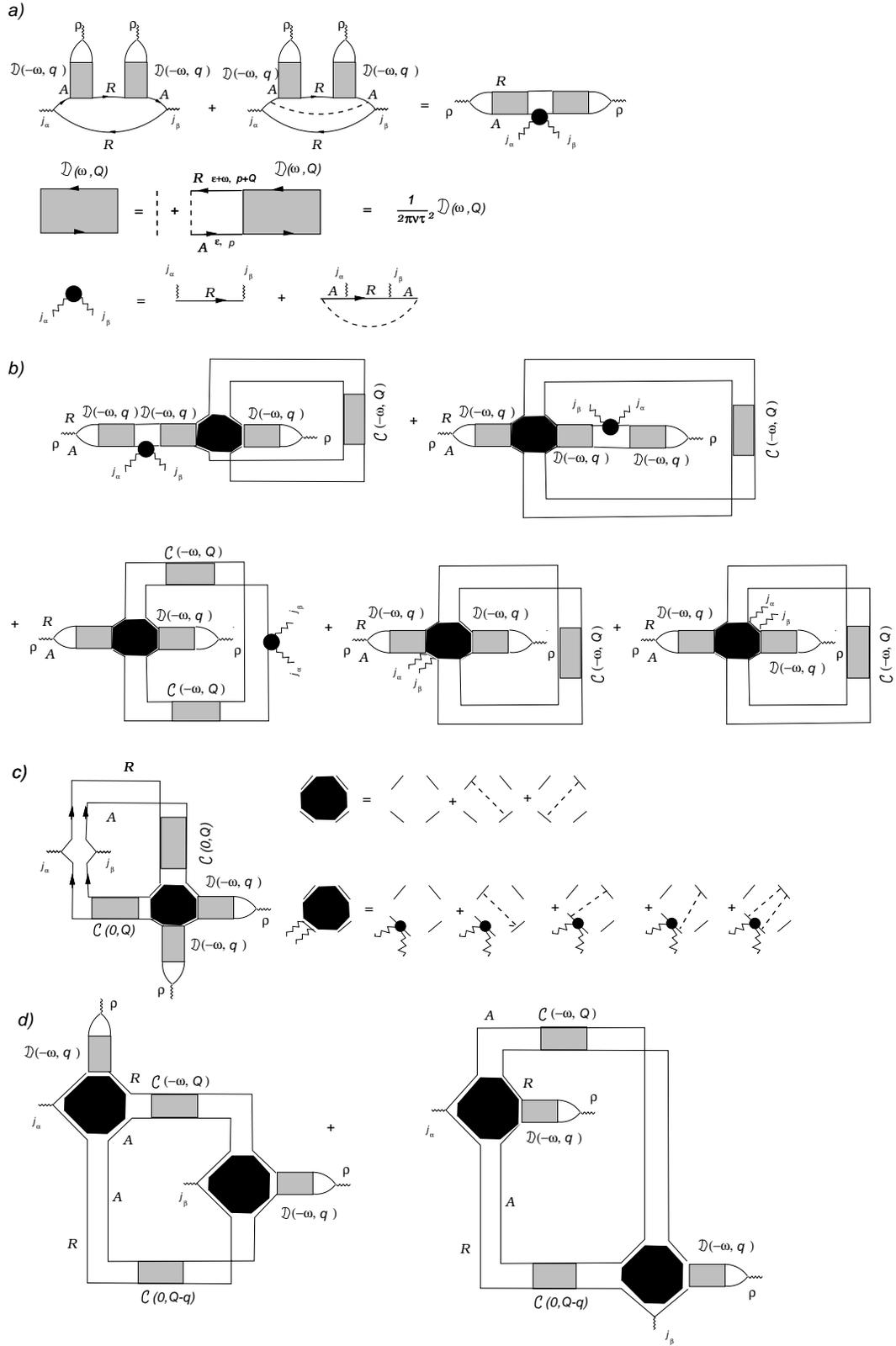}}
\vspace{0.8cm}
\caption{Diagrams for the averaging of products 
$\langle\langle
\hat{j}_\alpha
G^R_{12}
\hat{j}_\beta
G^A_{23}
G^R_{34}
G^A_{41}\rangle\rangle_{34}$.}
\label{Fig7}
\end{figure}

\begin{figure}
\vspace*{0.2cm}
\epsfxsize=8.0cm
\centerline{\epsfbox{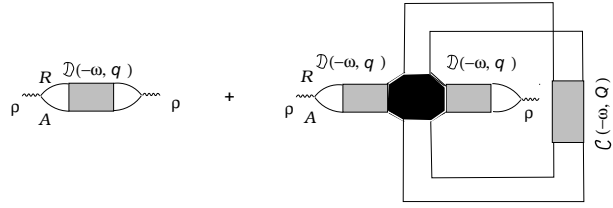}}
\vspace{0.8cm}
\caption{Diagrams for the averaging of  density-density
correlation function
$\langle\langle
G^R_{12}
G^A_{21}
\rangle\rangle_{12}$.}
\label{Fig8}
\end{figure}

\begin{figure}
\vspace*{0.2cm}
\epsfxsize=7.0cm
\centerline{\epsfbox{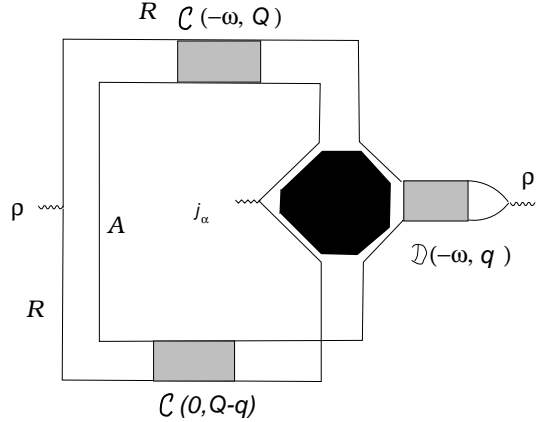}}
\vspace{0.8cm}
\caption{Diagrams for the averaging of product  
$\langle\langle G^R_{31}\hat{j}_\alpha
G^A_{12}
G^R_{23}\rangle\rangle_{23}
$
.}
\label{Fig9}
\end{figure}

Diagrams for the averaging of the terms entering into
Eq.~(\ref{eq:5.14b}) are a little bit more complicated, see
Figs.~\ref{Fig7} - \ref{Fig9}. The analytic expression for the
diagrams in Fig.~\ref{Fig7} is
\begin{eqnarray}
\langle\langle
\hat{j}_\alpha
G^R_{12}\left(\epsilon\right)
&\hat{j}_\beta&
G^A_{23}\left(\epsilon\right)
G^R_{34}\left(\epsilon - \omega\right)
G^A_{41}\left(\epsilon\right)\rangle\rangle_{34}=
2\pi i \sigma\delta_{\alpha\beta}
\frac{\partial}{\partial \omega} {\cal D}\left(- \omega,q\right)+
i\frac{2\sigma}{\nu}\delta_{\alpha\beta}\frac{\partial}{\partial \omega} 
\left[Dq^2\ {\cal D}^2\left(- \omega,q\right)\int 
\frac{d^dQ}{\left(2\pi\right)^d}{\cal C}\left(-\omega,Q\right)
 \right]
\nonumber\\
&&-\delta_{\alpha\beta}\frac{2\sigma}{\nu}
\int\frac{d^dQ}{\left(2\pi\right)^d}
\left[
{\cal C}^2(0, \mbox{\boldmath $Q$})
{\cal D}(-\omega, \mbox{\boldmath $q$})
+
{\cal C}(0, \mbox{\boldmath $Q$})
{\cal D}^2 (-\omega, \mbox{\boldmath $q$})
\right] 
\nonumber \\
&& + \frac{4\sigma}{\nu}\
{\cal D}^2 (-\omega, \mbox{\boldmath $q$})
\int\frac{d^dQ}{\left(2\pi\right)^d}
{\cal C}(-\omega, \mbox{\boldmath $Q$})
{\cal C}(0, \mbox{\boldmath $Q$}-\mbox{\boldmath $q$})
\left(Dq_\alpha Q_\beta+ DQ_\alpha q_\beta\right)
,
\label{eq:5.16}.
\end{eqnarray}
where Cooperon propagator ${\cal C}(-\omega, \mbox{\boldmath $Q$})$
is defined by Eq.~(\ref{eq:2.42}) and the
diffuson is given by
\begin{equation}
{\cal D}\left(\omega,Q\right) = \frac{1}{-i\omega + DQ^2}.
\label{eq:5.17}
\end{equation}
The first term in Eq.~(\ref{eq:5.16}) is the leading classical
approximation, Fig.~\ref{Fig7}a and all the other terms come from the
interference corrections which are shown term by term in 
Fig.~\ref{Fig7}b-d.

The third term in brackets in Eq.~(\ref{eq:5.14b}) is the density
-- density correlation function with the well known answer, see
Fig.~(\ref{Fig8})
\begin{equation}
\langle\langle G^A_{12}\left(\epsilon\right)
G^R_{21}\left(\epsilon - \omega\right)\rangle\rangle_{12}
=4\pi\nu {\cal D}(-\omega,q)
+4D q^2 {\cal D}^2(-\omega,q)\int
\frac{d^dQ}{\left(2\pi\right)^d}{\cal C}\left(-\omega,Q\right),
\label{eq:5.18}
\end{equation}
where the last term is nothing but the weak-localization correction to
the diffusion constant.

Finally, the last term in brackets in Eq.~(\ref{eq:5.14b}) is averaged
according to Fig.~\ref{Fig9}. The analytic expression for this
diagram is
\begin{equation}
\langle\langle G^R_{31}\left(\epsilon\right)\hat{j}_\alpha
G^A_{12}\left(\epsilon\right)
G^R_{23}\left(\epsilon - \omega\right)\rangle\rangle_{23}=
4 e Dq_\alpha {\cal D}\left(- \omega,q\right)
\int
\frac{d^dQ}{\left(2\pi\right)^d}{\cal C}\left(0,Q\right)
{\cal C}\left(-\omega,\mbox{\boldmath $Q$} + 
\mbox{\boldmath $q$} \right).
\label{eq:5.19}
\end{equation}

Now, we are ready to calculate the interaction correction
(\ref{eq:5.14b}). First we take into account  
leading $1/g$ approximation only, i.e. retain the first terms
in Eq.~(\ref{eq:5.16}) and (\ref{eq:5.18}), neglect terms
(\ref{eq:5.15a}) and (\ref{eq:5.19}), and use the interaction
propagator ${\cal L}^A$ from (\ref{eq:5.8}) with the bare diffusion
constant $D$. It immediately gives the known result, Eq.~(\ref{eq:2.25}).

In order to find the cross-term of weak interaction and weak
localization, we have to take into account not only all the remaining
terms in Eqs.~(\ref{eq:5.15a}), (\ref{eq:5.16}), (\ref{eq:5.18}) and
(\ref{eq:5.19}) but also  the weak localization
correction to the interaction propagator (\ref{eq:5.8}) in the leading
approximation
\begin{equation}
{\cal L}^A\left(\omega,q\right)
 \approx - \frac{1}{2\nu}\left[\frac{i\omega
+Dq^2}{D q^2}\right] -
 \frac{i\omega}{2\nu D q^2}
\left(\frac{1}{\pi\nu}\right) \int\frac{d^dQ}{\left(2\pi\right)^d}
{\cal C}(-\omega, \mbox{\boldmath $Q$}),
\label{eq:5.20}
\end{equation}
where the last term is the weak localization correction to the
diffusion constant.

Substituting Eqs.~(\ref{eq:5.15a}), (\ref{eq:5.16}) and (\ref{eq:5.18}) --
(\ref{eq:5.20}) into Eq.~(\ref{eq:5.14b}) and keeping only terms of
the second order in $1/g$, i.e. only those involving two integrations
over the wavevectors $Q$, $q$, we obtain
\begin{eqnarray}
\sigma^{\rm CWL}_{\alpha\beta}
&=&\displaystyle{\frac{\sigma}{2\pi\nu^2}{\rm Im}
\int
\frac{d\omega }{2\pi} 
\left[\frac{d}{d\omega}\left(\omega
\coth \frac{\omega}{2T}\right)\right] 
\int \frac{d^dqd^dQ}{\left(2\pi\right)^{2d}}}
\label{eq:5.21}
\\
&&\displaystyle{\times
\left\{
\frac{\delta_{\alpha\beta}}{Dq^2}
\left[{\cal C}^2(0, \mbox{\boldmath $Q$})
\left(\frac{{\cal C}(-\omega , \mbox{\boldmath $Q$}+ \mbox{\boldmath
$q$})}{{\cal D}(-\omega , \mbox{\boldmath $q$})}
-1\right) 
-  {\cal C}(0, \mbox{\boldmath $Q$})
{\cal D}(-\omega , \mbox{\boldmath $q$})
+ {\cal C}(-\omega , \mbox{\boldmath $Q$})
{\cal D}(-\omega , \mbox{\boldmath $q$})+
2 {\cal C}(-\omega , \mbox{\boldmath $Q$}+ \mbox{\boldmath
$q$})
{\cal C}(0, \mbox{\boldmath $Q$})
\right]
\right.} 
\nonumber\\
&&\displaystyle{\left.
+\frac{2q_\alpha}{D q^2}\frac{\partial}{\partial q_\beta}
{\cal C}(-\omega , \mbox{\boldmath $Q$}+ \mbox{\boldmath
$q$})
{\cal C}(0, \mbox{\boldmath $Q$})- 4\frac{q_\alpha q_\beta}{q^2}
{\cal D}(-\omega , \mbox{\boldmath
$q$})
{\cal C}(-\omega , \mbox{\boldmath $Q$}+ \mbox{\boldmath
$q$})
{\cal C}(0, \mbox{\boldmath $Q$})
\right.}\nonumber\\
&&\displaystyle{\left.
+2\frac{q_\alpha Q_\beta + Q_\alpha q_\beta}{q^2}
{\cal D}(-\omega , \mbox{\boldmath
$q$})
{\cal C}(0 , \mbox{\boldmath $Q$}- \mbox{\boldmath
$q$})
{\cal C}(-\omega, \mbox{\boldmath $Q$})
+\delta_{\alpha\beta}{\cal D}(-\omega , \mbox{\boldmath
$q$})
{\cal C}^2(-\omega, \mbox{\boldmath $Q$}) +
8Dq_\alpha q_\beta
{\cal D}^3(-\omega , \mbox{\boldmath
$q$}){\cal C}(-\omega, \mbox{\boldmath $Q$})
\right\}}.
\nonumber
\end{eqnarray}

Derivation of Eq.~(\ref{eq:4.7}) from Eq.~(\ref{eq:5.21}) is now just
a matter of a lengthy albeit straightforward calculation. We will
write most important intermediate formulas below. Integration over the
wavevectors is performed first. To carry out this integration, it is
convenient to represent each Cooperon and diffuson propagator by additional
time integral
\begin{equation}
{\cal C}(\omega, \mbox{\boldmath $Q$})
= \int_0^\infty dt \exp\left(- \frac{t}{{\cal C}(\omega, \mbox{\boldmath
$Q$})}\right)
\label{eq:5.22}
\end{equation}
After this manipulation, the integrals over wavevectors become
Gaussian and can be immediately calculated. We find
\begin{mathletters}
\label{eq:5.23}
\begin{eqnarray}
I_1 &\equiv&
\displaystyle{
 \int \frac{d^dqd^dQ}{\left(2\pi\right)^{2d}}
\frac{1}{Dq^2}
\left\{{\cal C}^2(0, \mbox{\boldmath $Q$})
\left[\frac{{\cal C}(-\omega , \mbox{\boldmath $Q$}+ \mbox{\boldmath
$q$})}{{\cal D}(-\omega , \mbox{\boldmath $q$})}
-1\right] 
+ \left[{\cal C}(-\omega , \mbox{\boldmath $Q$})
-  {\cal C}(0, \mbox{\boldmath $Q$})\right]
{\cal D}(-\omega , \mbox{\boldmath $q$})
+
2 {\cal C}(-\omega , \mbox{\boldmath $Q$}+ \mbox{\boldmath
$q$})
{\cal C}(0, \mbox{\boldmath $Q$})
\right\}}
\nonumber \\
&=&\frac{\Gamma^2\left(2-\frac{d}{2}\right)}{\left(4\pi D\right)^d}
\left\{
\frac{4}{2-d}\left(i\omega \right)^{d/2-2}
\left(\frac{1}{\tau_H}\right)^{d/2-1} -
\left[\frac{2}{2-d}\right]^2
\left[1 + \left(1-d\right)
\frac{\Gamma (3-d) \Gamma (d/2)}{\Gamma (2-d/2)}
\right]\left(i\omega \right)^{d-3}
\right\},
\label{eq:5.23a}
\end{eqnarray}
where $\Gamma(x)$ is the Gamma function. 
Equation (\ref{eq:5.23a}) is written with an accuracy up to term of the
order
of $\omega^{d-3}\left({\omega\tau_H}\right)^{-d/2}$. This accuracy
however is sufficient for our purposes, since all the Cooperons and
diffusons
here are functions analytic in the complex semiplane ${\rm
Im}\omega < 0$, and therefore integration over $\omega$ is determined
by $\omega \gtrsim T$. On the other hand, we are considering the case
$T\tau_H \gg 1$, and all the corrections to Eq.~(\ref{eq:5.23}) are
smaller by this parameter. Other integrals calculated in the same
approximation are given by
\begin{eqnarray}
I_2 &\equiv&
 \int \frac{d^dqd^dQ}{\left(2\pi\right)^{2d}}
\frac{q_\alpha}{D q^2}\frac{\partial}{\partial q_\beta}
{\cal C}(-\omega , \mbox{\boldmath $Q$}+ \mbox{\boldmath
$q$})
{\cal C}(0, \mbox{\boldmath $Q$})=
\nonumber \\
&=&
-\frac{2\delta_{\alpha\beta}}{d}
\frac{\Gamma^2\left(2-\frac{d}{2}\right)}{\left(4\pi D\right)^d}
\left\{
\frac{2}{2-d}\left(i\omega \right)^{d/2-2}
\left(\frac{1}{\tau_H}\right)^{d/2-1} -
\frac{2}{2-d}
\frac{\Gamma (3-d) \Gamma (d/2)}{\Gamma (2-d/2)}
\left(i\omega \right)^{d-3}
\right\};
\end{eqnarray}
\begin{eqnarray}
I_3&\equiv &
 \int \frac{d^dqd^dQ}{\left(2\pi\right)^{2d}}
\frac{q_\alpha q_\beta}{q^2}
{\cal D}(-\omega , \mbox{\boldmath
$q$})
{\cal C}(-\omega , \mbox{\boldmath $Q$}+ \mbox{\boldmath
$q$})
{\cal C}(0, \mbox{\boldmath $Q$})
\nonumber\\
&=&\frac{\delta_{\alpha\beta}}{d}
\frac{2}{2-d}
\frac{\Gamma^2\left(2-\frac{d}{2}\right)}{\left(4\pi D\right)^d}
\left[
\left(i\omega \right)^{d/2-2}\left(\frac{1}{\tau_H}\right)^{d/2-1}
-\frac{1}{3-d}\left(i\omega \right)^{d-3}
\right]
\end{eqnarray}
\begin{eqnarray}
I_4 &\equiv&
 \int \frac{d^dqd^dQ}{\left(2\pi\right)^{2d}}
\frac{q_\alpha Q_\beta}{q^2}
{\cal D}(-\omega , \mbox{\boldmath
$q$})
{\cal C}(0, \mbox{\boldmath $Q$}- \mbox{\boldmath
$q$})
{\cal C}(-\omega, \mbox{\boldmath $Q$}) \nonumber\\
&=&\frac{\delta_{\alpha\beta}}{d}
\frac{2}{2-d}
\frac{\Gamma^2\left(2-\frac{d}{2}\right)}{\left(4\pi D\right)^d}
\left[
\left(i\omega \right)^{d/2-2}\left(\frac{1}{\tau_H}\right)^{d/2-1}
-\frac{1}{3-d}\left(i\omega \right)^{d-3}
\right]
\label{eq:5.23d}\\
&-&\frac{\delta_{\alpha\beta}}{d}
\frac{\left(i\omega \right)^{d-3}}{\left(4\pi D\right)^d}
\int_0^{\infty}\frac{dx \ dy \ dz}{\left(xy+xz+yz\right)^{d/2}}
\left(\frac{y}{y+z}\right)e^{-x-y}
\nonumber
;
\end{eqnarray}
\begin{equation}
I_5 \equiv
 \int \frac{d^dqd^dQ}{\left(2\pi\right)^{2d}}
{\cal D}(-\omega , \mbox{\boldmath
$q$})
{\cal C}^2(-\omega, \mbox{\boldmath $Q$})=
\frac{2}{2-d}
\frac{\Gamma^2\left(2-\frac{d}{2}\right)}{\left(4\pi D\right)^d}
\left(i\omega \right)^{d-3};
\end{equation}
\begin{equation}
I_6 \equiv
 \int \frac{d^dqd^dQ}{\left(2\pi\right)^{2d}}
Dq_\alpha q_\beta{\cal D}^3(-\omega , \mbox{\boldmath
$q$})
{\cal C}(-\omega, \mbox{\boldmath $Q$})=
\frac{\delta_{\alpha\beta}}{4\left(2-d\right)}
\frac{\Gamma^2\left(2-\frac{d}{2}\right)}{\left(4\pi D\right)^d}
\left(i\omega \right)^{d-3}.
\end{equation}
\end{mathletters}
Substituting Eqs.~(\ref{eq:5.23}) into Eq.~(\ref{eq:5.21}),
retaining only most singular at $d\to 2$ coefficients in front of
$\omega^{d-3}$,  
and omitting indices $\alpha,\beta$, we find
\begin{equation}
\delta\sigma_{\rm CWL}
=\frac{\sigma}{2\pi\nu^2}
\frac{1}{\left(4\pi D\right)^d}
{\rm Im}
\int
\frac{d\omega }{2\pi} 
\left[\frac{d}{d\omega}\left(\omega
\coth \frac{\omega}{2T}\right)\right] 
\left[-\frac{4\Gamma^2(2-d/2)}{d}
\left(i\omega \right)^{d/2-2}
\left(\frac{1}{\tau_H}\right)^{d/2-1}
+\frac{4}{2-d}\left(i\omega \right)^{d-3}\right].
\label{eq:5.24}
\end{equation}
Using
formula
\[
{\rm Im} \int_{-\infty}^\infty\frac{d x}{(i x)^\alpha}
\frac{d}{dx}\left(x \coth\frac{x}{2}\right)=
-\frac{2\alpha}{\left(2\pi\right)^{\alpha-1}}
\zeta(\alpha),
\]
with $\zeta(x)$ being Riemann zeta-function, we obtain
Eq.~(\ref{eq:4.7}).

\section{Discussion of the results}
\label{sec:6}
\subsection{Theory}

In Sec.~\ref{sec:5}, we demonstrated explicitly that dephasing can be
caused only by e-e collisions with the energy transfer smaller or of
the order of temperature, see Eq.~(\ref{eq:4.5}).  This conclusion
clearly contradicts the results of Ref.~\onlinecite{Zaikin2}.  It
means that the procedure proposed in Ref.~\onlinecite{Zaikin2} fails
already on the level of the first order perturbation theory in the e-e
interactions.  This calculation is sufficient to make the conclusion
that the ``old'' rather than ``new'' theory describes the
temperature dependence of the dephasing rate correctly and thus to end
the pointless discussion.  However, the errors in Ref.~\onlinecite{Zaikin2}
originate from a quite typical misuse of the semiclassical
approximations.  For this reason we would like to highlight these
errors and to discuss them in more detail.  

According to Eqs.~(\ref{eq:2.42}), (\ref{eq:4.5}) and (\ref{eq:3.3}),
dephasing can be found by determining the coefficient in front of the
term $\left[{\cal C}(\mbox{\boldmath $Q$},0)\right]^2$.  The $Q
\rightarrow O$ and $\tau \rightarrow \infty$ limit of this coefficient
has a meaning of the inverse dephasing time $1/\tau_{\phi}$, while its
$Q$ and $H$-dependences represent renormalization of the diffusion
constant $D$ and of the residue of the Cooper pole respectively. These
renormalizations contribute eventually to the interaction correction
to the conductivity (\ref{eq:4.7}).

Therefore we can focus only on the first term in brackets
in Eq.~(\ref{eq:5.21}):
\begin{equation}
\sigma^{?}
=-\frac{\sigma}{2\pi\nu^2}{\rm Im}
\int
\frac{d\omega }{2\pi} 
\left[\frac{d}{d\omega}\left(\omega
\coth \frac{\omega}{2T}\right)\right] 
\int \frac{d^dqd^dQ}{\left(2\pi\right)^{2d}}
{\cal L}^A(\omega,q)
{\cal C}^2(0, \mbox{\boldmath $Q$})
\left[{{\cal C}(-\omega , \mbox{\boldmath $Q$}+ \mbox{\boldmath
$q$})}
-{{\cal D}(-\omega , \mbox{\boldmath $q$})}\right].
\label{eq:6.1}
\end{equation}

Note that in the corresponding equation of Ref.~\onlinecite{Zaikin2}, 
see Eq.~(\ref{eq:3.3}), only
the first Cooperon term in the brackets is present.  The second
diffuson term changes the result dramatically since at $Q
\rightarrow O$ and $\tau_{H} \rightarrow \infty$ it simply
cancels the first one. 

In order to understand why the diffuson term and thus the
cancellation are missed in Ref.~\onlinecite{Zaikin2}, 
let us examine step by step the derivation proposed in that paper.
\begin{enumerate}
\item
 Authors of Ref.~\onlinecite{Zaikin2} used Keldysh technique and decoupled the 
interaction by Hubbard-Stratonovich transformation.  This step is rather 
standard and the result in the perturbation theory should be
consistent 
with our 
Eq.~\ref{eq:5.12}.

\item
 Their next step was to write down the formally exact path integral
expression for Green functions in a disordered potential
\begin{eqnarray}
G^R\left(\epsilon; \mbox{\boldmath $r$}_1, \mbox{\boldmath
$r$}_2\right)&=&
\int_0^\infty dt e^{i\epsilon t} 
\int_{\mbox{\boldmath $r$}(0)=
\mbox{\boldmath $r$}_1}^{\mbox{\boldmath $r$}(t)=\mbox{\boldmath $r$}_2}
{\cal D}\left[\mbox{\boldmath $r$}(t)\right]
\exp\left(\frac{i}{\hbar}{\cal S}\left[\mbox{\boldmath $r$}(t)\right]
\right)\nonumber\\
G^A\left(\epsilon; \mbox{\boldmath $r$}_1, \mbox{\boldmath
$r$}_2\right)&=&
\int_{-\infty}^0 dt e^{i\epsilon t} 
\int_{\mbox{\boldmath $r$}(0)=
\mbox{\boldmath $r$}_1}^{\mbox{\boldmath $r$}(t)=\mbox{\boldmath $r$}_2}
{\cal D}\left[\mbox{\boldmath $r$}(t)\right]
\exp\left(-\frac{i}{\hbar}{\cal S}\left[\mbox{\boldmath $r$}(t)\right]
\right)\nonumber\\
\label{eq:6.2}
\end{eqnarray}
where   ${\cal S}\left[\mbox{\boldmath $r$}(t)\right]$ is the
classical action over the trajectory $\mbox{\boldmath $r$}(t)$.

\item
Afterwards an averaging of products of Green functions was performed.
During this average only some paths were selected by hand.  
It is this selection that, 
as far as we understand, causes all the errors.  
 
\end{enumerate}

\begin{figure}
\vspace{0.2 cm}
\epsfxsize=7.7cm
\centerline{\epsfbox{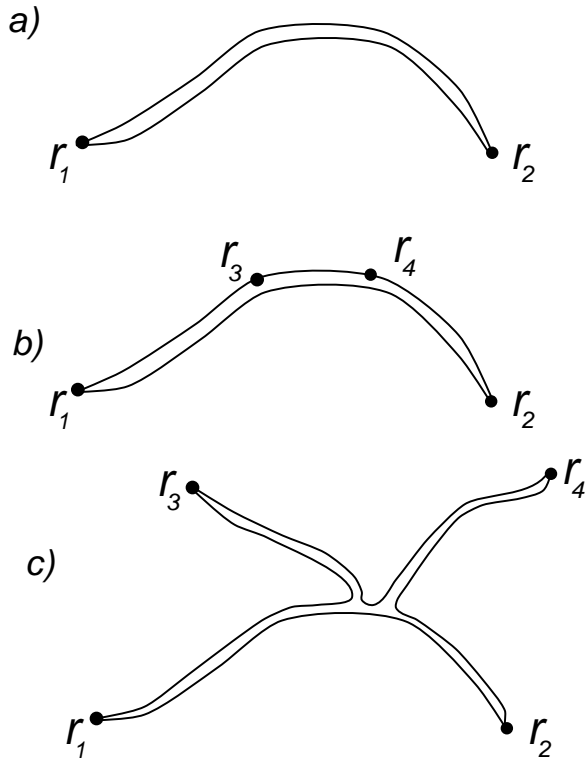}}
\vspace{0.8cm}
\caption{Example of the classical paths contributing to the averages
of products of the Green functions.}
\label{Fig10}
\end{figure}

Indeed, consider
the average 
\begin{equation}
\langle\langle
G^R(\epsilon; \mbox{\boldmath $r$}_1, \mbox{\boldmath $r$}_2)
G^A(\epsilon; \mbox{\boldmath $r$}_1, \mbox{\boldmath $r$}_2)
\rangle\rangle=2\pi\nu
{\cal D}(0;\mbox{\boldmath $r$}_1, \mbox{\boldmath $r$}_2),
\quad
{\cal D}(\omega;\mbox{\boldmath $r$}_1, \mbox{\boldmath $r$}_2)
=\int \frac{d^dQ}{(2\pi)^d}\frac{
e^{i\mbox{\boldmath $Q$}
(\mbox{\boldmath $r$}_1- \mbox{\boldmath $r$}_2)}}{-i\omega + DQ^2}
\label{eq:6.3}
\end{equation}
Here, for simplicity, we consider the case of zero magnetic field and
do not distinguish between diffusons and Cooperons.  This average can
be obtained semiclassically by pairing the paths shown in
Fig.~\ref{Fig10}a and it is reproduced correctly in
Ref.~\onlinecite{Zaikin2}. The same is true for the simple
average involving four coordinates, see Fig.~\ref{Fig10}b:
\begin{equation}
\langle\langle
G^A(\epsilon; \mbox{\boldmath $r$}_1, \mbox{\boldmath $r$}_2)
G^R(\epsilon; \mbox{\boldmath $r$}_1, \mbox{\boldmath $r$}_3)
G^R(\epsilon-\omega; \mbox{\boldmath $r$}_3, \mbox{\boldmath $r$}_4)
G^R(\epsilon; \mbox{\boldmath $r$}_4, \mbox{\boldmath $r$}_2)
\rangle\rangle=- 2\pi\nu
{\cal D}(0;\mbox{\boldmath $r$}_1, \mbox{\boldmath $r$}_3)
{\cal D}(-\omega;\mbox{\boldmath $r$}_3, \mbox{\boldmath $r$}_4)
{\cal D}(0;\mbox{\boldmath $r$}_4, \mbox{\boldmath $r$}_2)
.
\label{eq:6.4}
\end{equation}
This expression corresponds to the diagram (\ref{Fig6})a.

However, the other higher order products involve intersection of the
classical paths, see Fig.~\ref{Fig10}c.  Such configurations
contribute in, e.g., average of the type
\begin{eqnarray}
\langle\langle
G^R(\epsilon; \mbox{\boldmath $r$}_1, \mbox{\boldmath $r$}_2)
&G^A&(\epsilon; \mbox{\boldmath $r$}_1, \mbox{\boldmath $r$}_3)
G^R(\epsilon-\omega; \mbox{\boldmath $r$}_3, \mbox{\boldmath $r$}_4)
G^A(\epsilon; \mbox{\boldmath $r$}_4, \mbox{\boldmath $r$}_2)
\rangle\rangle
\label{eq:6.5}
\\
&=&2\pi\nu
\int d \mbox{\boldmath $r$}_5
\left\{i\omega
{\cal D}(0;\mbox{\boldmath $r$}_1, \mbox{\boldmath $r$}_5)
{\cal D}(-\omega;\mbox{\boldmath $r$}_5, \mbox{\boldmath $r$}_3)
{\cal D}(-\omega;\mbox{\boldmath $r$}_5, \mbox{\boldmath $r$}_4)
{\cal D}(0;\mbox{\boldmath $r$}_2, \mbox{\boldmath $r$}_5)
\right.
\nonumber
\\
&+&
D
\left[\mbox{\boldmath $\nabla$}_5
{\cal D}(0;\mbox{\boldmath $r$}_1, \mbox{\boldmath $r$}_5)
{\cal D}(-\omega;\mbox{\boldmath $r$}_5, \mbox{\boldmath $r$}_4)
\right] 
\left[\mbox{\boldmath $\nabla$}_5
{\cal D}(-\omega;\mbox{\boldmath $r$}_5, \mbox{\boldmath $r$}_3)
{\cal D}(0;\mbox{\boldmath $r$}_2, \mbox{\boldmath $r$}_5)
\right]
\nonumber\\
&+&2
D
\left[\mbox{\boldmath $\nabla$}_5
{\cal D}(0;\mbox{\boldmath $r$}_1, \mbox{\boldmath $r$}_5)
\right]
\left[\mbox{\boldmath $\nabla$}_5
{\cal D}(-\omega;\mbox{\boldmath $r$}_5, \mbox{\boldmath $r$}_4)
\right]
{\cal D}(-\omega;\mbox{\boldmath $r$}_5, \mbox{\boldmath $r$}_3)
{\cal D}(0;\mbox{\boldmath $r$}_2, \mbox{\boldmath $r$}_5)
\nonumber\\
&+&2D
\left.
{\cal D}(0;\mbox{\boldmath $r$}_1, \mbox{\boldmath $r$}_5)
{\cal D}(-\omega;\mbox{\boldmath $r$}_5, \mbox{\boldmath $r$}_4)
\left[\mbox{\boldmath $\nabla$}_5
{\cal D}(-\omega;\mbox{\boldmath $r$}_5, \mbox{\boldmath $r$}_3)
\right]
\left[\mbox{\boldmath $\nabla$}_5
{\cal D}(0;\mbox{\boldmath $r$}_2, \mbox{\boldmath $r$}_5)
\right]
\right\},
\nonumber
\end{eqnarray}
see diagram in Fig.~\ref{Fig7}c. The point $\mbox{\boldmath $r$}_5$
has the meaning of the intersection point between classical
trajectories. These intersections were not taken into account at all
in Ref.~\onlinecite{Zaikin2}.

After performing the Fourier transforms over the coordinate
differences $\mbox{\boldmath $r$}_3 - \mbox{\boldmath $r$}_4$ and 
$\mbox{\boldmath $r$}_1 - \mbox{\boldmath $r$}_2$, one finds
\begin{eqnarray*}
&&\left(\ref{eq:6.4}\right) \to
-\frac{2\pi\nu}{\left[DQ^2\right]^2\left[i\omega +
D\left(\mbox{\boldmath $Q$}-\mbox{\boldmath $q$}\right)^2\right]}\\
&&\left(\ref{eq:6.5}\right) \to
\frac{2\pi\nu}{\left[DQ^2\right]^2\left[i\omega +
Dq^2\right]}
+
\frac{2\pi\nu}{\left[DQ^2\right]\left[i\omega +
Dq^2\right]^2}
- \frac{4\pi\nu\ D\mbox{\boldmath $Q$}\mbox{\boldmath $q$}}
{\left[DQ^2\right]^2\left[i\omega +
Dq^2\right]^2}
\end{eqnarray*} 
where the last term will vanish after the angular integration.
One can see that signs of Eqs.~(\ref{eq:6.3}) and
(\ref{eq:6.4}) are opposite, which eventually leads to the
cancellation in Eq.~(\ref{eq:6.1}).
In fact, taking into account the switches between intersection of the
classical paths is very crucial: mistreatment of such switches leads,
e.g. to the violation of the conservation of number of particles and
all the theoretical predictions of similar quality! In the
diagrammatic technique such switches are described by so-called Hikami
boxes (the shaded boxes in Fig.~\ref{Fig7} - \ref{Fig9}). Semiclassical
approach usually has difficulties in the description of the
transitions between the classical trajectories. Physical reasons for
this difficulty were first discussed by Larkin and
Ovchinnikov\cite{LarkinOvchinnikov} in application to the
superconductivity, and  in
Ref.~\onlinecite{AleinerLarkin} in application to the weak localization.

\subsection{Experiment}

Let us compare the available experimental results with predictions of the
theory by Golubev and Zaikin\cite{Zaikin2}. 
Authors concluded that
``according to Eqs.~(81), [of Ref.~\onlinecite{Zaikin2}] in 2d and 3d
systems the decoherence time becomes independent on $T$ already at
relatively high temperatures. In the 3d case such temperatures can be
of the order of the inverse transport time, while for 2d systems we
have
$T_q \sim \left(2\tau \ln (p_F^2al)\right)^{-1}$ ''. 
If we now recall that the momentum
relaxation time $\tau $ in the weak localization experiments with
disordered metals is typically $10^{-16}-10^{-14}\sec $,
we will have to believe that the phase-breaking time due to electron-electron
scattering should be $T$-independent in 2d and 3d conductors 
at all temperatures
below the
melting point! This estimate for $\tau _\varphi $ is also much smaller
than the electron-phonon scattering time $\tau _{e-ph}$ in the temperature
range $T<20K$, which is typical for the experimental studies of the quantum
corrections to the conductivity. In this situation, the $T$-dependent weak
localization corrections to the resistance of 2d and 3d conductors should be 
\textit{non-observable}. For 1d and 2d conductors, the strong-localization
regime should be \textit{non-observable } either. Both corollaries of
Eqs.~(81) of Ref.~\onlinecite{Zaikin2}
are in a profound contradiction with the bulk of experimental data,
which we review below.

\subsubsection{Three-dimensional conductors.}

\begin{figure}
\vspace{0.2 cm}
\epsfxsize=10.7cm
\centerline{\epsfbox{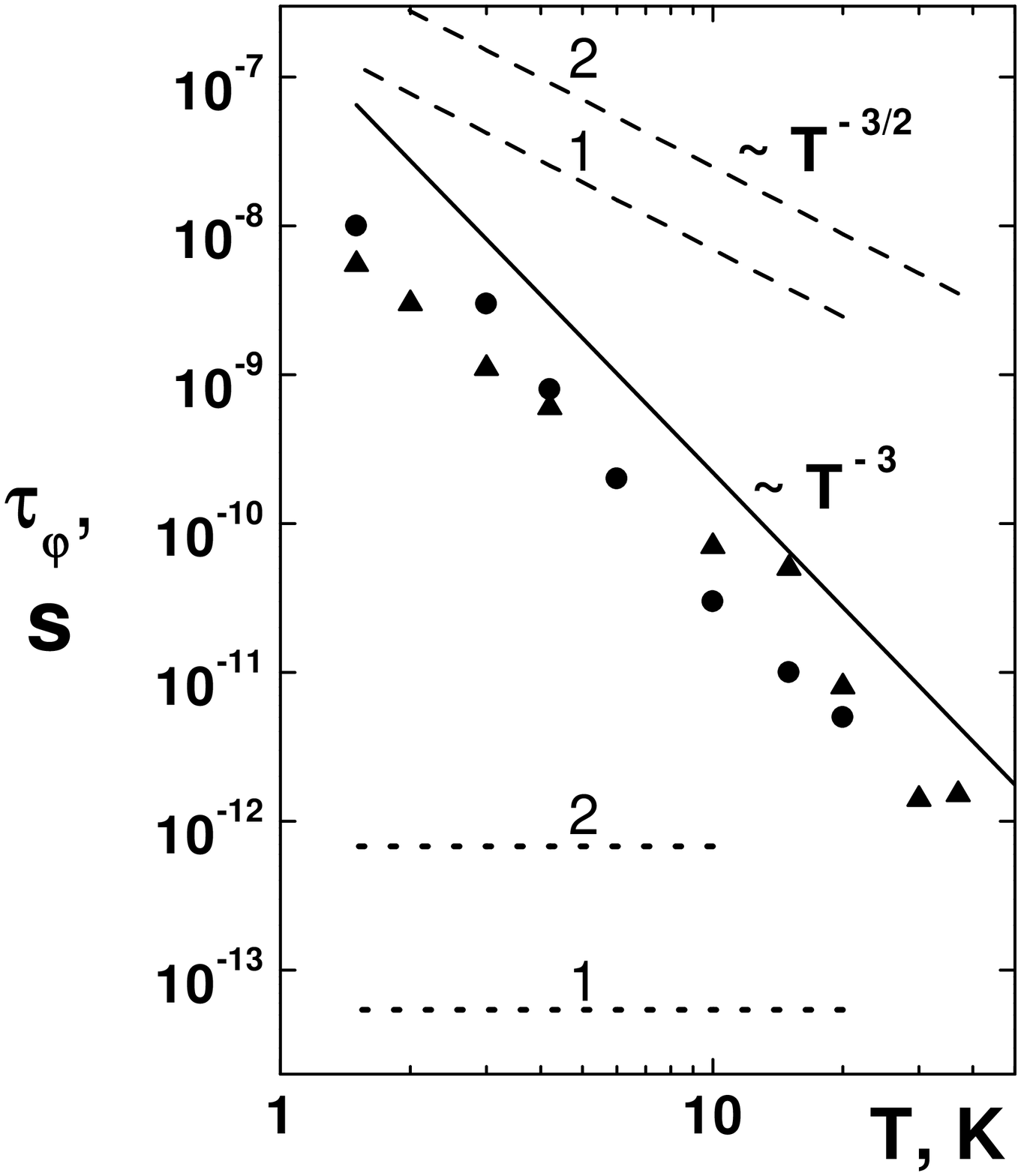}}
\vspace{0.2cm}
\caption{ 
The temperature dependence of $\tau_\varphi $ in $3d$ metal
films: $ \bigcirc $ - disordered Cu films ($1$) with 
$\rho =6.6\cdot 10^{-5}\Omega \cdot cm$, 
$D=3.9\ cm^2/s$, $\tau =4.8\cdot 10^{-16}s$, 
Ref.~\protect\onlinecite{AGZ}; 
$\bigtriangleup$ ---  $Cu_{0.9}Ge_{0.1}$ films ($2$) with
 $\rho =2.8\cdot 10^{-5}\Omega \cdot cm$, $D=9.3 \ cm^2/s$, $\tau =1.1\cdot
10^{-15}s$, Ref.~\protect\onlinecite{gey}.  
The solid line: $\tau_{e-ph}(\mbox{pure
Cu})=2.2\cdot 10^{-7}/T^3s$. The dashed lines: $\tau _{ee}(3d)$,
Eq.~(\ref{tee3D}) for these samples
$1$ and $2$ respectively. The dotted lines: the maximum
$\tau_\varphi\ (3d)$ according to Eq.~(\protect\ref{tGZ3D}).  
}
\label{Fig3d}
\end{figure}

The study of the WL corrections to the conductivity of 3d systems has
been focused mostly on thick ($a\gg L_\varphi$) 
disordered metal films, metal glasses,
and heavily doped semiconductors (for reviews, see
Refs.~\onlinecite{AAGS,dugdale,pol}). For all these systems, the
well-pronounced temperature dependence of the WL corrections has been
observed. From the analysis of the WL magnetoresistance, the
temperature dependence of the phase-breaking time has been
extracted. The typical $\tau _\varphi (T)$ dependences, obtained for
3d disordered films of Cu \cite{AGZ} and Cu-Ge \cite{gey}, are shown
in Fig.~\ref{Fig3d}.  The electron-phonon scattering governs the
phase-relaxation in these $3d$ conductors over the whole temperature
range $T=1.5-40K$. The solid line $ \tau _{e-ph}\propto T^{-3}$
corresponds to the electron-phonon relaxation time due to the
longitudinal phonons in pure Cu (see, e.g., Table 3 from
Ref.~\onlinecite{AAGS}). 
The agreement with the experimental data can be further
improved if one takes into account the contribution of transverse
phonons, which dominates in disordered metals \cite{Reyzer,ptit}. The
temperature dependence of $\tau _\varphi (3d)$ due to the
electron-electron interactions with the small momentum transfer
\cite{AA}
\begin{equation}
\frac{\hbar }{\tau _{ee}}=\frac{\sqrt{2}}{12\pi ^2\nu }
\left(\frac {k_BT}{\hbar D}\right)^{3/2}  \label{tee3D}
\end{equation}
($\nu $ is the density of states) is shown in Fig.~\ref{Fig3d} 
with the dashed lines.
The upper limit of $\tau _\varphi $ according to 
Ref.~\onlinecite{Zaikin2}
\begin{equation}
\frac{\hbar }{\tau_{GZ}^{(3d)}}=\frac{e^2}{3\pi ^2\sigma \sqrt{2D}
\tau ^{3/2}}  \label{tGZ3D}
\end{equation}
($\sigma $ is the conductivity, and $\tau $ is the elastic scattering
time) is shown in Fig.~\ref{Fig3d} with the dotted lines; this
estimate is much smaller than the experimentally measured time. The
discrepancy here is as large as $5$ (!) orders of magnitude.

Similar discrepancy between the experimental values of $\tau _\varphi $
and predictions of Golubev and Zaikin \cite{Zaikin2} 
has been observed for metal
glasses (see,
e.g. Refs.~\onlinecite{dugdale,sahnoune,bieri,mayeya}). 
For example,
 for
$(Ca_{80}Al_{20})_{97.1}Au_{2.9}$ with $\rho =1.3\cdot 10^{-4\text{
}}\Omega \cdot cm$ and $D\simeq 1$ $cm^2/s$ the experimental value of  
 $\tau _\varphi
(2.5K)
 = 7.5\cdot
10^{-10}s$ was observed\cite{mayeya}.
On the other hand, from  Eq.~(\ref{tGZ3D}) one could expect $
\tau _\varphi <4\cdot 10^{-14}s$. For $Y_{80}{Si}_{20}$ with $\rho
=5\cdot 10^{-4\text{ }}\Omega \cdot cm$, $\tau \simeq 1\cdot
10^{-16}s$ and $ D\simeq 0.7$ $cm^2/s,$ Bieri \textit{et
al}.\cite{bieri} reported $\tau _\varphi (4.2K)\simeq 2\cdot
10^{-11}s$, whereas Eq.~(\ref{tGZ3D}) gives $\tau _\varphi \simeq
3\cdot 10^{-16}s.$

\subsubsection{Two-dimensional conductors.}

\begin{figure}
\vspace{0.2 cm}
\epsfxsize=10.7cm
\centerline{\epsfbox{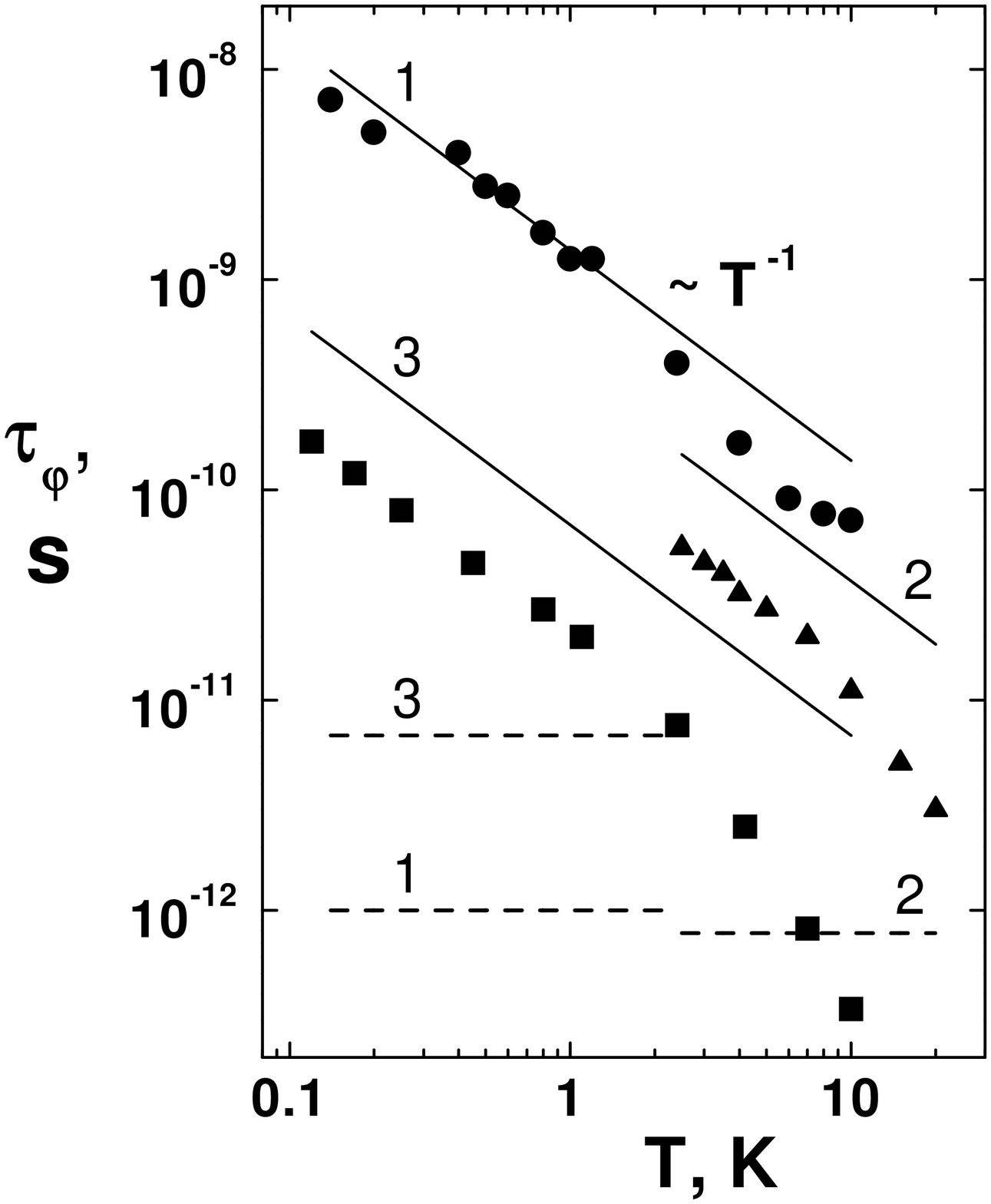}}
\vspace{0.2cm}
\caption{ The temperature dependence of $\tau _\varphi $ in $2d$ metal
films: 1 $(\bigcirc )$ - $Mg$ film with $R_{\Box }=22.3$ $\Omega $,
$D=3.8$ $cm^2/s,$ $\tau =4.6\cdot 10^{-16}s$; \protect\cite{white} 2
$(\bigtriangleup )$ - $Al$ film with $R_{\Box }=112$ $\Omega $,
$D=10.4$ $cm^2/s,$ $\tau =1.7\cdot 10^{-15}s$; \protect\cite{GGZ} 3
$(\Box )-Bi$ film, $R_{\Box }=630$ $\Omega $, $D=8$ $cm^2/s,$ $ \tau
=2.8\cdot 10^{-14}s$.\protect\cite{komori}  The solid lines: the temperature
dependences of $\tau_\varphi(2d)$ [Eq.~(\protect\ref{eq:2.35b})] for these
samples. The dashed lines: the maximum $\tau_{GZ} (2d)$ according to 
Eq.~(\protect\ref{tGZ2D}).  }
\label{Fig2df}
\end{figure}

\begin{figure}
\vspace{0.2 cm}
\epsfxsize=10.7cm
\centerline{\epsfbox{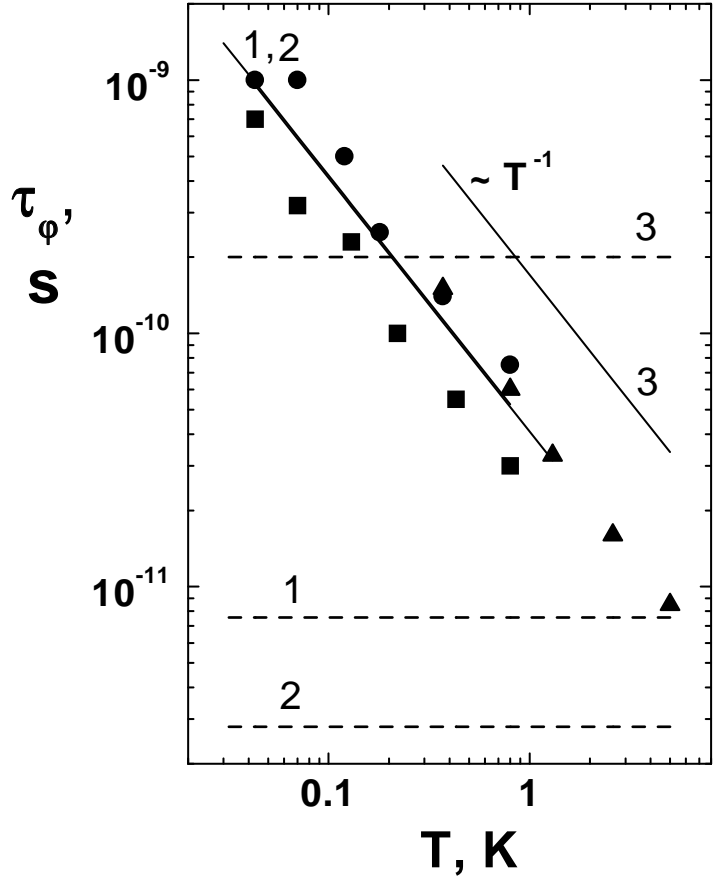}}
\vspace{0.2cm}
\caption{ The temperature dependence of $\tau_\varphi $ in 2d $
 GaAs-Al_xGa_{1-x}As$ heterostructures \protect\cite{lin}: 1 $(\Box )$
 - $n=0.87\cdot 10^{11}cm^{-2}$; $R_{\Box }=4.3$ $k\Omega $, $D=51$
 $cm^2/s,$ $\tau =6.3\cdot 10^{-13}s$ ; 2 $(\bigcirc )$ - $n=2.86\cdot
 10^{11}cm^{-2}$; $ R_{\Box }=4.0$ $k\Omega $, $D=55$ $cm^2/s,$ $\tau
 =2.1\cdot 10^{-13}s$; 3 $ (\bigtriangleup )-n=7.11\cdot
 10^{11}cm^{-2}$; $R_{\Box }=300$ $\Omega $, $ D=753$ $cm^2/s,$ $\tau
 =1.1\cdot 10^{-12}s$. The solid lines: the temperature dependences
 of $\tau _N(2D)$ [Eq.~(\ref{eq:2.35b})] for these samples.  The
 dashed lines: the maximum $\tau_\varphi (2d)$ according to
 Eq.~(\protect\ref{tGZ2D}).  }
\label{Fig2dGaAs}
\end{figure}

{\em The weak localization regime.}--- Figures~\ref{Fig2df} and
\ref{Fig2dGaAs} show the typical temperature dependences of 
$\tau_\varphi $ in $2d$ metal films \cite{white,komori,GGZ} and 
$GaAs-AlGaAs$
heterostructures \cite{lin}, correspondingly. At low temperatures
($T<1-5K)$, the phase-breaking is governed by the quasi-elastic
electron-electron scattering (the Nyquist phase breaking). The
experimental values of $\tau _\varphi $ are usually half as large as
the estimate of $\tau_\varphi(2d)$, given by Eq.~(\ref{eq:2.35b}).

On the other hand, the experimental data for $\tau_\varphi$ exceed by far
the upper limit of $\tau_\varphi (2d)$ according to Eq.~(81) of Ref.~
\onlinecite{Zaikin2}
\begin{equation}
\frac{\hbar }{\tau_{GZ}(2D)}= \frac{e^2}{4\pi \sigma _2\tau }.
\label{tGZ2D}
\end{equation}
The  discrepancy of Eq.~(\ref{tGZ2D}) with the experiment is
especially pronounced for the most disordered samples
(4 orders of magnitude for cryo-deposited $Mg$ films at $T=0.1K$, 
Ref.~\onlinecite{white}).

{\em The strong localization regime.}  With increasing the sheet
resistance of $2d$ samples up to $\simeq 30$ $k\Omega $, the crossover
to the strong localization regime has been observed for both
quasi-$2d$ metal films and $2d$ electron gas in semiconductor
structures (see, f.i., Refs.~\onlinecite{bishop,imry,liu,hsu,dahm}).
A little extension of the Golubev-Zaikin logic leads one to the
conclusion that the ``zero point motion'' also suppresses the
interaction correction to the conductivity since the latter is also
caused by the interference, see Sec.~\ref{sec:2.2}. As the result, the
conductivity should be temperature-independent at $T \lesssim T_{GZ}
=\hbar/\tau_{GZ}(2d)$. For all the samples studied, $T_{GZ}$ is very
large; for example, the samples of Refs.~\onlinecite{imry,liu,hsu} are
characterized with $T_{GZ} \simeq 10^5K$.  On the other hand, all the
crossovers to the strong localization were observed at temperatures of
the order of $1K$---$10K$.  These experimental data rule out the
statement by Golubev and Zaikin that ``\dots strong localization does
not take place at all and the $1d$ and $2d$ metals do not become
insulators even at $T=0$''.

\subsubsection{One-dimensional conductors.}
In Ref.~\onlinecite{Zaikin3}, Golubev and Zaikin compared their
results with the experiment by Khavin {\em et al.}\cite{khavin} and
claimed a convincing agreement. We will show below that, in fact,
 the agreement exists only in the temperature region where their
result coincides with the conventional one\cite{AAK}. Outside this
region the comparison becomes meaningless because of the observed
crossover to the strong localization regime.

{\em The weak localization regime.}---
In one dimension, the expression for $\tau_\varphi (1d)$ obtained by
Golubev and Zaikin \cite{Zaikin1,Zaikin2}, see Eq.~(77) of 
Ref.~\onlinecite{Zaikin2} or Eq.~(10) of Ref.~\onlinecite{Zaikin1}, 
is 
\begin{equation}
\frac{\hbar }{\tau_{GZ} (1d)}=\frac{e^2}{\pi \sigma _1}
\sqrt{
\frac{2D}{\tau }}
\left[ 1+\frac{2k_BT}{\hbar }\sqrt{\tau_{GZ} \tau }\right].
   \label{tGZ1D}
\end{equation}
In the temperature range $k_B{T}\gg\hbar /\sqrt{\tau_\varphi \tau }$,
this expression coincides up to a numerical factor of the order of
unity with the Nyquist phase-breaking time in 1d,\cite{AAK} see
Eq.~(\ref{eq:2.35a}).
 
It has been established experimentally that the phase breaking in 1d
conductors (both metal wires and semiconductor structures) is well
described by $\tau_\varphi(1d)$ over a wide temperature range (typically,
below several Kelvin) (see, e.g.
Refs.~\onlinecite{thornton,wind,lin,pooke,echternach,khavin}). Thus,
no wonder that the expression (\ref{tGZ1D}) is in accord with the
experimental data in the regime where the ``new theory'' coincides with the
old-fashioned Eq.~(\ref{eq:2.35a}).  We would like to point out again
that the explicit comparison of $1/\tau_{GZ}$ with the experiment is
meaningful only if the analytic expression for the 
 magnetoresistance is provided.
Since Golubev and Zaikin were not able
to obtain such formula, similar to
Eq.~(\ref{eq:2.39b}), the procedure of comparison of their
results with $\tau_\varphi$ extracted with the help of
Eq.~(\ref{eq:2.39b}) is at least problematic.

On the other hand, in the temperature range $k_B{T} \ll
\hbar\sqrt{\tau_\varphi \tau } $,  Golubev and Zaikin predict
saturation of the phase breaking length:
\begin{equation}
\frac{L_{GZ} }{\xi }=\left(\frac{3}{2\pi^2}\right)^{1/4}
\frac{1}{\sqrt{N}}  
\label{lfai}
\end{equation}
where $N$ is the number of the transverse channels in the wire,
$L_{GZ}=\sqrt{D\tau_{GZ}}$, and $\xi$ is the 1d localization
length. Once again, we do not know what the numerical factor in this
formula means, since Golubev and Zaikin did not derive the explicit
formula for the conductance.  [In fact, Eq.~(\ref{tGZ1D}) seems to be
erroneous even within the Golubev-Zaikin logic: the contribution to
the zero-temperature term comes from the wave-vectors of the order of
the inverse elastic mean free path $1/l$. Therefore, to be consistent
with the ``new theory'', one ``should'' use for a wire with the
cross-sectional dimensions greater than the elastic mean free path
either two- or three- dimensional formulas (\ref{tGZ2D}) and
(\ref{tGZ3D}), depending on the structure of the wire.]

Anyway, the authors of Ref.~\onlinecite{Zaikin3} use Eq.~(\ref{tGZ1D})
for the analysis of the data. In the regime where deviations from the
conventional formula (\ref{eq:2.35a}) can be expected from
Eq.~(\ref{tGZ1D}), the information on $\tau_\varphi$ can not and was not
obtained from the experiment, because the samples were already in the
strong localization regime.

{\em The strong localization regime.}-- It has been shown recently
that the 1d wires fabricated from Si $\delta $ -doped GaAs,
demonstrate the crossover from weak to strong localization with
decreasing the temperature\cite{gersh,khavin2}. This crossover is
driven by both localization and interaction effects\cite{khavin}; in
particular, the importance of the localization effects has been
demonstrated by observation of (1) the shift of the crossover by the
weak magnetic field; (2) the orbital magnetoresistance on the
insulating side of the crossover, and (3) the doubling of the
localization length in strong magnetic fields. These experimental
facts  argue against the conclusion of Golubev and Zaikin 
``strong localization does not take place at all and the 1d and 2d
metals do not become insulators even at $T=0$''.

\section{Conclusion}

This paper is devoted to the quantum transport of electrons in disordered
conductors in the weak localization regime. 
The motivation to write a paper on the subject that was developed almost
15 years ago was caused by recent experimental and theoretical attempts to
reconsider the existing theory of destruction of the one-particle phase
coherence (dephasing) in a system of degenerate interacting fermions. It
was proposed that, in contrast to basic conclusions of this theory, the
dephasing rate remains finite even at zero temperature due to the
zero-point oscillations of the electric field.

We criticized this proposal in details from several points of view. 
\begin{enumerate}
\item
We
recalled and discussed in details the physical picture of the quantum
corrections to the properties of the disordered conductors and tried to
demonstrate that such a revision looks absurd from the qualitative point
of view. 

\item
We made a straightforward calculation of the first order in the e-e
interaction correction to the conductivity at low dimensions in strong
magnetic fields, taking into account all the relevant
diagrams. Results of this calculation, see Eqs.~(\ref{eq:4.13}), are in
a perfect agreement with the ``old'' theory and explicitly contradict
the zero-point dephasing calculations.

\item
We  highlighted the sources of the mistakes of the
new ``theory''. The errors were found to be  due to the
misuse  of the semiclassical approximation in disorder averaging. 

\item 

We reviewed existing experiments on magnetoresistance in $3d$, $2d$, and
$1d$ dimensions, and demonstrated that the measured dephasing rate is
sometimes more than five orders of magnitude smaller than the limit,
which follows from the zero-point-motion ``theory''. 
\end{enumerate}

We really hope that
arguments presented are sufficiently strong to stop the ``theoretical''
discussion of the zero temperature dephasing.

Our paper also contains the new explicit formulas for the effect of the
interaction on the weak localization due to the elastic
scattering. Such effects, are shown to be $1/g^2$ corrections which
are smaller than the main terms considered by the conventional theory
everywhere except the vicinity of the metal-insulator crossover. We
should warn the practitioners of the field that in addition to the
extra terms in Eq.~(\ref{eq:4.13}) the first order correction to the
conductivity in the Cooper channel\cite{AA} and the second order weak
localization correction to the conductivity\cite{Gorkov79} has to be
taken into account for comparison with the experiment.

\acknowledgements 
Useful discussions with M.Yu. Reyzer are gratefully acknowledged.
We are thankful to ICTP Trieste for kind hospitality.
Part of the work performed in
Ruhr-Universit\"{a}t-Bochum 
was supported  by SFB 237
``Unordnung und grosse Fluktuationen''. 
I.A. is A.P. Sloan research fellow. 

\appendix
\section{Derivation of identities (\protect\ref{ID})}
\label{ap:1}
Let us perform the gauge transformation of the Green function
\begin{equation}
G^{R}\left(\epsilon, \mbox{\boldmath $r$}_1,
\mbox{\boldmath $r$}_2 \right)
\to 
e^{ie \mbox{\boldmath $A$} \cdot ( \mbox{\boldmath $r$}_1-
\mbox{\boldmath $r$}_2) }
G^{R}\left(\epsilon, \mbox{\boldmath $r$}_1,
\mbox{\boldmath $r$}_2 \right)
\label{a1}
\end{equation}
where {\boldmath $A$} is an arbitrary vector independent of the
coordinates.  Under this transformation, the exact one-electron
Hamiltonian changes according to the rules
\begin{equation}
\hat{H} \to \hat{H} - \hat{\mbox{\boldmath $j$}}\mbox{\boldmath $A$}
+ \frac{e^2\mbox{\boldmath $A$}^2}{2m}.
\label{a2}
\end{equation}
where $\hat{\mbox{\boldmath $j$}}$ is the current operator
(\protect\ref{eq:5.10}).

Therefore, it follows from the definition of the Green function that
the relation
\begin{equation}
e^{ie \mbox{\boldmath $A$} \cdot ( \mbox{\boldmath $r$}_1-
\mbox{\boldmath $r$}_2) } \left[ \frac{1}{\epsilon - \hat{H}+i0 }
\right]_{12} = \left[ \frac{1}{\epsilon - \hat{H}
+\hat{\mbox{\boldmath $j$}}\mbox{\boldmath $A$} -
\frac{e^2\mbox{\boldmath $A$}^2}{2m}+i0 } \right]_{12}
\label{a3}
\end{equation}
holds. Expanding both sides of Eq.~(\ref{a3}) up to the second order
in  {\boldmath $A$}, we obtain Eqs.~(\protect\ref{ID}).


\begin{references}
\bibitem{AA}B.L. Altshuler and A.G. Aronov, in {\em
Electron-Electron Interaction in Disordered Systems}, edited
by A.L. Efros and M. Pollak (North-Holland, Amsterdam, 1985).
\bibitem{Altshuler87} B.L. Altshuler, A.G. Aronov, M.E. Gershenson,
and Yu. V. Sharvin, Sov. Sci, A. Phys, {\bf 9}, 223 (1987).
\bibitem{LeeRama} P.A. Lee and T.V. Ramakrishnan, Rev. Mod. Phys.,
{\bf 57}, 287 (1985).
\bibitem{Bergman84} G. Bergman, Physics Reports, {\bf 107}, 1 (1984).
\bibitem{Altshuler82} B.L. Altshuler, A.G. Aronov, D.E. Khmelnitskii,
and A.I. Larkin, in {\em Quantum theory of solids}, edited by
I.M. Lifshitz, (Mir Publishers, Moscow, 1982).
\bibitem{Mohanty1}P. Mohanty, E.M.Q. Jarivala, and R.A. Webb,
  Phys.Rev.Lett, {\bf 78}, 3366 (1997).
\bibitem{Mohanty2} P.Mohanty and R.A. Webb, Phys. Rev. {\bf B55}, 13452
(1997).
\bibitem{Zaikin1}D.S. Golubev and A.D. Zaikin, Phys. Rev. Lett,
{\bf 81} 1074 (1998).
\bibitem{Zaikin2}D.S. Golubev and A.D. Zaikin, cond-mat/9712203.
\bibitem{Altshuler98}
B.L. Altshuler, M.E. Gershenson, and I.L. Aleiner, cond-mat/9803125.
\bibitem{Zaikin3} D.S. Golubev and A.D. Zaikin, cond-mat/9804156. 
\bibitem{MRhistory}R.A. Chentsov, Zh. Exp. Teor. Fiz., {\bf1 8}, 374
(1948).
\bibitem{ALee}B.L. Altshuler and P.A. Lee, Physics Today, {\bf 41}, 36
(1988).
\bibitem{Gorkov79}L.P. Gorkov, A.I. Larkin and D.E. Khmelnitskii,
Pis'ma Zh. Eksp. Teor. Fiz. {\bf 30}, 248 (1979) [JETP Lett. {\bf 30},
248 (1979)].
\bibitem{AB}Y. Aharonov and D. Bohm,
Phys. Rev. {\bf 115}, 485 (1959).
\bibitem{Altshuler80} B.L. Altshuler, D.E. Khmelnitskii, A.I. Larkin,
and P.A. Lee, Phys. Rev. B {\bf 22}, 5142 (1980).
\bibitem{footnote} In order to avoid confusion, we emphasize that the
term ``one dimensional diffusive system'' means that transverse size of the
system $a$ is much larger than the Fermi wavelength $\lambda_F$ but
much smaller than the magnetic length $\lambda_H$ and the phase
breaking length $L_\varphi$. Notion of Luttinger liquid is not applicable
here at all.
\bibitem{Gutzwiller} M.C. Gutzwiller, {\em Chaos in Classical and
Quantum Mechanics}, (Springer-Verlag, New York, 1990).
\bibitem{Thouless}D.J. Thouless, Phys. Rev. Lett. {\bf 39}, 1167 (1977).
\bibitem{Schmid} A.Schmid, Z. Phys. {\bf 271}, 251 (1973).
\bibitem{Reyzer} M. Reyzer and A.V. Sergeev,
Sov. Phys. JETP, {\bf 65}, 616 (1986).
\bibitem{inelastic}B.L. Altshuler and A.G. Aronov, JETP Lett., 
{\bf 30}, 482 (1979).
\bibitem{AGKL} B.L. Altshuler, Y. Gefen, A. Kamenev, L.S. Levitov,
Phys. Rev. Lett. {\bf 78}, 2803 (1997).
\bibitem{AP}V. Prigodin and B.L. Altshuler, preprint cond-mat/9703071.
\bibitem{Aronov} U. Sivan, Y. Imry and A. Aronov,
Europhys. Lett. {\bf 28}, 115 (1994).
\bibitem{AndersonLee}E. Abrahams, P.W. Anderson, P.A. Lee, and
T.V. Ramakrishnan, Phys. Rev. {\bf B24}, 6783 (1981).
\bibitem{AAK} B.L. Altshuler, A.G. Aronov, and D.E. Khmelnitskii,
J. Phys. C {\bf 15}, 7367 (1982).
\bibitem{Stern}A.~Stern, Y.~Aharonov, and Y.~Imry, Phys. Rev. A ,
 {\bf 41}, 3436 (1990).
\bibitem{footnote2}The difference in numerical factor of $2$ in 
Eqs.~(\ref{eq:2.35a}) and (\ref{eq:2.39b}) in comparison with 
Eqs.~(4.33) and (4.35) of Ref.~\onlinecite{AA} is caused by an
algebraic error in the latter reference. Note, that the experimental
data were fitted without taking this into account.
\bibitem{Chakravarty} S. Chakravarty and A. Schmid,  Phys. Rep.
{\bf 140}, 193 (1986).
\bibitem{Reyzerphi} In fact, the appearance of 
$\left(\coth\frac{\omega}{2T}+\tanh\frac{\epsilon-\omega}{2T}\right)$ in the
expression for $1/\tau_\varphi$ is well
known, contrary to the statement of Ref.~\onlinecite{Zaikin2},
see 
H. Fukuyama and E. Abrahams, Phys. Rev. B {\bf 27}, 5976 (1983);
M.Y. Reyzer, Phys. Rev. B {\bf 45}, 12949 (1992). 
\bibitem{Keldysh}L.V. Keldysh, Zh. Eksp. Teor. Fiz. {\bf 47}, 1945
(1964) [Sov. Phys. JETP, {\bf 20}, 1018 (1964)]; we use here the
electron Keldysh matrices in notation of A.I. Larkin and
Yu.N. Ovchinnikov in {\em Nonequilibrium Superconductivity}, edited by
D.N. Langenberg and A.I. Larkin, (Elsevier, Amsterdam, 1986).
\bibitem{AGD} A.A. Abrikosov, L.P. Gorkov, and I.E. Dzyaloshinskii,
{\em Methods of Quantum Field Theory in Statistical Physics,}
(Prentice-Hall, Englewood Cliffs, NJ, 1963).
\bibitem{Efetov} K.B. Efetov, {\em Supersymmetry in Disorder and
Chaos}, (Cambridge University Press, New York, 1997).
\bibitem{LarkinOvchinnikov}
A.I. Larkin and Yu.N. Ovchinnikov, Zh. Eksp. Teor. Fiz, {\bf 55}, 2262
(1968) [Sov. Phys. JETP {\bf 28}, 1200 (1969)].
\bibitem{AleinerLarkin} I.L. Aleiner and A.I. Larkin, Phys. Rev. B,
{\bf 54}, 14423 (1996).
\bibitem{AAGS}  B. L. Altshuler, A. G. Aronov, M. E. Gershenson, and Yu. V.
Sharvin, in {Sov. Phys. Rev.} A\textbf{9}, 223 (1987).

\bibitem{dugdale}  J. S. Dugdale, The electrical properties of disordered
metals, Cambridge University Press, 1995.

\bibitem{pol}  T. A. Polyanskaya and Yu. V. Shmartsev, {Sov. Phys. -
Semicond}. \textbf{23}, 1 (1989).

\bibitem{AGZ}  A. G. Aronov, M. E. Gershenson, and Yu. E. Zhuravlev, {%
Sov. Phys. - JETP} \textbf{60}, 554 (1984).

\bibitem{gey}  W. Eschner, W. Gey, and P. Warnecke, in {Proc. of the
17-th Int. Conf. on Low Temp}., eds. U. Eckern, A. Schmid, W. Weber and H.
Wuhl, p. 497, Elsevier, 1984.


\bibitem{ptit}  N. G. Ptitsina \textit{et al}., {Phys. Rev.} B 
\textbf{56}, 10089 (1997).


\bibitem{sahnoune}  A. Sahnoune, J. O. Strom-Olsen, and H. E. Fisher, 
{Phys. Rev.} B \textbf{46}, 10035 (1992).

\bibitem{bieri}  J. B. Bieri, A. Fert, G. Creuzet, and A. Schuhl,{\
J. Phys. }F \textbf{16}, 2099 (1986).

\bibitem{mayeya}  F. M. Mayeya and M. A. Howson, {J. Phys. Condens.
Matter} \textbf{4}, 9355 (1992).


\bibitem{lin}  B. J. F. Lin, M. A. Paalanen, A. C. Gossard, and D. C. Tsui, 
{Phys. Rev.} B \textbf{29}, 927 (1984).

\bibitem{white}  A. E. White, R. C. Dynes, and J. P. Garno, {Phys. Rev%
}. B \textbf{29}, 3694 (1984).

\bibitem{komori}  F. Komori, S. Kobayashi, and W. Sasaki, {J. Phys.
Soc. Jpn}. \textbf{52}, 4306 (1983).

\bibitem{GGZ}  M. E. Gershenson, V. N. Gubankov, and Yu. E. Zhuravlev, 
{Sov. Phys. - JETP} \textbf{58}, 167 (1983).

\bibitem{bishop}  D. J. Bishop, D. C. Tsui, and R. C. Dynes, {Phys.
Rev. Lett}. \textbf{44}, 1153 (1980).

\bibitem{imry}  Z. Ovadyahu and Y. Imry,{\ J. Phys.} C \textbf{16},
L471 (1983).

\bibitem{liu}  Y. Liu, B. Nease, K. A. McGreer, and A. M. Goldman, {%
Europhys. Lett.} \textbf{19}, 409 (1992).

\bibitem{hsu}  S.-Y. Hsu and J. M. Valles,{\ Phys. Rev. Lett.} 
\textbf{74}, 2331 (1995).

\bibitem{dahm}  F. W. Van Keuls, H. Mathur, H. W. Jiang, and A. J. Dahm, 
{Phys. Rev.} B \textbf{56}, 13263 (1997).

\bibitem{thornton}  T. J. Thornton, M. Pepper, H. Ahmed, D. Andrews, and G.
J. Davies,{\ Phys. Rev. Lett}. \textbf{56}, 1198 (1986).

\bibitem{wind}  S. Wind, M. J. Rooks, V. Chandrasekhar, and D.E. Prober,
Phys. Rev. Lett. 57, 633 (1986).

\bibitem{lin}  J. J. Lin and N. Giordano, {Phys. Rev}. B \textbf{33},
1519 (1986).

\bibitem{pooke}  D. M. Pooke, N. Paquin, M. Pepper, and A. Gundlach.{%
\ J. Phys. Condens. Matter} \textbf{1}, 3289 (1989).

\bibitem{echternach}  P. M. Echternach, M. E. Gershenson, H. M. Bozler, A.
L. Bogdanov, and B. Nilsson. {Phys.Rev.} B \textbf{48}, 11516 (1993).

\bibitem{khavin}  Yu. B. Khavin, M. E. Gershenson, and A. L. Bogdanov, 
{Phys. Rev. Lett.} {\bf 81}, 1066 (1998).

\bibitem{gersh}  M. E. Gershenson, Yu. B. Khavin, A. G. Mikhalchuk, H. M.
Bozler, and A. L. Bogdanov, {Phys. Rev. Lett.}\textbf{79}, 725 (1997).

\bibitem{khavin2}  Yu. B. Khavin, M. E. Gershenson, and A. L. Bogdanov, 
{Phys. Rev.} B (1998).

\end{references}
\end{document}